\documentclass[twocolumn, tighten]{aastex6}

\usepackage{graphicx}
\usepackage{subfigure}

\slugcomment{}

\shorttitle{Optical Follow-up to the COBRA Survey}
\shortauthors{Golden-Marx et al.}

\begin{document}

\title{The High-Redshift Clusters Occupied by Bent Radio AGN (COBRA) Survey: Follow-up Optical Imaging}

\author{Emmet Golden-Marx\altaffilmark{1}, E.\,L. Blanton\altaffilmark{1}, R. Paterno-Mahler\altaffilmark{2}, M. Brodwin\altaffilmark{3}, M.\,L.\,N. Ashby\altaffilmark{4}, B.C. Lemaux\altaffilmark{5}, L.M. Lubin\altaffilmark{5}, R.R. Gal\altaffilmark{6}, A.R. Tomczak\altaffilmark{5}} 

\email{emmetgm@bu.edu}

\altaffiltext{1}{Department of Astronomy and The Institute for Astrophysical Research, Boston University, 725 Commonwealth Avenue, Boston, MA 02215, USA}
\altaffiltext{2}{WM Keck Science Center, 925 N. Mills Avenue, Claremont, CA  91711, USA}
\altaffiltext{3}{Department of Physics \& Astronomy, University of Missouri-Kansas City, 5110 Rockhill Road, Kansas City, MO 64110, USA}
\altaffiltext{4}{Center for Astrophysics $|$ Harvard \& Smithsonian, 60 Garden Street, Cambridge, MA 02138, USA}
\altaffiltext{5}{Department of Physics, University of California, Davis, One Shields Avenue, Davis, CA 95616, USA}
\altaffiltext{6}{University of Hawai'i, Institute for Astronomy, 2680 Woodlawn Drive, Honolulu, HI 96822, USA}

\begin{abstract}
Here we present new red sequence overdensity measurements for 77 fields in the high-$z$ Clusters Occupied by Bent Radio AGN (COBRA) survey, based on $r$- and $i$-band imaging taken with Lowell Observatory's Discovery Channel Telescope.  We observe 38 COBRA fields in $r$-band and 90 COBRA fields in $i$-band.  By combining the $r$- and $i$-band photometry with our 3.6\,$\mu$m and 4.5\,$\mu$m $Spitzer$ IRAC observations, we identify 39 red sequence cluster candidates that host a strong overdensity of galaxies when measuring the excess of red sequence galaxies relative to a background field.  We initially treat the radio host as the cluster center and then determine a new cluster center based on the surface density of red sequence sources.  Using our color selection, we identify which COBRA cluster candidates have strong red sequence populations.  By removing foreground and background contaminants, we more securely determine which fields include cluster candidates with a higher significance than our single-band observations.  Additionally, of the 77 fields we analyze with a redshift estimate, 26 include newly estimated photometric redshifts.      
\end{abstract}


\section{Introduction}

Forming at the deepest potential wells in the matter distribution of the early universe, galaxy clusters are a tool for studying galaxy evolution and the cosmological make-up and dark matter distribution of the universe.  Because we search for high-$z$ galaxy clusters, especially at $z$ $>$ 1, we see clusters with a variety of different characteristics, from large-scale cluster mergers, to clusters dominated by populations of star forming and red early-type galaxies \citep[e.g.,][]{McGee2009, Brodwin2013,Cooke2016, Hennig2017}.  These observations yield many exciting questions: how does the dynamical state of the cluster change over time?  At high-$z$, are clusters merging or are they dynamically relaxed?  How do cluster galaxies evolve and merge over time?  Additionally, because clusters are comprised of tens to thousands of similarly aged galaxies, some of which host active galactic nuclei (AGN), observing a sample of clusters across redshift space allows us to gain a better overall understanding of how these cluster galaxies change over time.  

There are thousands of low-$z$ ($z$ $<$ 0.25) spectroscopically-confirmed galaxy clusters, but the number of confirmed high-$z$ clusters diminishes dramatically at $z$ $>$ 1.  In the era of high-$z$ astrophysics, the methods for identifying galaxy clusters have evolved.  Some of the earliest recognized galaxy clusters were identified purely by large optical overdensity surveys \citep{Abell1958}.  Difficulties due to the overcounting of foreground and background galaxies make this method problematic, especially for high-$z$ searches where field contamination is prevalent.  However, infrared (IR) overdensity searches, similar to those performed by \citet{Abell1958}, but also taking advantage of multiple longer wavelength bands, are excellent for identifying high-$z$ clusters.  At $z$ $>$ 0.5, the optical peak of a typical galaxy's spectral energy distribution (SED) shifts from the optical into the IR.  At these wavelengths, the galaxy population should be brighter than the typical foreground contaminants, making it easier to identify high-$z$ galaxies.  By searching the IR sky with satellites such as the $Spitzer$ $Space$ $Telescope$ and the $Wide$-$field$ $Infrared$ $Survey$ $Explorer$ ($WISE$) and taking advantage of multi-wavelength photometry, new, $z$ $>$ 1 clusters and cluster candidates have been identified \citep[e.g.,][]{Eisenhardt2008,Muzzin2009,Wilson2009,Demarco2010,Brodwin2011,Muzzin2013,Stanford2014,Gonzalez2018}.  

With the aid of multi-band photometry, single band optical and IR overdensity searches can be dramatically improved.  One of the most recognizable components of evolved galaxy clusters are dense cores of early-type galaxies.  Since early-type galaxies are characterized by little to no star formation and populations of older, redder stars, these early-type galaxies populate the red sequence \citep{Gladders2000}.  The red sequence, a prominent evolutionary track on color magnitude diagrams (CMDs), includes large populations of similarly colored galaxies across a range of magnitudes and has been studied thoroughly across multiple wavebands in, for example, the Coma cluster \citep[e.g.,][]{Brodwin2006,Eisenhardt2007}.  These red galaxies not only identify clusters, but also can be used to estimate a cluster's redshift.  The similar color of red sequence galaxies yields reliable photometric redshift estimates for galaxy clusters that lack the spectroscopic redshifts of individual  cluster galaxies needed to confirm a cluster \citep{Eisenhardt2008}.  Although \citet{Gladders2000} first identified the red sequence in clusters out to $z$ $\approx$ 1, dense populations of red galaxies exist in clusters out to $z$ $\approx$ 1.5 - 1.8 \citep[e.g.,][]{Andreon2014, Cerulo2016}.  Unlike traditional overdensity searches, which are dramatically hampered by field contamination, the level of red sequence contamination is low \citep{Cerulo2016}. 

Instead of searching for galaxy populations across large survey fields, AGN targeting can also be used to find clusters across a wide range of redshifts with a broad range of masses \citep[e.g.,][]{Galametz2012, Wylezalek2013, Cooke2015, Paterno-Mahler2017}.  The Clusters Around Radio-Loud AGN (CARLA) survey find that radio-loud AGN (RLAGN) at 1.3 $<$ $z$ $<$ 3.2 preferentially live in environments with positive excesses of galaxy counts above an average background in $\sim$ 92$\%$ of their sample and in denser, 2$\sigma$ environments, such as clusters or groups, 55.3$\%$ of the time \citep[e.g.,][]{Wylezalek2013, Wylezalek2014}.  CARLA galaxy cluster candidates include both mature and young galaxy populations, as well \citep{Cooke2016}.  Additionally, \citet{Castignani2014} reinforce this result by showing that $\approx$ 70$\%$ of low-luminosity RLAGN at 1 $<$ $z$ $<$ 2 are found in overdense environments.  Furthermore, using the CARLA survey, \citet{Hatch2014} find a significant correlation between environment and radio power; when looking at similarly massive host galaxies, on average RLAGN reside in more significantly overdense structures than their radio quiet counterparts.  \citet{Hatch2014} also estimate the spatial distribution of RLAGN in the universe and find a similar distribution to protoclusters, possibly implying that all high-$z$ cluster progenitors at 1.3 $<$ $z$ $<$ 3.2 experience an epoch of powerful AGN feedback, although there are clear counterexamples \citep[e.g.,][]{Cucciati2014}.  

The Clusters Occupied by Bent Radio AGN (COBRA) survey was compiled to provide another way to identify galaxy clusters.  Each COBRA bent, double-lobed radio source was observed as a part of the Very Large Array Faint Images of the Radio Sky at Twenty-Centimeters (VLA FIRST) survey \citep{Becker1994}.  The VLA FIRST survey consists of observations of $\approx$ 10,000 square degrees of the radio sky primarily around the North Galactic Cap, but also including the South Galactic Cap.  It was designed to cover approximately the same area of the sky as the Palomar Sky Survey.  This same region of the sky was further observed as part of the Sloan Digital Sky Survey (SDSS).  Each low- and high-$z$ COBRA source was selected from larger samples of visual-bent and auto-bent samples described in \citet{Wing2011}.  The low-$z$ sample was assembled by identifying the host galaxies of each radio source in SDSS $r$-band images to a limit of $m_{r}$ = 22.0\,mag within a search radius set to ensure a 95$\%$ accuracy \citep{Wing2011}, while the high-$z$ sample consists of sources where no SDSS host galaxy was identified.  The visual-bent sample consists of 384 sources identified by eye from a portion ($\approx$ 3000 square degrees) of the VLA FIRST survey to have a bent morphology \citep{Blantonphd}.  Of those, 112 lack an SDSS host.  The auto-bent sample was selected using a pattern recognition program \citep{Proctor2006} to locate three-component radio sources (nominally a core and two lobes).  From the 1546 auto-bent sources selected from the VLA FIRST survey, 541 sources were identified as being at high-$z$.  Both the visual- and auto-selected samples limit the distance between each radio component to be no greater than 60$\arcsec$.  The 653 sources lacking optical hosts were placed in the high-$z$ sample.  Based on the limiting magnitude of SDSS, any source in the high-$z$ sample should be at $z$ $\gtrsim$ 0.5. 

Bent, double-lobed radio sources, such as those in the COBRA survey, offer a highly efficient method for selecting clusters \citep[e.g.,][]{Blanton2000, Blanton2001, Blanton2003, Wing2011}.  The distinctive ``c" shapes of bent radio sources, caused by the ram pressure of radio jets as their host galaxies and the ICM in which they are embedded move relative to each other \citep[e.g.,][]{Owen1976, ODonoghue1993}, act as beacons that reveal otherwise concealed high-$z$ clusters.  The host galaxies of these bent radio sources are often luminous giant elliptical galaxies in the centers of clusters.  In the low-$z$ universe, $z$ $<$ 0.5, 40$\%$ - 80$\%$ of bent sources have been shown to reside in cluster environments depending on different richness cutoffs and samples \citep{Wing2011}.  In clusters where the central galaxies may not have the necessary peculiar velocity to create the observed bent lobes, the bent morphology may result from large-scale fluid flows of the surrounding gas, as can be found in cluster-cluster mergers \citep[e.g.,][]{Burns1990, Roettiger1996, Burns1996, Douglass2011}.  Bent sources can also reside in relatively relaxed systems, where ``sloshing spirals" of hot ICM gas are observed in the X-ray \citep[e.g.,][]{Paterno-Mahler2013}.  Here, the bent shape results from the displacement of the hot ICM in the main galaxy cluster due to an off-axis encounter with a nearby subcluster.  When the subcluster approaches the main cluster, the main cluster's hot ICM is pulled toward it; when the subcluster recedes, the hot ICM falls back, creating a sloshing pattern.  \citet{Wing2013} find that bent, double-lobed radio sources are found in a range of dynamical environments.  Based on an optical substructure analysis of low-$z$ bent sources, \citet{Wing2013} find that bent sources are no more likely to be found in major merging environments than non-bent AGN.  Additionally, when not in clusters, bent radio sources can be found in other environments with surrounding gas, including groups, fossil groups, and large-scale filaments \citep[e.g.,][]{Edwards2010}.  

The first results of the high-$z$ COBRA survey are presented in \citet{Blanton2015} and \citet{Paterno-Mahler2017}.  Of the 653 sources in the original sample, 646 were successfully observed with $Spitzer$ in 3.6\,$\mu$m with the Infrared Array Camera (IRAC; \citealp{Fazio2004}).  Additionally, 135 of these sources were observed with $Spitzer$ in 4.5\,$\mu$m.  These observations were designed such that our exposure times led to a S/N of 5.0 in the 3.6\,$\mu$m observations.  Although the sample was designed to exclude sources at $z$ $<$ 0.5, when we carefully compare our $Spitzer$ hosts with SDSS, we find that 119 of the 646 high-$z$ sources have SDSS photometric redshift estimates at $z$ $<$ 0.5.  The low-$z$ objects result from uncertainty as to which radio component is the core, thus resulting in previous misidentification of the lack of a radio host in the early automated analysis of our sources, and a range of absolute magnitudes for the host galaxies.  Of the 646 sources, there are 41 quasars, each with SDSS spectroscopic redshifts, ranging from 0.708 $\leq$ $z$ $\leq$ 2.943.  

\citet{Paterno-Mahler2017} find that 190 of the high-$z$ COBRA candidates are in cluster environments, defined as a 2$\sigma$ overdensity in the number of galaxies found within either a 1$\arcmin$ or 2$\arcmin$ region centered on each radio source when compared to galaxy counts in the $Spitzer$ Ultra Deep Survey (SpUDs, PI: J. Dunlop) field.  Overall, 530 of the total 646 sources ($\approx$ 82$\%$) are in regions that have a higher density of sources than the mean of the SpUDs field.  \citet{Paterno-Mahler2017} additionally directly compare the COBRA selection methodology to that of the CARLA sample of RLAGN \citep{Wylezalek2013}.  In doing a direct comparison, \citet{Paterno-Mahler2017} find that COBRA and CARLA identify regions with a similar overdensity, although CARLA identifies a higher percentage of rich clusters.  The differences between the two surveys when compared at the same depth (29$\%$ for COBRA versus 44$\%$ for CARLA) may result from bent, double-lobed radio sources being found in a larger range of cluster and group environments than the CARLA sources, which have a higher average radio power.  

In this paper, we present results from optical follow-up imaging of the high-$z$ COBRA survey.  A summary of our observations and data analysis are presented in \textsection{2}.   Our photometric redshift estimates are discussed in \textsection{3}.  Our initial measurements of red sequence galaxy overdensity are presented in \textsection{4}.  Further analysis of the red sequence overdensity is presented in \textsection{5}.   Our results are discussed in \textsection{6} and our conclusions are presented in \textsection{7}.  We assume a $\Lambda$CDM cosmology with H$_{0}$ = 70 km s$^{-1}$ Mpc$^{-1}$, $\Omega$$_{\Lambda}$ = 0.7, and $\Omega$$_{M}$ = 0.3.  All magnitudes presented are AB magnitudes unless otherwise specified.  

\section{Observations}

The observations for the high-$z$ COBRA survey include IR observations taken with the $Spitzer$ $Space$ $Telescope$ via a $Spitzer$ Snapshot Proposal (PID 80161, PI Blanton) using IRAC and optical follow-up observations taken at Lowell Observatory's 4.3\,m Discovery Channel Telescope (DCT) using the Large Monolithic Imager (LMI; \citealp{Massey2013}).  

\subsection{Optical Imaging}
\begin{figure}
\centering
\figurenum{1}
\epsscale{1}
\includegraphics[scale=0.5,trim={0.45in 0.1in 0.0in 0.4in},clip=True]{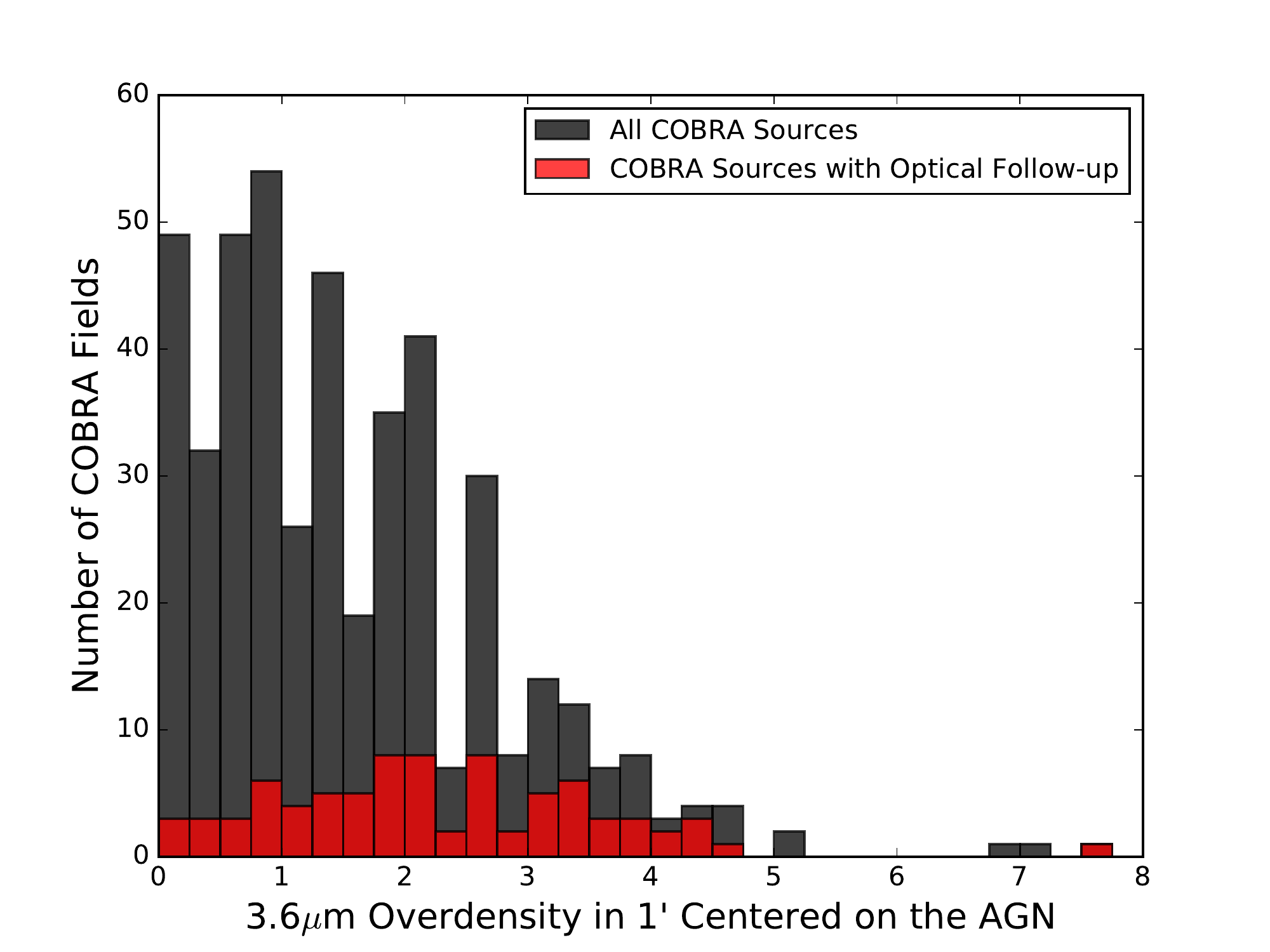}

\caption{Histogram showing the overdensity of all sources within 1$\arcmin$ of the radio source in 3.6\,$\mu$m for all COBRA sources (shown in black) relative to the optical follow-up sample (shown in red).  These measurements are based on the values reported in \citet{Paterno-Mahler2017}.\label{Fig:Allsources}}
\end{figure}

The optical follow-up for the high-$z$ COBRA survey was performed at Lowell Observatory's 4.3m DCT using the 12$\farcm$3 $\times$ 12$\farcm$3 field-of-view (FOV) LMI.  The LMI has a 0$\farcs$24 per pixel scale (when binned by a factor of two, typical for LMI).  Fields were observed on 26.5 nights over a four year span from 2013 through 2017.  The typical seeing was $\approx$ 0$\farcs$8, but ranged from 0$\farcs$6 to 1$\farcs$9.  We observed 38 fields in the SDSS $r$-band and 90 fields in the SDSS $i$-band.  Each field observed in $r$-band was also observed in $i$-band (see Table \ref{tb:1} for information on the observations).  Since our optical follow-up began before the analysis presented in \citet{Paterno-Mahler2017} was completed, the earlier DCT observed fields were chosen based on radio observations and morphology.  Specifically, we targeted fields that included either particularly distinct bent radio sources or quasars.  More recently observed fields were additionally chosen based on strong IR overdensities ($>$ 2$\sigma$) determined in \citet{Paterno-Mahler2017}, although we avoid fields with known low-$z$ clusters in them.  The 3.6\,$\mu$m overdensity in 1$\arcmin$ of all COBRA sources relative to the sources with optical follow-up is shown in Figure\,\ref{Fig:Allsources}.  As can be seen, we have follow-up observations of a larger fraction of fields with stronger overdensities.  Of the 90 fields observed, 57 are classified as clusters in \citet{Paterno-Mahler2017}.  These fields are noted in Table\,\ref{tb:2}.  Examples of 2$\arcmin$ $\times$ 2$\arcmin$ cutouts of COBRA fields observed at the DCT in $r$- and $i$-band are shown in Figure\,\ref{Fig:6}.  

The majority of our observations are either 3$\times$600\,s (3$\times$900\,s) with the average depth reached being $m_{r}$ = 24.5\,mag ($m_{r}$ = 25.0\,mag for the longer exposure) and $m_{i}$ = 24.0\,mag ($m_{i}$ = 24.5\,mag), respectively.  The average magnitude of an L* galaxy, which we model using EzGal \citep{Mancone2012} and discuss fully in \textsection{3}, at $z$ = 1 is 24.6\,mag in $r$-band and 23.5\,mag in $i$-band.  Our deeper $r$-band imaging is sufficiently sensitive to detect objects as faint as L* + 0.5\,mag, and our deeper $i$-band imaging reaches L* + 1.0\,mag.  Of the 38 $r$-band targets, 6 host quasars, while 20 of the 90 $i$-band targets host quasars.  The quasars in our sample have been labeled in Table\,\ref{tb:1}.

\subsection{Data Reduction}

\subsubsection{DCT Observations}

We reduce our DCT data using standard IRAF \citep{Tody1993} reduction methods.  We correct each object image for bias and flat fields.  Typically, twilight flat fields are used from the night of the observation.  In some instances, we use either dome flats or twilight flats from the nearest possible night if no flats are available from the night of the observation.  Because we dither between each exposure, we align the images before combining them.  The final image for each field is a combination of two to four frames using an exposure time weighted average.  The world coordinate system (wcs) solutions are calculated for the combined images using positions of SDSS objects in the field.

\begin{figure*}
\begin{center}
\figurenum{2}
\subfigure{\includegraphics[scale=0.7,trim={1.65in 2.8in 2.0in 3.3in},clip=true]{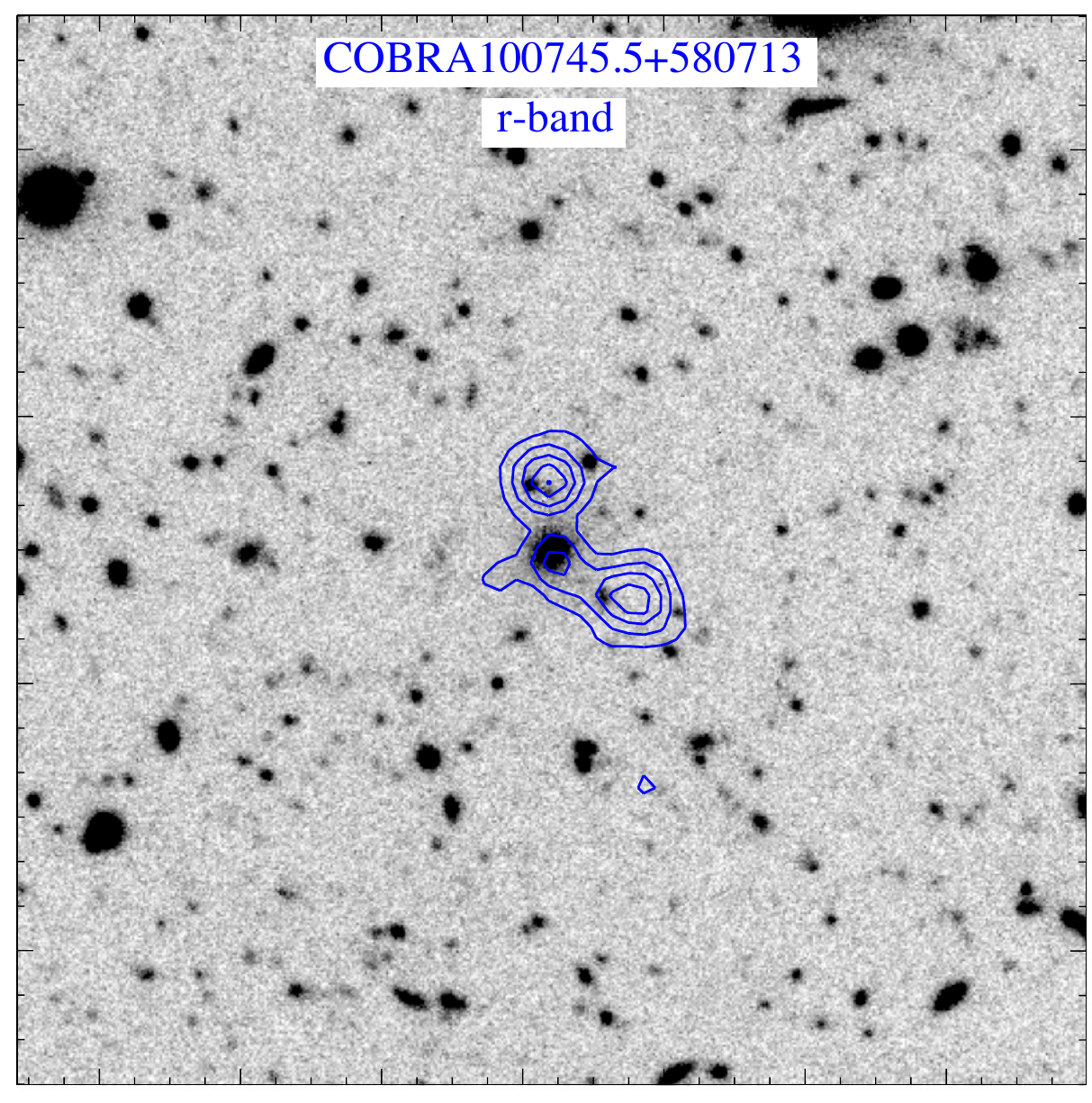}}
\subfigure{\includegraphics[scale=0.7,trim={1.65in 2.8in 2.0in 3.3in},clip=true]{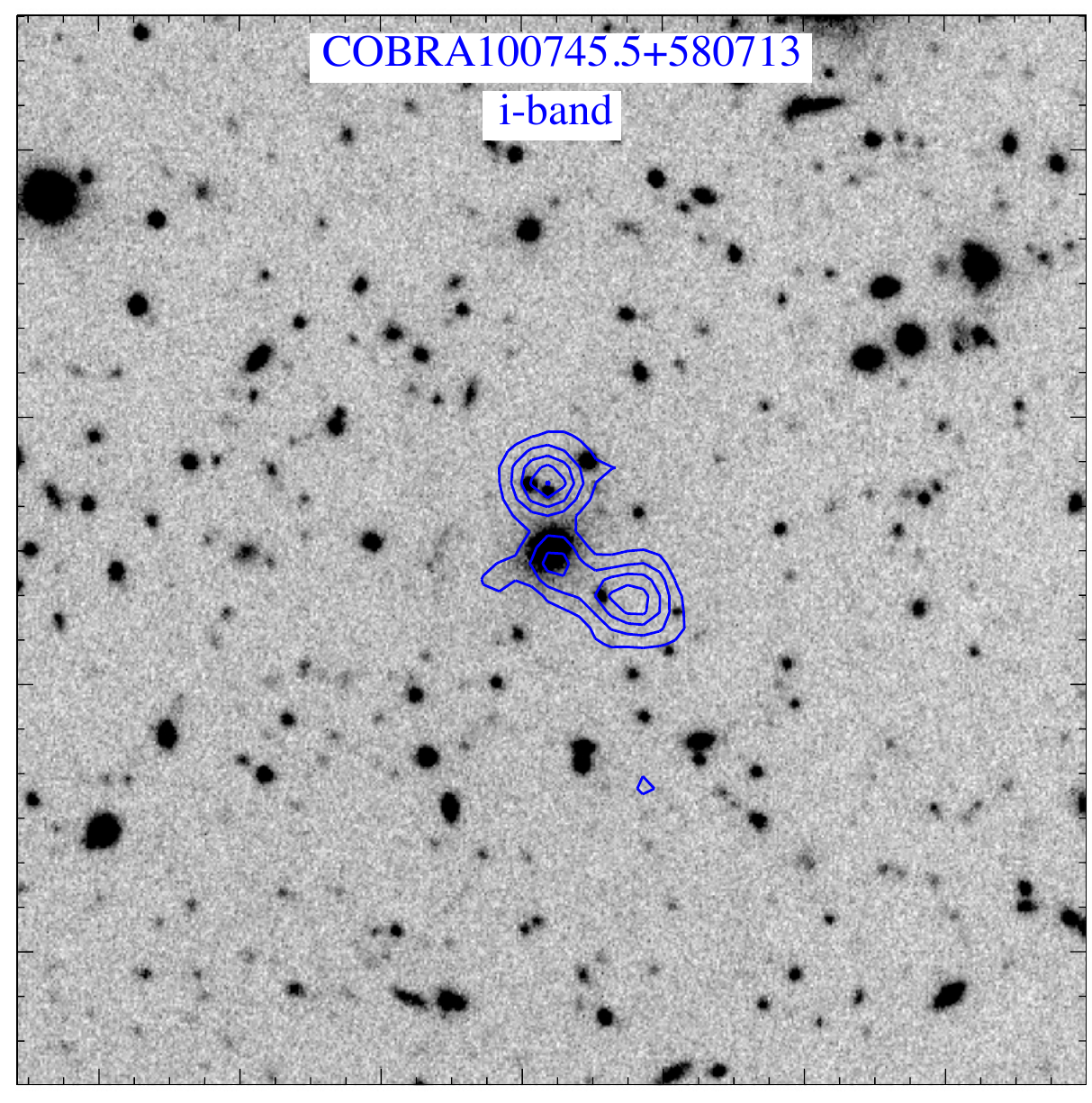}}
\subfigure{\includegraphics[scale=0.7,trim={1.65in 2.8in 2.0in 3.3in},clip=true]{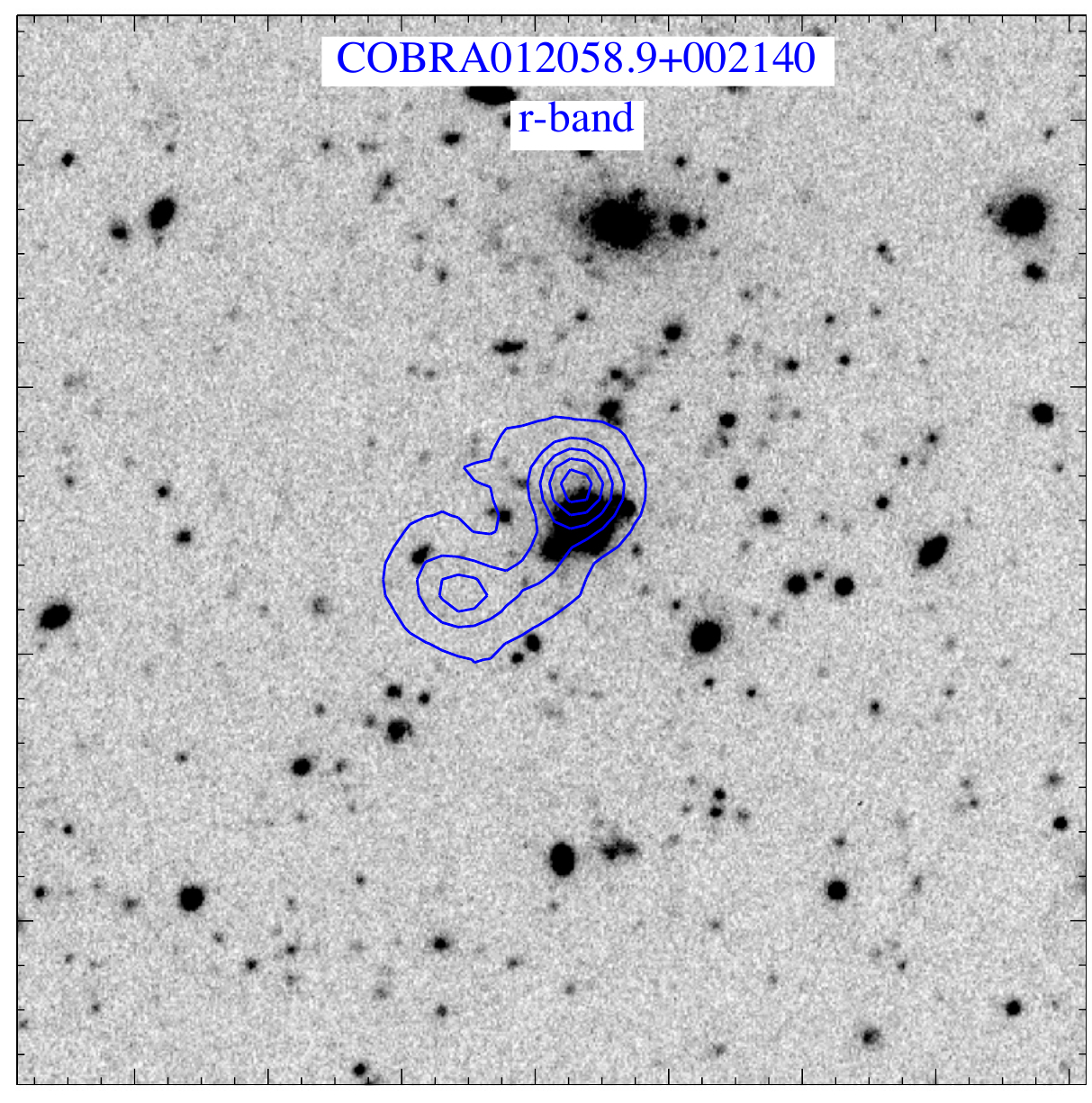}}
\subfigure{\includegraphics[scale=0.7,trim={1.65in 2.8in 2.0in 3.3in},clip=true]{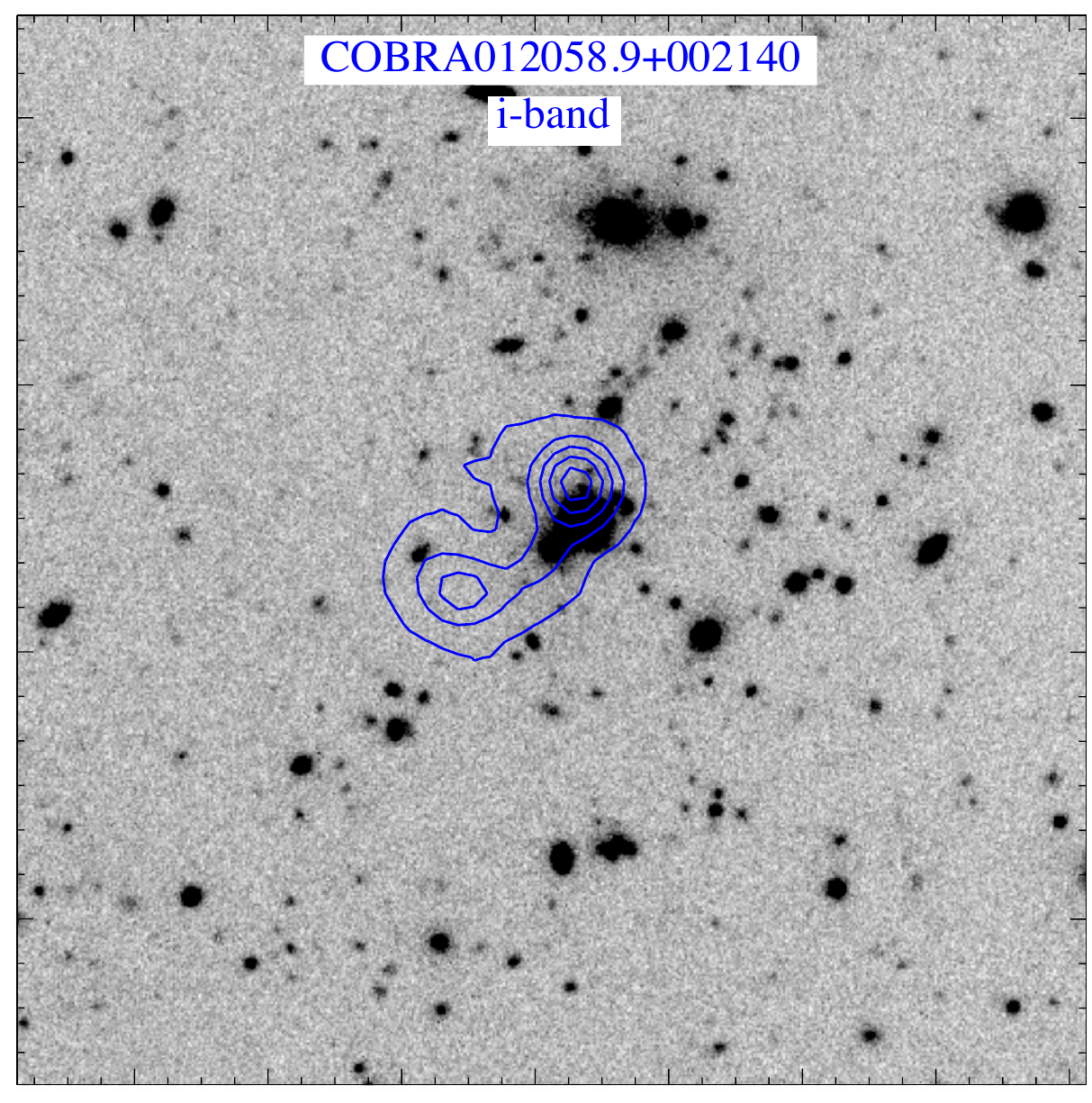}}

\caption{DCT $r$- and $i$-band 2\arcmin$\times$2\arcmin cutouts of COBRA fields.  The $r$-band images are in the left column, while $i$-band images are in the right column.  Each image is centered on the radio host.  The blue contours reflect the 20\,cm VLA FIRST imaging.  COBRA100745.5+580713 is a strong cluster canetectionidate (3.1$\sigma$ in $i - [3.6]$ and 6.5$\sigma$ in $r$ $-$ $i$ in 1$\arcmin$) with a spectroscopically confirmed host galaxy at $z$ = 0.656.  COBRA012058.9+002140 is another cluster candidate (2.5$\sigma$ in $i - [3.6]$ and 5.9$\sigma$ in $r$ $-$ $i$ in 1$\arcmin$) at $z$ $\approx$ 0.75.    These are typical examples; overdensity measurements for all COBRA fields are given in Table\,\ref{tb:2}.}
\label{Fig:6}
\end{center}
\end{figure*}

\begin{figure*}
\begin{center}
\figurenum{2}
\subfigure{\includegraphics[scale=0.7,trim={1.65in 2.8in 2.0in 3.3in},clip=true]{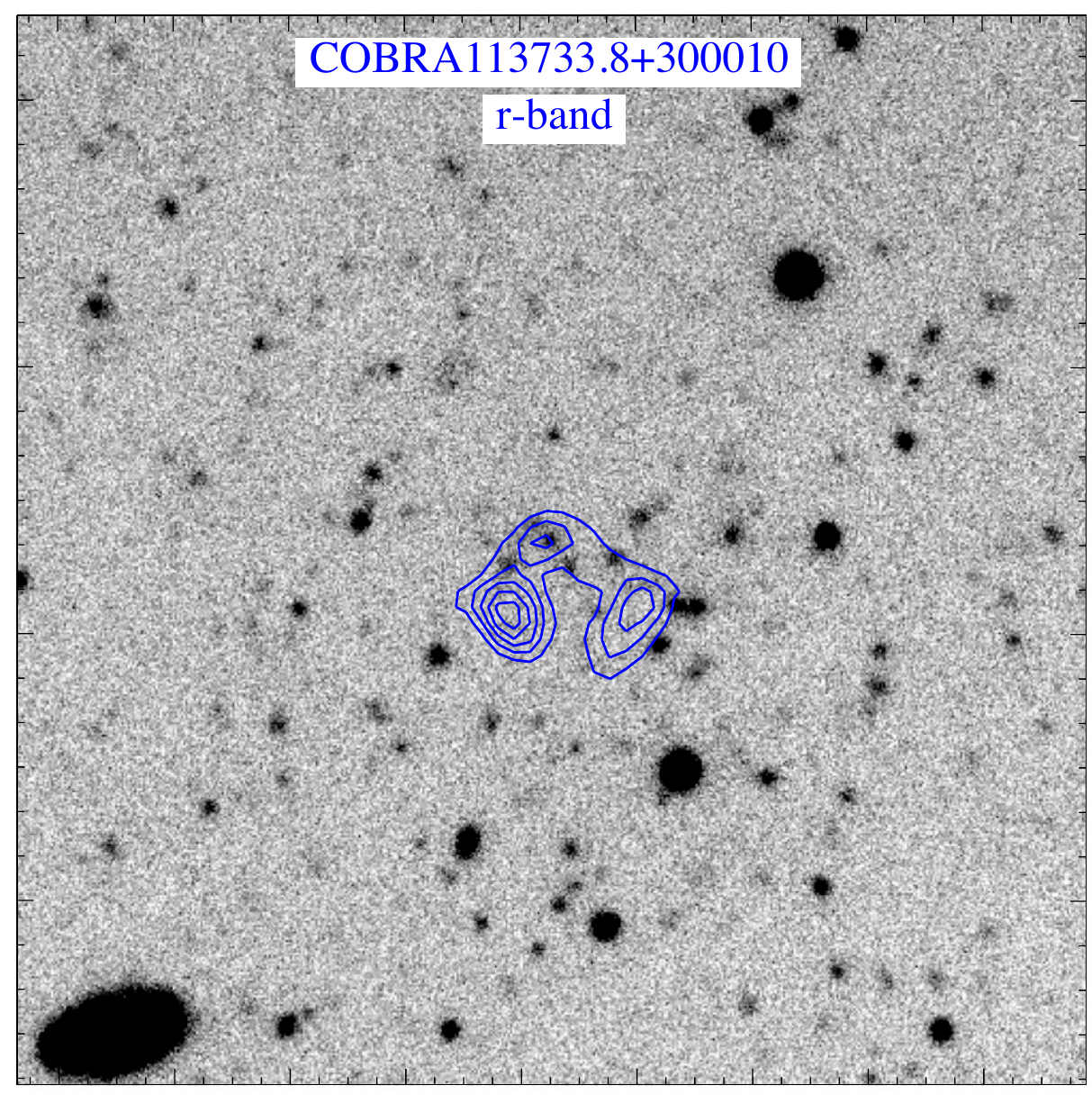}}
\subfigure{\includegraphics[scale=0.7,trim={1.65in 2.8in 2.0in 3.3in},clip=true]{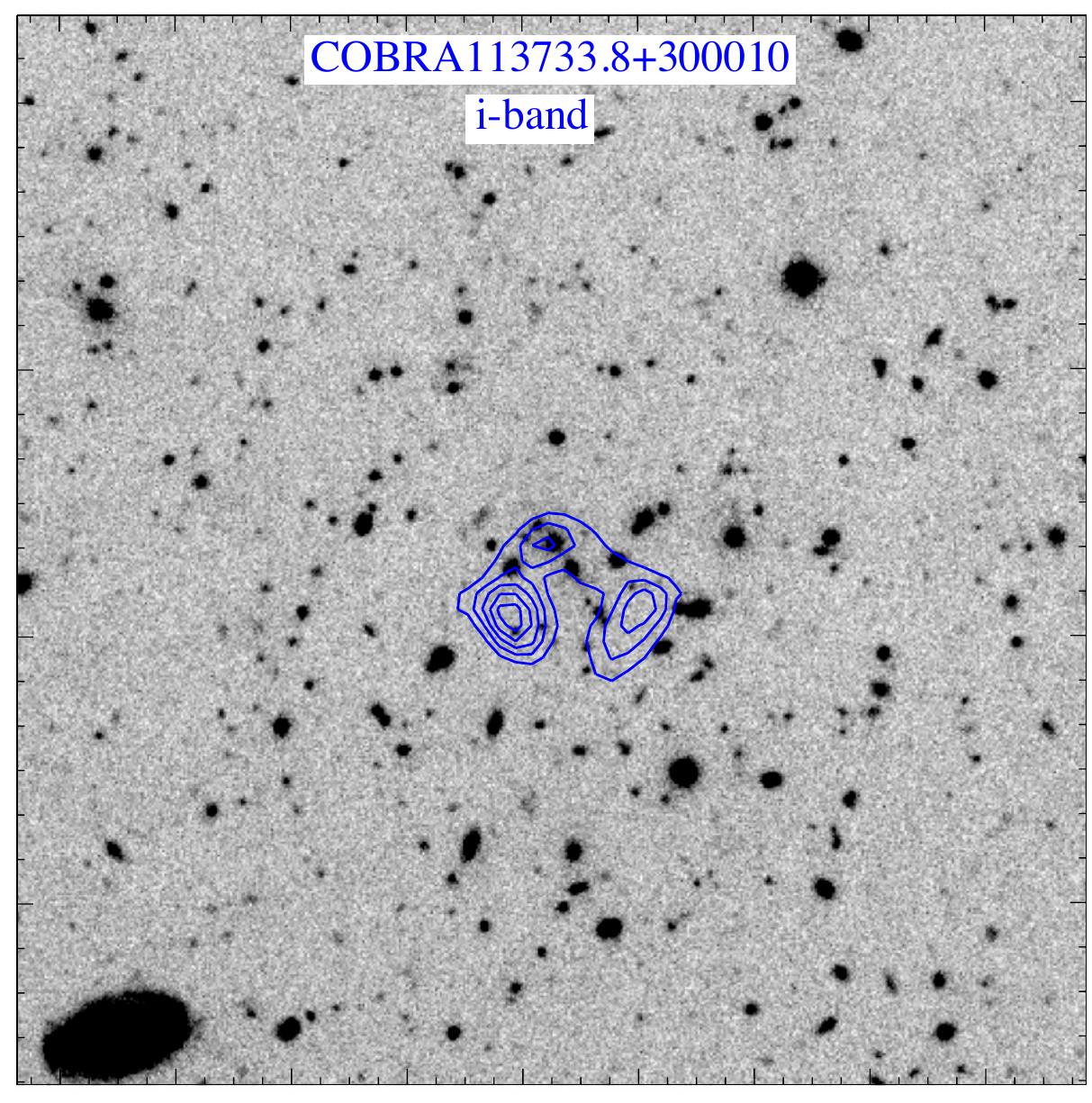}}
\subfigure{\includegraphics[scale=0.7,trim={1.65in 2.8in 2.0in 3.3in},clip=true]{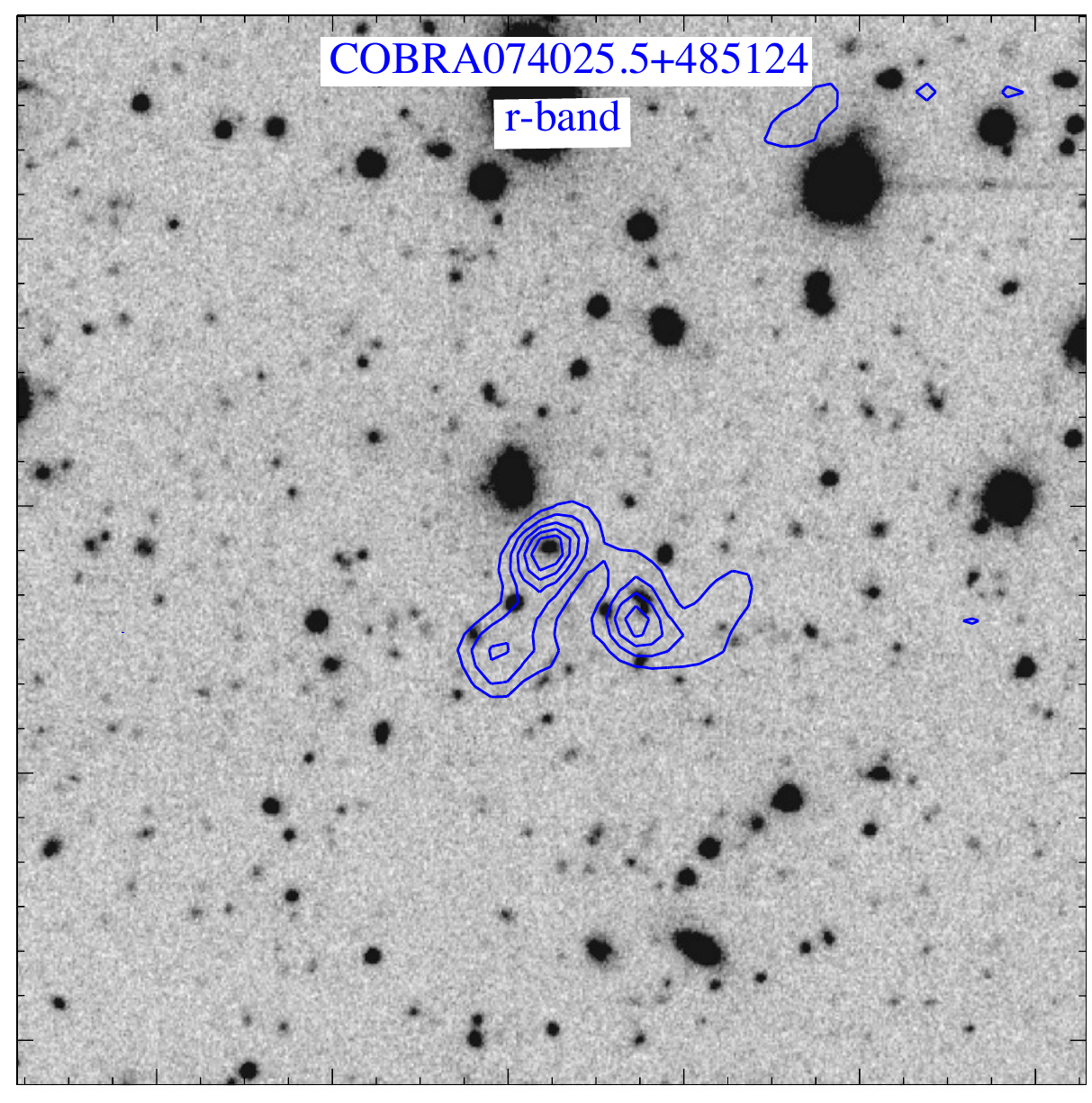}}
\subfigure{\includegraphics[scale=0.7,trim={1.65in 2.8in 2.0in 3.3in},clip=true]{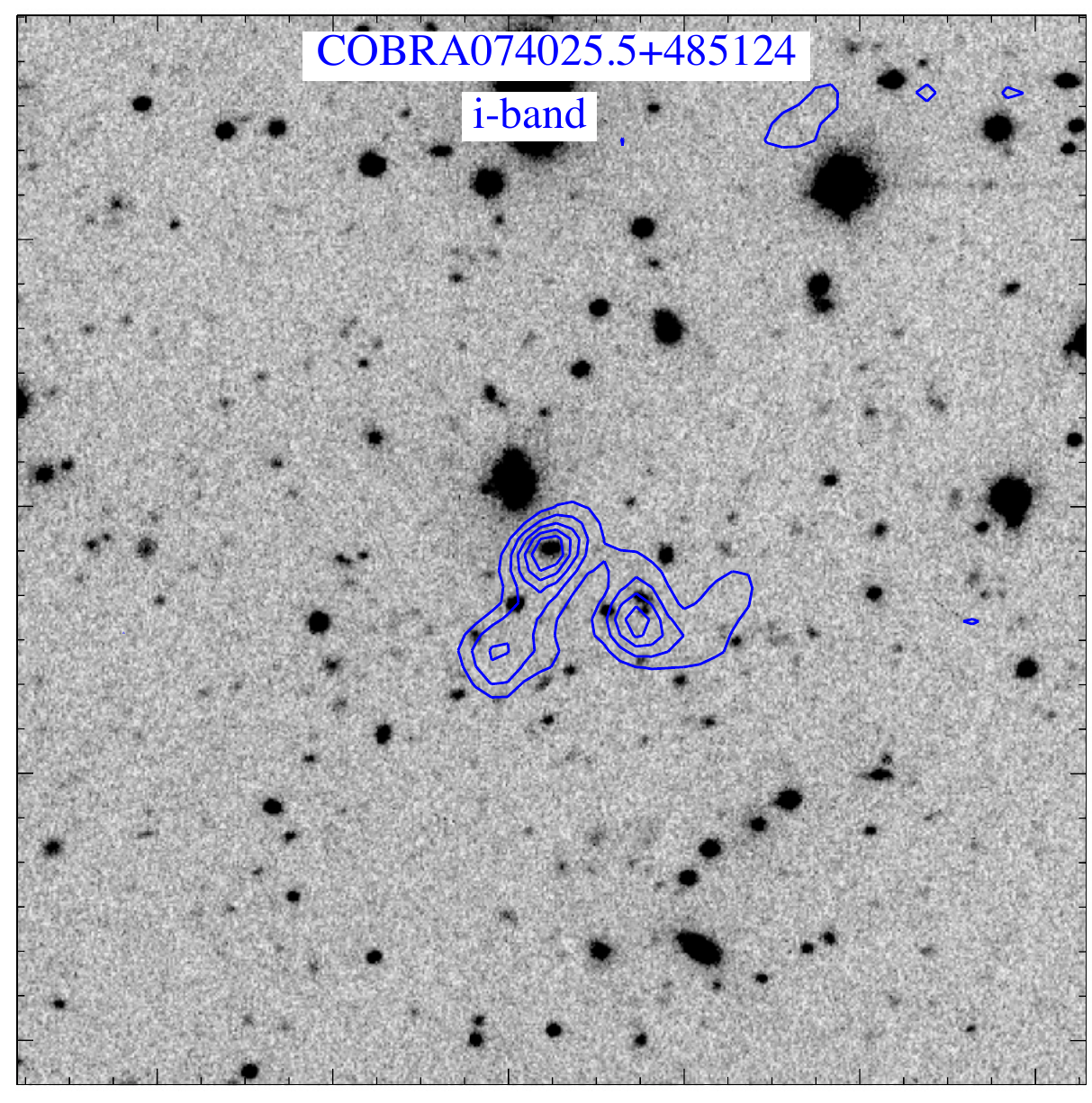}}

\caption{(continued) COBRA113733.8+300010 is the only COBRA cluster candidate with spectroscopic redshift confirmation of several cluster galaxies \citep{Blanton2003}.  It shows one of the stronger overdensity measurements (4.6$\sigma$ in $i - [3.6]$ and 3.7$\sigma$ in $r$ $-$ $i$ in 1$\arcmin$) and is at $z$=0.96.   COBRA074025.5+485124 is a high-$z$ cluster candidate in the magnitude-limited sample (3.6$\sigma$ in $i - [3.6]$ in 1$\arcmin$) at $z$ $\approx$ 1.10.  }
\end{center}
\end{figure*}

\begin{figure*}
\begin{center}
\figurenum{2}
\subfigure{\includegraphics[scale=0.7,trim={1.65in 2.8in 2.0in 3.3in},clip=true]{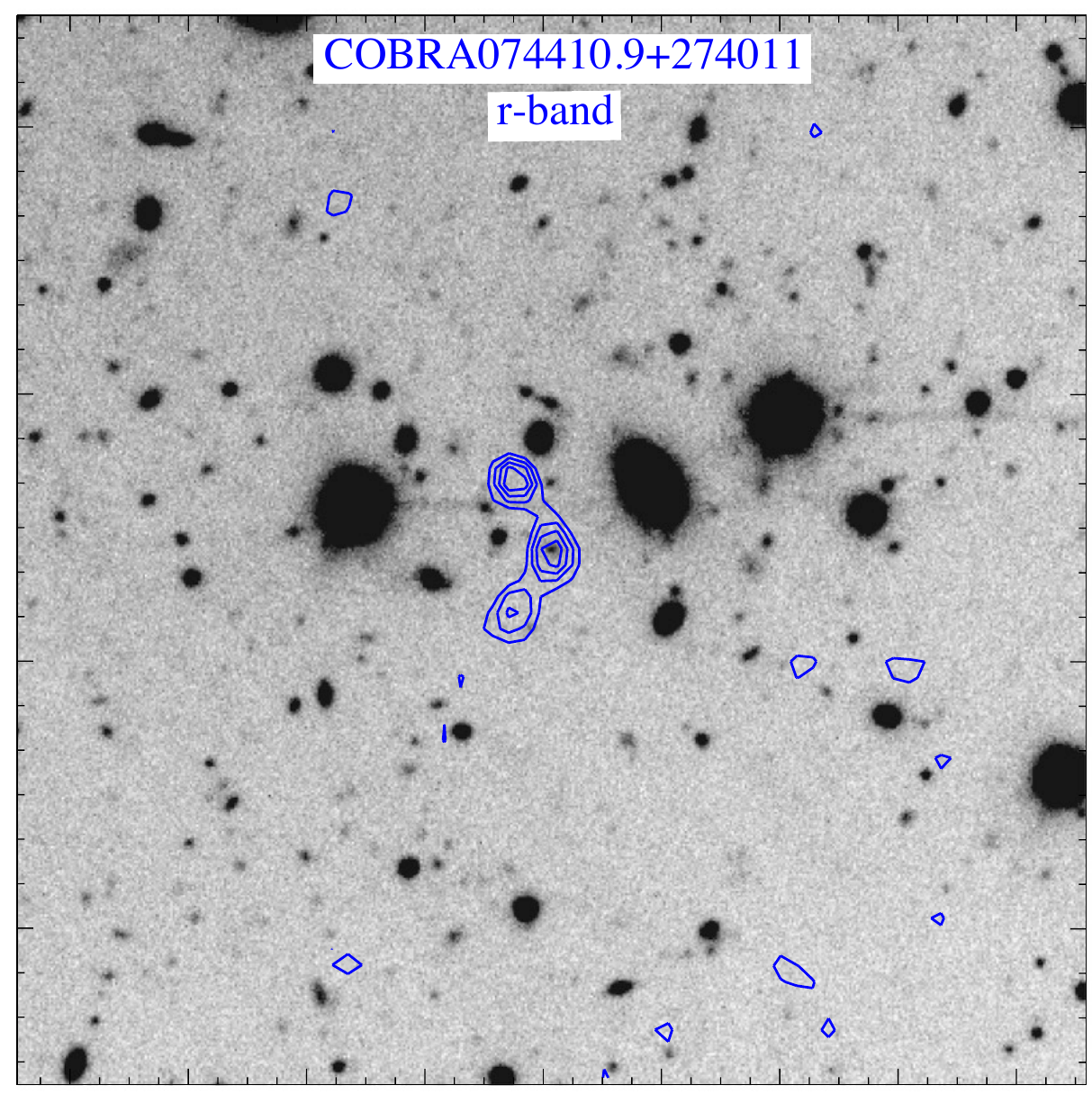}}
\subfigure{\includegraphics[scale=0.7,trim={1.65in 2.8in 2.0in 3.3in},clip=true]{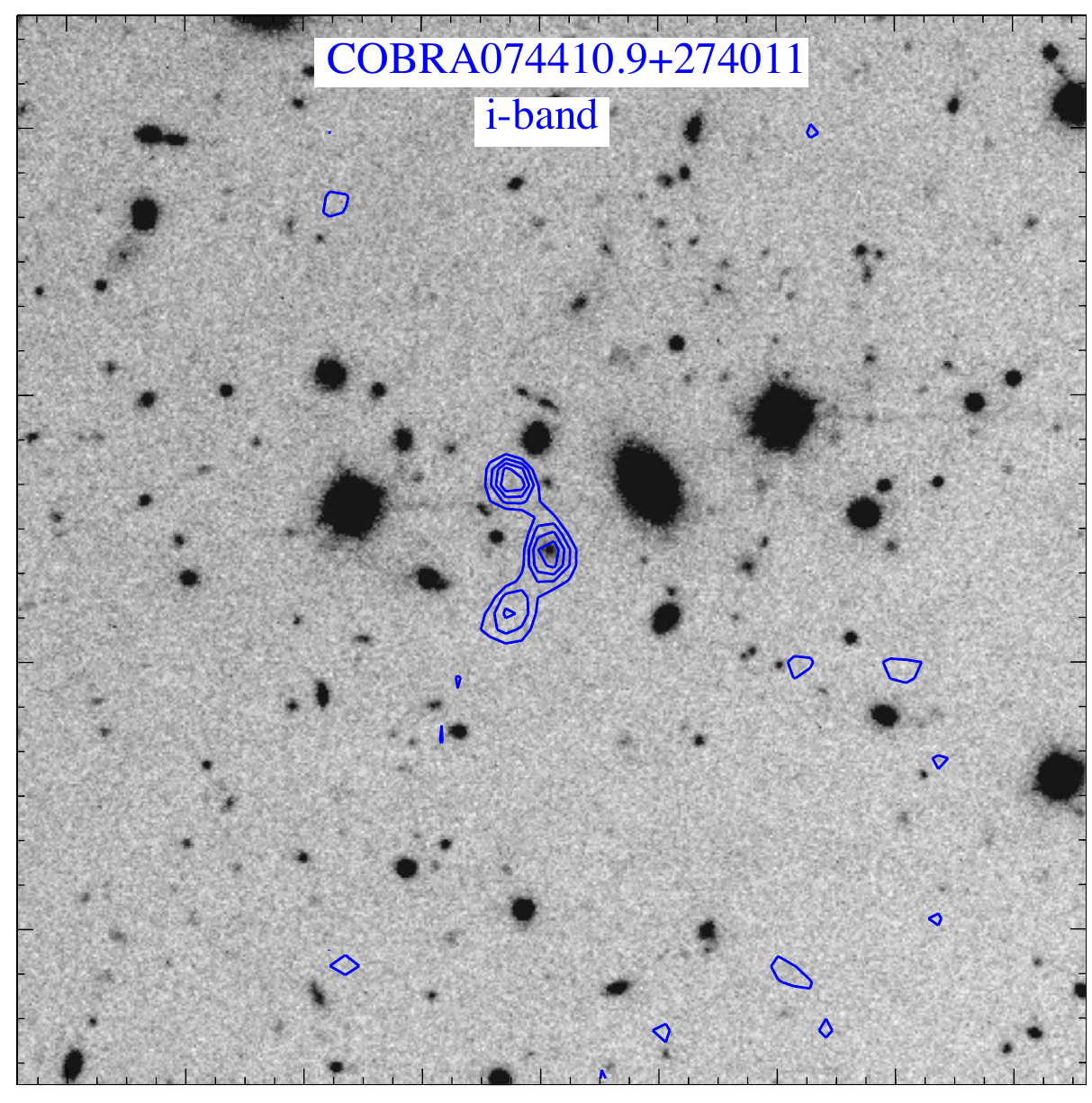}}
\subfigure{\includegraphics[scale=0.7,trim={1.65in 2.8in 2.0in 3.3in},clip=true]{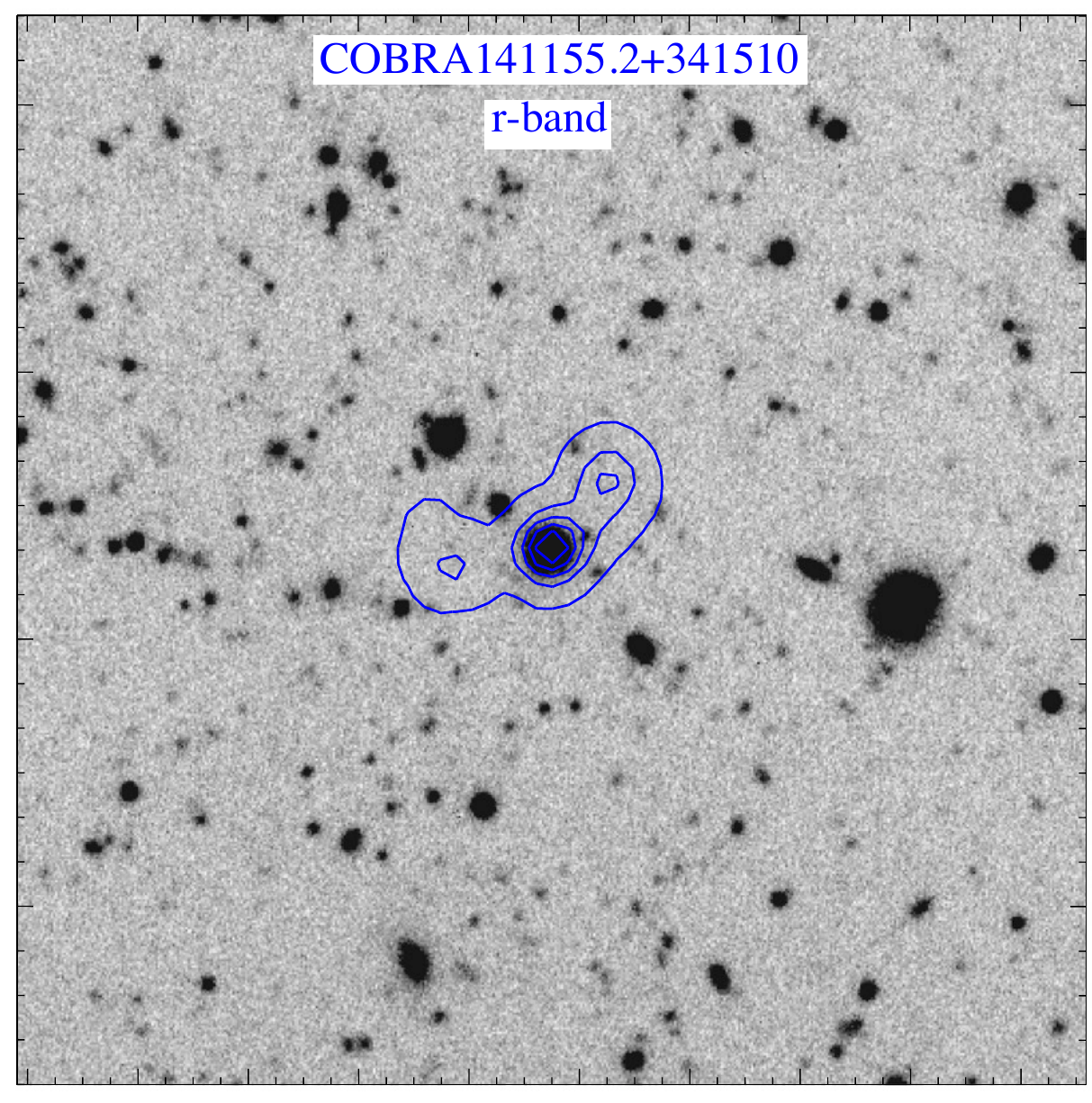}}
\subfigure{\includegraphics[scale=0.7,trim={1.65in 2.8in 2.0in 3.3in},clip=true]{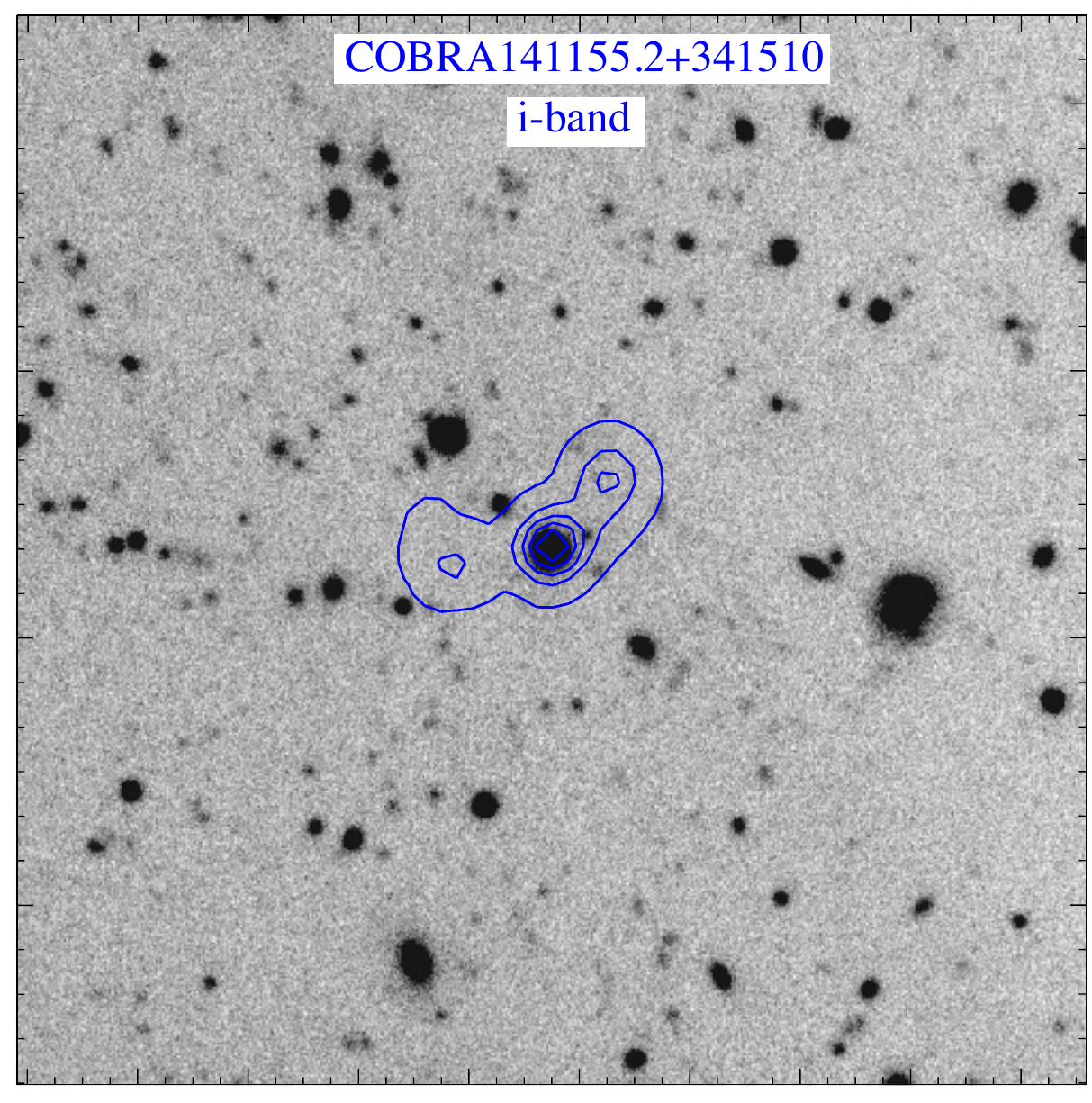}}
\caption{(continued)   COBRA074410.9+274011 is a high-$z$ cluster candidate in the magnitude-limited sample (4.0$\sigma$ in $i - [3.6]$ in 1$\arcmin$) at $z$ $\approx$ 1.30.  COBRA141155.2+341510 is an IRAC cluster candidate (2.7$\sigma$ in 3.6\,$\mu$m in 1$\arcmin$ in \citep{Paterno-Mahler2017}) that hosts a quasar at $z$ = 1.818.  Because of its high-$z$ nature, we are unable to measure the significance of the measurement with our optical observations, but can with our dual-band $Spitzer$ observations (0.6$\sigma$ measurement in $[3.6] - [4.5]$ in 1$\arcmin$ when centered on the radio source, but a 2.2$\sigma$ measurement when centered on the peak of the distribution of red sources).}
\end{center}
\end{figure*}

\begin{figure}
\figurenum{3}

\includegraphics[scale=0.48,trim={0.35in 0in 0in 0.2in},clip=true]{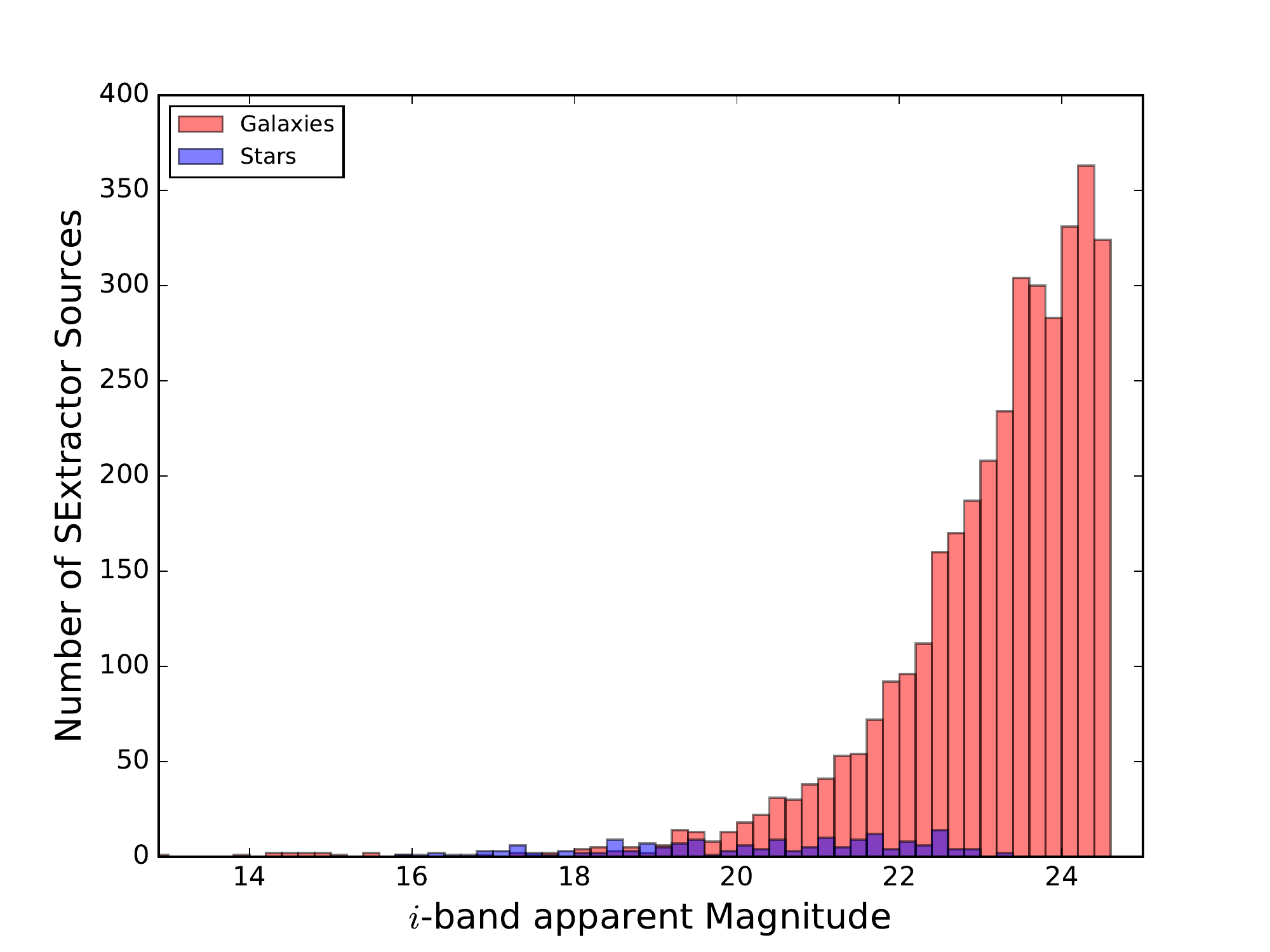}
\caption{A histogram showing the sources identified as stars and galaxies in COBRA012058.9+002140.  The plot divides stars and galaxies at a CLASS$\_$STAR value of 0.95.  The red histograms show all sources identified as galaxies, while the blue show all sources identified as stars.} \label{Fig:starcheck}  

\end{figure}

We perform photometry on each field with SExtractor \citep{Bertin1996} in single-image mode.  We apply the tophat$\_$2.0$\_$3x3.conv filter, a 3$\times$3 convolution mask of a top-hat PSF with a diameter of 2 pixels, to better detect low surface brightness sources.  To deblend our sources, we set the minimum contrast to deblend two objects, DEBLEND$\_$MINCONT, to 0.0001 to accurately separate the closely spaced sources common in our denser fields.  We use the SDSS catalog to determine a unique zeropoint for each field by comparing the magnitudes of non-saturated stars in our fields to the SDSS magnitudes, allowing us to determine standard magnitudes for nights that are not photometric.  Our field star magnitudes are measured using the MAG$\_$AUTO measurement in SExtractor that uses a flexible elliptical aperture to capture the flux surrounding each object.  We determine color corrections for the magnitudes, and find that these corrections are $\approx$ 0.02\,mag, and are therefore neglected.  Additionally, we use the CLASS$\_$STAR parameter in SExtractor to examine our fields for stellar contaminants.  The parameter uses the measured SEEING$\_$FWHM to differentiate between point sources and extended sources and identifies galaxies as closer to 0.0 and stars as closer to 1.0.  \citet{Melchior2015} identify galaxies in their sample of four Dark Energy Survey galaxy clusters as having a CLASS$\_$STAR value $\le$ 0.95 in their approximately 1$\farcm$1 seeing $r$- and $i$-band observations.  We follow this methodology for our COBRA sources (see Figure\,\ref{Fig:starcheck}).  We find a strong excess of extended sources out to $\approx$ 23\,mag in $i$-band.  At this magnitude, the star/galaxy separation appears to break down because there are obviously stars at fainter magnitudes (even though the classifier doesn't detect them).  Because of this breakdown and the very small contribution of stars to the overall object counts in general, we do not exclude sources based on morphology (star vs galaxy).  The exception may be at brighter magnitudes, where the relative contribution of stars is more significant.  However, the absolute number is small and the impact on our statistics is minimal.

\subsubsection{$Spitzer$ Observations}

We reduce the $Spitzer$ observations following the procedures described in \citet{Paterno-Mahler2017}.  For photometry, we use SExtractor in single-image mode for all fields, unlike \citet{Paterno-Mahler2017}, with the same parameters as the optical images for each field (described in \textsection{2.2.1}).  We use single-image mode to best capture the full extent of each source for our color analysis since the pixel scale of IRAC and the LMI differs by over a factor of two.  The magnitude limit of the $Spitzer$ observations is 21.4\,mag in both 3.6\,$\mu$m and 4.5\,$\mu$m.  The average magnitude of an EzGal modeled L* galaxy at $z$ = 1 is 20.3\,mag in 3.6\,$\mu$m and 20.6\,mag in 4.5\,$\mu$m.  Thus we can detect $\approx$ 1.0\,mag fainter than an L* galaxy in 3.6\,$\mu$m and $\approx$ 0.8\,mag fainter than an L* galaxy in 4.5\,$\mu$m.  For our analysis, we match our output $i$-band catalogues with our $Spitzer$ 3.6\,$\mu$m catalogues to measure the colors of the surrounding galaxies.  This methodology is described in \textsection{4}. 

\section{Photometric Redshift Estimates}

Our new optical photometry allows us to refine the first-order redshift estimates presented in \citet{Paterno-Mahler2017}.  The photometric redshift of each field is estimated based on the color of the AGN host galaxy identified in \citet{Paterno-Mahler2017}.  In both \citet{Paterno-Mahler2017} and this work, we assume that each bent radio source is hosted by a normal early-type galaxy with an SED not strongly affected by the AGN, as is typical of these sources \citep{Wing2011}.  Since quasars are generally significantly bluer than typical early-type galaxies at these redshifts and all quasars in our sample have spectroscopic redshifts from SDSS, we remove them from the sample for estimating photometric redshifts.  To transform our host galaxy colors into photometric redshift estimates, we use EzGal \citep{Mancone2012}, a galaxy SED modeling program, to model a typical early-type galaxy with no AGN component.  We use a standard $\Lambda$CDM cosmology, a \citet{Bruzual2003} Stellar Population Synthesis (SPS) model, a Salpeter initial mass function (IMF), a single burst of star formation at a formation redshift of $z_{f}$ = 5.0, and normalize these magnitudes to the location of the knee of the luminosity function of the Coma Cluster\footnote{The values of our Coma normalization are standard input for EzGal and can be found at http://www.baryons.org/ezgal/model.php.}.  Throughout this paper, when we refer to an m* galaxy, we are referring to the magnitude of an m* galaxy calculated using EzGal at various redshifts based on our star formation history.  It should be noted that these values will differ slightly from the measured value as we are not accounting for galaxy mergers that are prevalent in cluster environments.  

Some of the initial conditions for EzGal differ from \citet{Paterno-Mahler2017} and result in different photo-$z$ estimates.  Specifically, the change in formation redshift from $z_{f}$ = 3.0 to $z_{f}$ = 5.0 better accounts for the existence of massive field galaxies at $z$ $>$ 5.0 and the populations of massive galaxies with AGN and large star formation rates seen at $z$ $\approx$ 2.0 in clusters \citep[e.g.,][]{Stark2016,Coogan2018,Shimakawa2018}.  Additionally, the switch from $z_{f}$ = 3.0 to $z_{f}$ = 5.0 allows us to better account for galaxies of our predicted color and redshift evolving onto the red sequence.  However, the difference in formation redshifts only affects the highest redshift sources in our sample.  At $z$ $<$ 1.2, the redshift estimates are unaffected.  We experimented with different IMFs and find identical results.  

\begin{figure}
\figurenum{4}
\epsscale{1.15}
\plotone{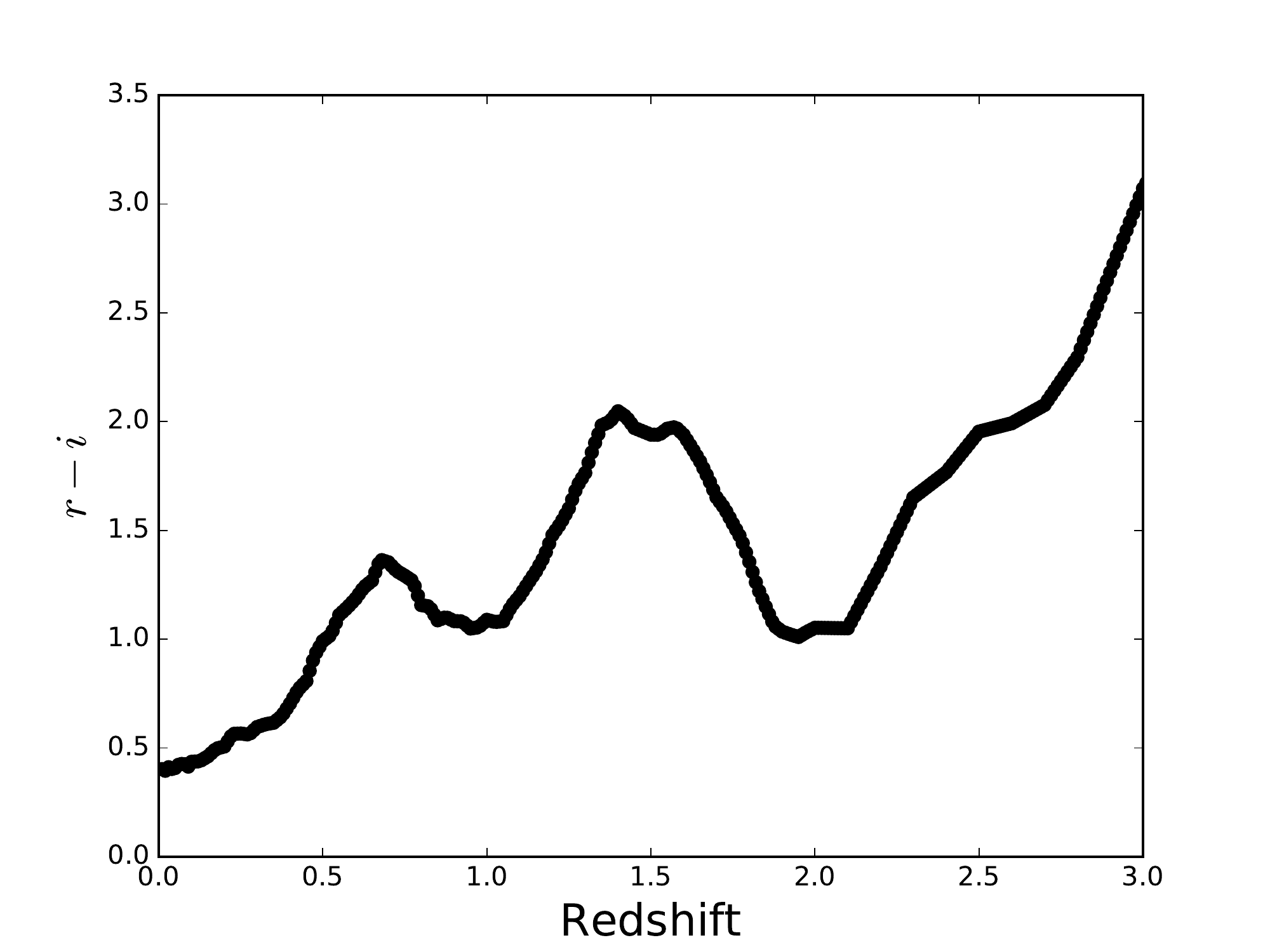}

\epsscale{1.15}
\plotone{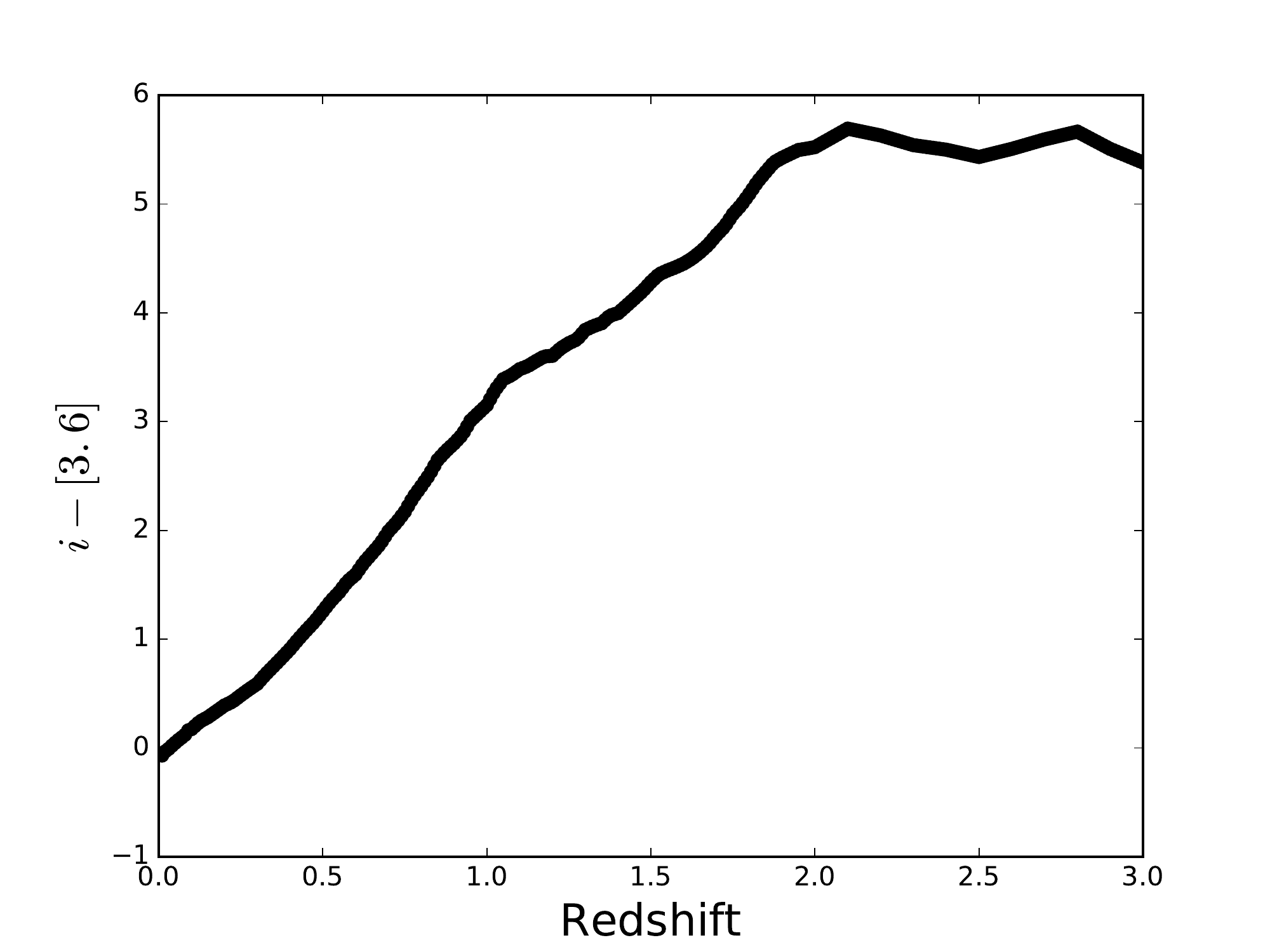}

\epsscale{1.15}
\plotone{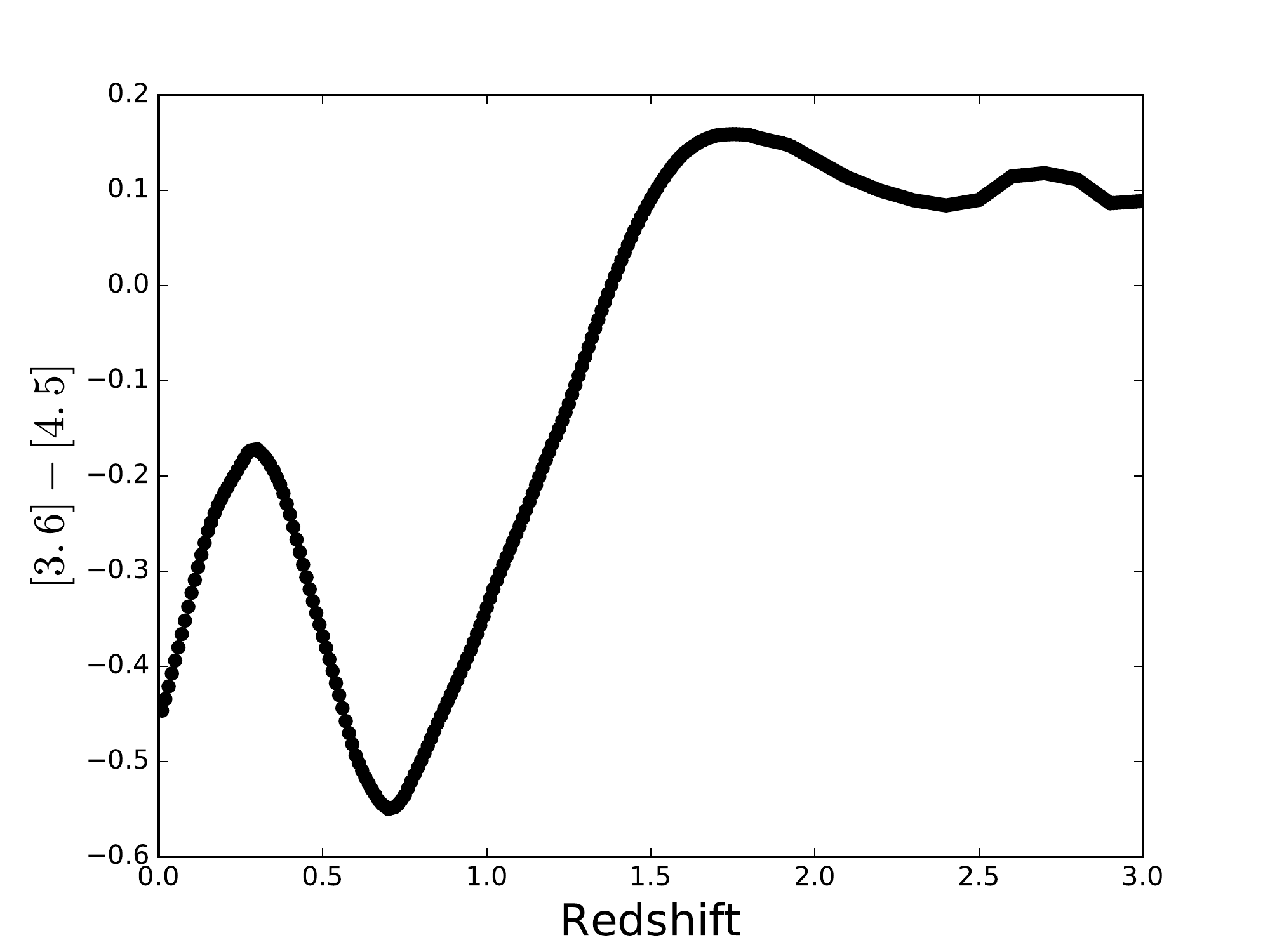}

\caption{EzGal \citep{Mancone2012} modeled early-type galaxy colors as a function of redshift in three colors.  Each plot uses standard $\Lambda$CDM Cosmology, a \citet{Bruzual2003} SPS Model, a Salpeter IMF, a single burst of star formation at $\sl{z}_{f}$ = 5.0, and is normalized to the Coma Cluster,\label{Fig:4} as described in the text.  The combination of $r - i$, $i - [3.6]$, and $[3.6] - [4.5]$ colors are used here to constrain the redshifts of our COBRA sources.} 
\end{figure}

To account for the lack of pre-existing SDSS photometric redshift estimates for most of the fields observed on the DCT, we examine each radio host's $r$ $-$ $i$, $i$ $-$ $[3.6]$, and $[3.6] - [4.5]$ colors when available.  To determine the redshifts of the COBRA candidates, we measure the difference between our measured host galaxy colors and the modeled EzGal colors in all available colors ($r$ $-$ $i$, $i$ $-$ $[3.6]$, $[3.6] - [4.5]$) at every redshift between 0.0 and 3.0 with a spacing of 0.01 in redshift (see Figure\,\ref{Fig:4}).  Since some relationships between redshift and color are degenerate, we identify the redshifts that minimize the difference between the measured and modeled colors.  In doing this, we compare possible redshifts across different bands, allowing us to break most degeneracies and verify the values reported in \citet{Paterno-Mahler2017}. 

We present photo-$z$ estimates for 73 fields based on host galaxy identification.  With our color-redshift relation from EzGal, we estimate redshifts for 45 fields (see Table\,\ref{tb:4} for the individual redshift estimates).  Additionally, four COBRA fields that aren't quasars in this sample have SDSS photo-$z$ estimates (12 other fields have SDSS photo-$z$'s that match our own photo-$z$ estimates and are included in the 45 fields above).  We use previously observed spectra from SDSS, with the exception of COBRA113733.8+300010, which was verified using Keck II LRIS spectra in \citet{Blanton2003}, to determine the redshift for 24 COBRA fields.  Of these 24 COBRA fields with spectroscopic host redshifts, four are non-quasar host galaxies.  As shown in Table\,\ref{tb:4}, each photo-$z$ estimate is within 0.05 - 0.1 of the spectroscopic redshift in at least one color.  Since this is within the expected uncertainty of our photometric redshifts, we are confident in our photometric redshifts without a spectroscopically confirmed host galaxy.  All four spectroscopically confirmed host galaxies that are not quasars are surrounded by red sequence cluster overdensities at greater than or equal to 2$\sigma$, making them red sequence cluster candidates.  For the analysis in this paper, we use the spectroscopic redshift when available as the host redshift.  We are unable to estimate a redshift for the remaining 17 fields using the color of the host galaxy either because \citet{Paterno-Mahler2017} identify no host, we identify no host with our $i$-band observations (despite a host being identified with our 3.6\,$\mu$m observations in \citep{Paterno-Mahler2017}), or the host's color does not agree with the EzGal models.  For 18 of the fields observed in at least three bands, the redshift estimates match and are within 0.05 in redshift space of one another.

\begin{figure}
\centering
\figurenum{5}
\epsscale{1}
\includegraphics[scale=0.5,trim={0.4in 0.1in 0.0in 0.3in},clip=True]{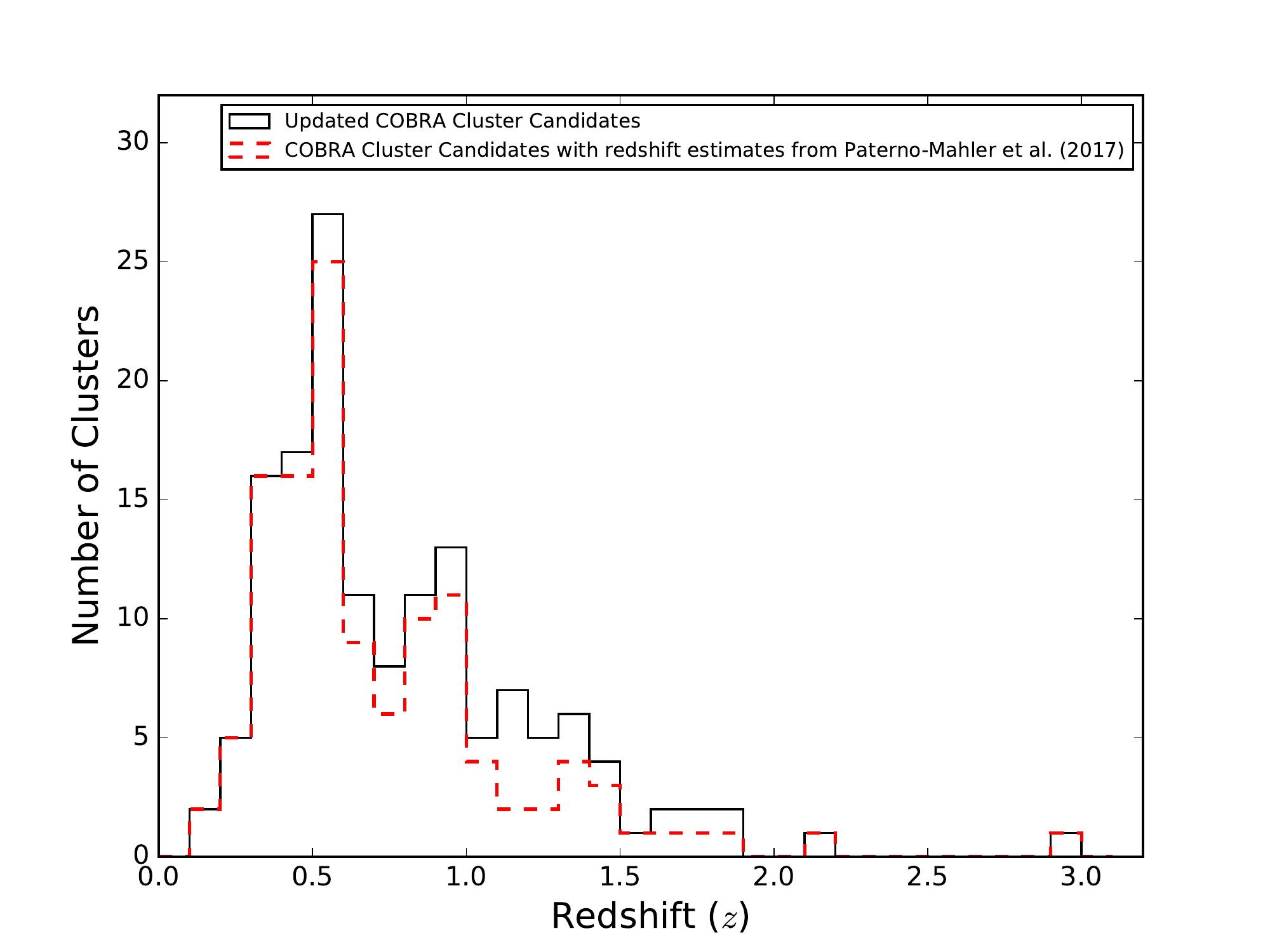}
\caption{The redshift distribution of COBRA cluster candidates with redshift estimates.  The black line represents all COBRA cluster candidates with redshift estimates, including new clusters and redshift estimates presented in this paper.  The dashed red line shows the redshift estimates for the cluster candidates presented in \citet{Paterno-Mahler2017}.  Both histograms exclude the 49 cluster candidates without redshift estimates. \label{Fig:10}}
\end{figure}

By combining our optical and IR imaging with our EzGal analysis, we measure new redshift estimates for 26 fields previously lacking such a measurement (this includes the additional four redshift estimates made in \textsection{4.1}).  Of these fields with new redshift estimates, 21 are cluster candidates (13 are red sequence cluster candidates), increasing the total number of COBRA clusters with a redshift estimate reported here or in \citet{Paterno-Mahler2017} from 125 to 146.  Within this sample of 21 cluster candidates with newly reported redshift estimates, all are at $z$ $>$ 0.5 (four are at 0.5 $\leq$ $z$ $<$ 0.75, five are at 0.75 $\leq$ $z$ $<$ 1.0, and twelve are at $z$ $\geq$ 1.0).  The highest new redshift estimate is at $z$ $\approx$ 1.80.  Of the cluster candidates with new redshift estimates, only two are newly identified cluster candidates based on this work (the remaining 19 were identified in \citet{Paterno-Mahler2017}; see \textsection{4} and \textsection{5} for a detailed discussion of how we determine cluster candidates).  

Though we increase the number of COBRA cluster candidates with a redshift estimate, we still only have redshift estimates for 146 of 195 cluster candidates in COBRA.  We have at least 49 additional COBRA cluster candidates reported in \citet{Paterno-Mahler2017} lacking redshift information.  We expect these fields to populate the high-$z$ portion of Figure\,\ref{Fig:10}.  We aim to observe these fields with the DCT in the coming years.  Although we have only observed a small subset of COBRA in the optical, these results emphasize that our cluster candidates without redshift estimates are likely at high-$z$ (see \citealp{Paterno-Mahler2017}).

\section{Red Sequence Analysis}
Clusters are expected to host overdensities of red sequence early-type galaxies \citep[e.g.,][]{Gladders2000, Rykoff2014, Andreon2014, Cooke2015,Cerulo2016}.  Thus, to better determine which bent, double-lobed radio sources are associated with high-$z$ galaxy clusters and improve on the single-band IR overdensities in \citet{Paterno-Mahler2017}, we identify which bent AGN are surrounded by red sequence galaxies.  In this section, we measure the red sequence overdensity for each field by comparing the number of red sequence sources within 1$\arcmin$ of the radio source to a background field.  We use similarities to the host galaxy color as the basis for the detection of a red sequence.  

To determine which galaxies are red, we compare the optical and IR SExtractor catalogues for each COBRA field and match sources within 1$\arcsec$ of each other (matching 3.6\,$\mu$m to $i$-band, 4.5\,$\mu$m to 3.6\,$\mu$m, and $i$-band to $r$-band).  Our matching routine is a nearest-neighbor matching routine, which determines the nearest source in one catalogue to another and removes sources without a match within 1$\arcsec$.  To verify the efficacy of our search region, we compare the number of matched sources in two bands to the number of sources that match when one of the catalogues is offset by 1$\arcmin$.  Since there are no real matches with the offset catalogue, we treat this number of matches as the fraction of non-real matches that we find.  The 1$\arcsec$ search radius yields $\approx$ 95$\%$ accuracy in real galaxy matches.  Any source without a match is removed since we cannot determine a color (this removes low stellar mass foreground objects that are not detected in the IR, but removes IR-bright, potentially dusty, galaxies that are not detected in the optical).  Using our catalogues of matched sources, we search for red sequence galaxies within 1$\arcmin$ of the bent radio source.  We focus on a 1$\arcmin$ ($\approx$ 430\,kpc at $z$ = 0.7, 480\,kpc radius at $z$ $=$ 1.0, and 503\,kpc at $z$ = 1.3 as compared to R$_{200}$ $\approx$ 800\,kpc for a 10$^{14}$M$\odot$ cluster \citep[e.g.,][]{Sifon2016}) region because we expect a dense core of red early-type galaxies near the cluster center.

\subsection{$i - [3.6]$ Analysis}
We focus our analysis on the $i - [3.6]$ color because it is monotonically redder with increasing redshift out to $z$ $\approx$ 2.0 (see Figure\,\ref{Fig:4}) and because we have images in both bands for all 90 fields.  Additionally, at $z$ $>$ 0.8, these bands straddle the 4000\,\AA\,\,break characteristic of an elliptical galaxy's spectrum, making this color a good identifier of red sequence galaxies at the redshifts of most of our COBRA cluster candidates.   

\begin{figure*}
\figurenum{6}
\subfigure{\includegraphics[scale=0.5,trim={0.4in 0.1in 0.6in 0.5in},clip=True]{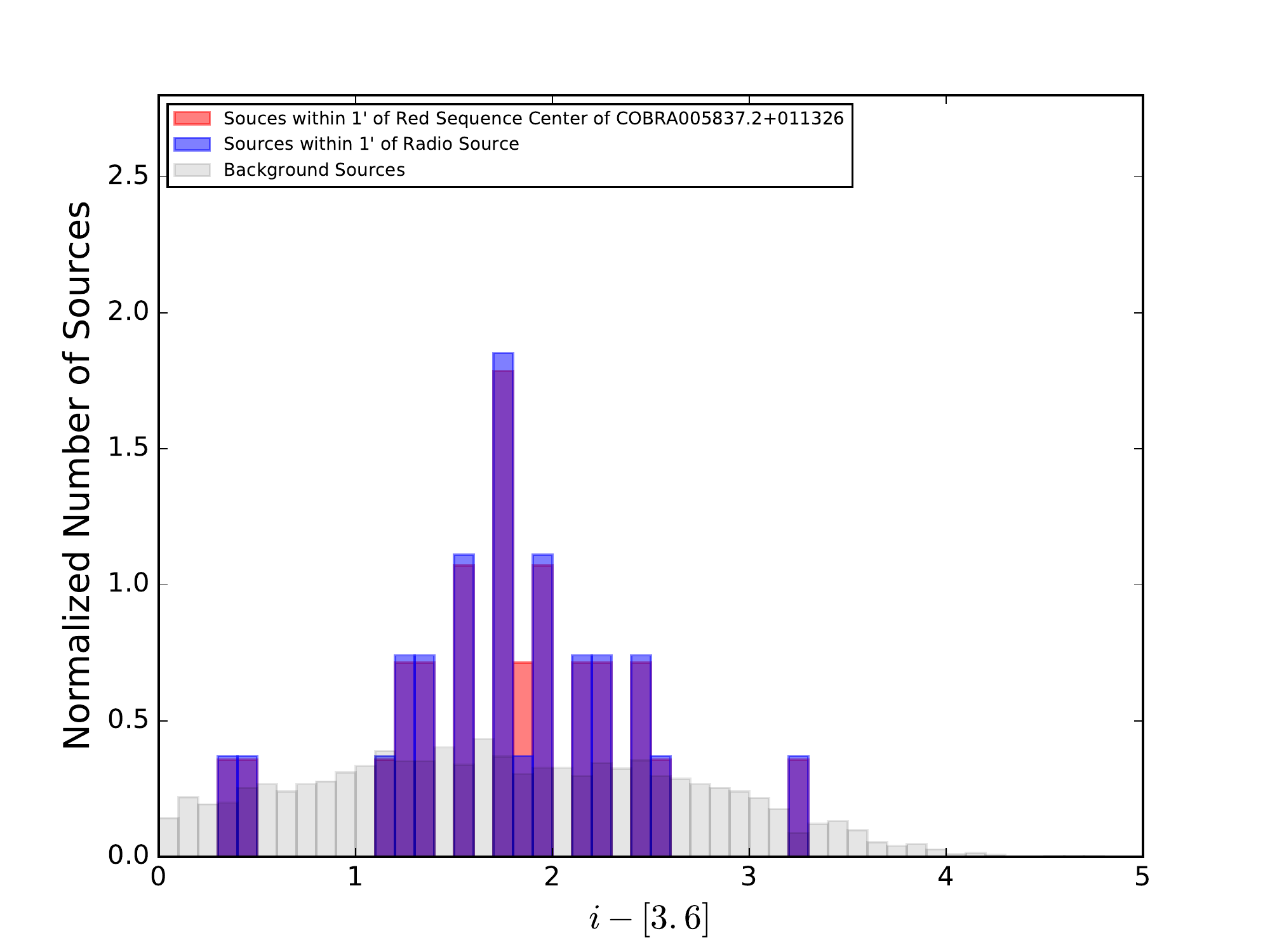}}
\subfigure{\includegraphics[scale=0.5,trim={0.4in 0.1in 0.6in 0.5in},clip=True]{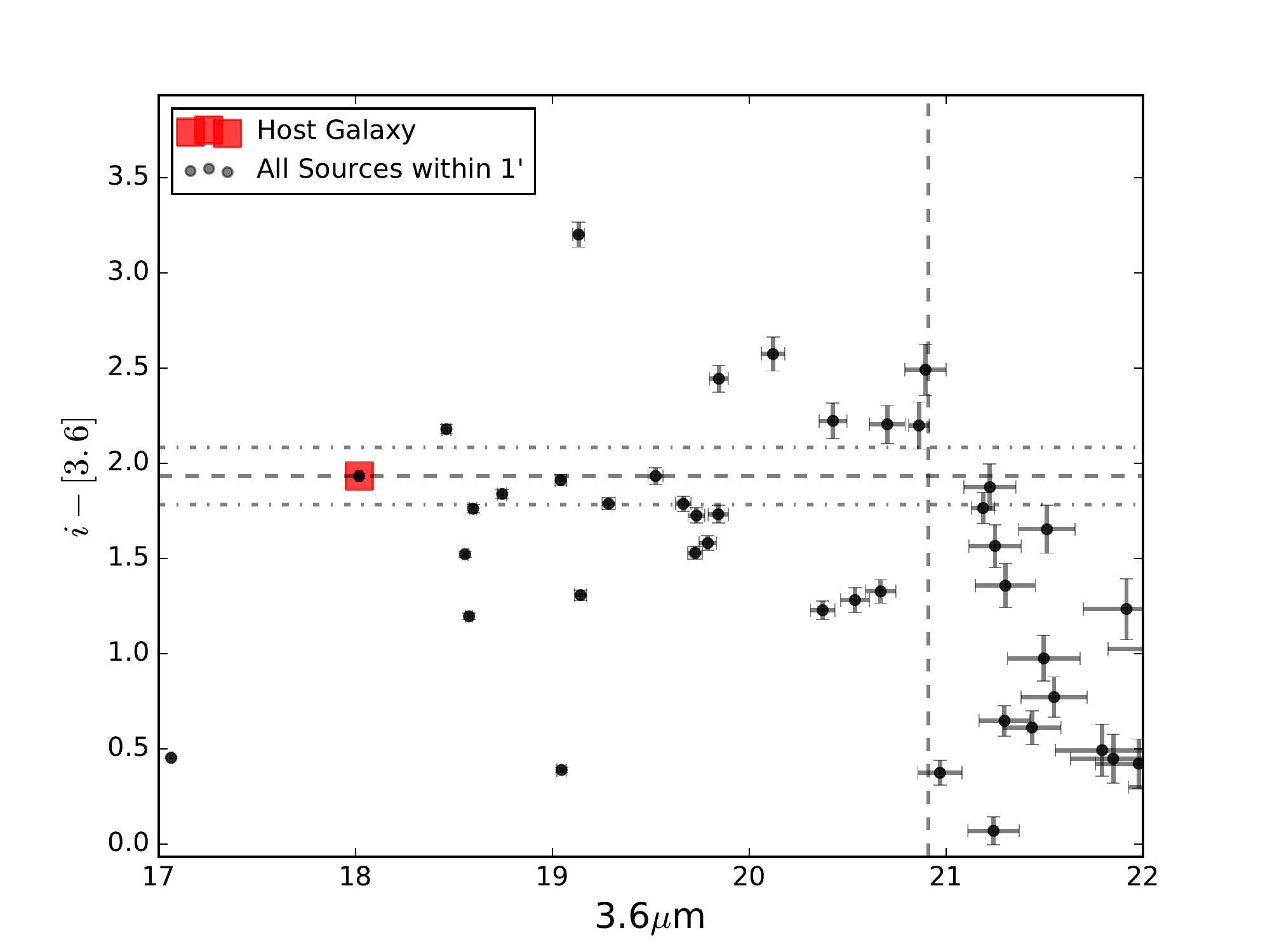}}
\subfigure{\includegraphics[scale=0.5,trim={0.4in 0.1in 0.6in 0.5in},clip=True]{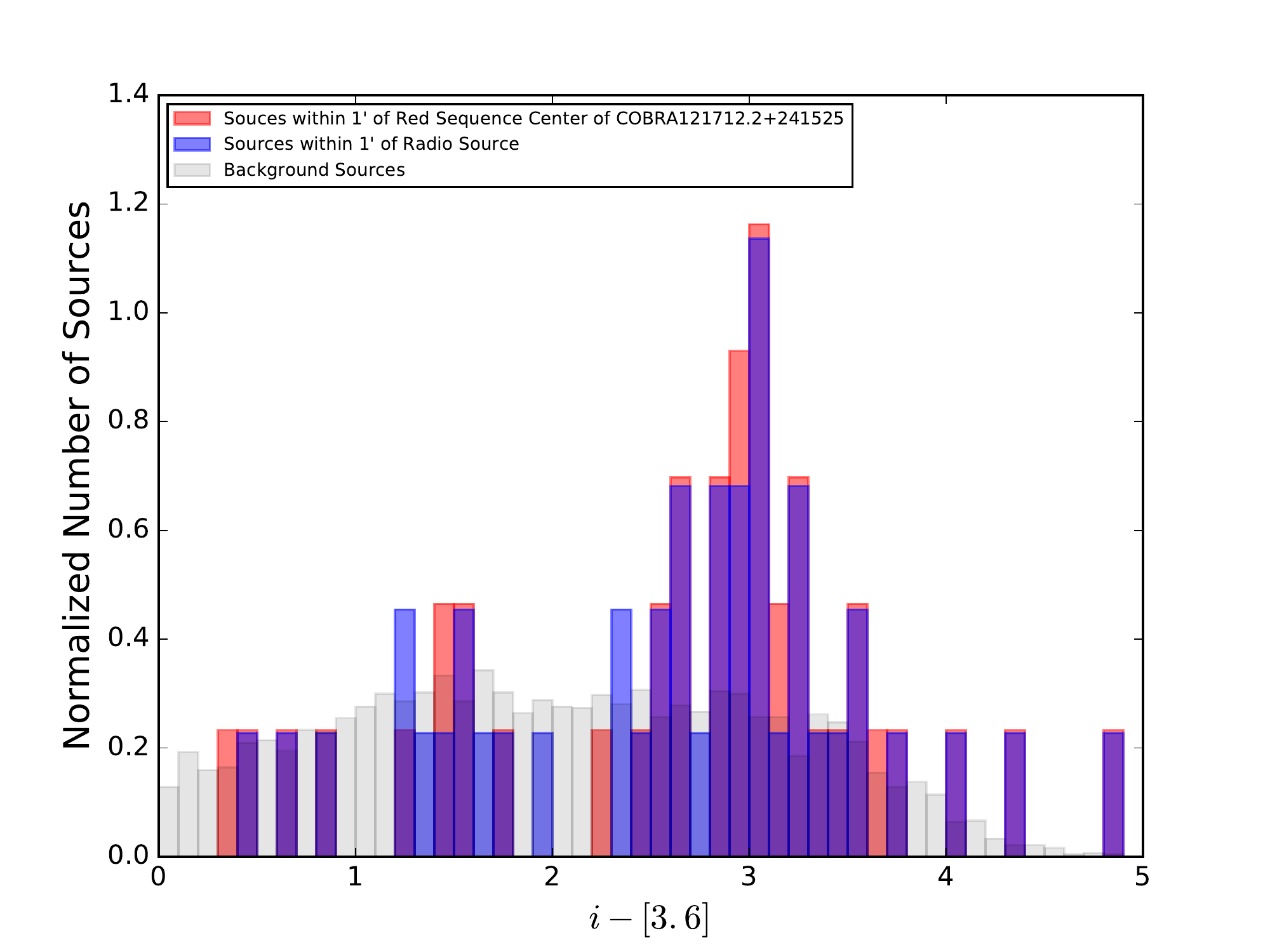}}
\subfigure{\includegraphics[scale=0.5,trim={0.4in 0.1in 0.6in 0.5in},clip=True]{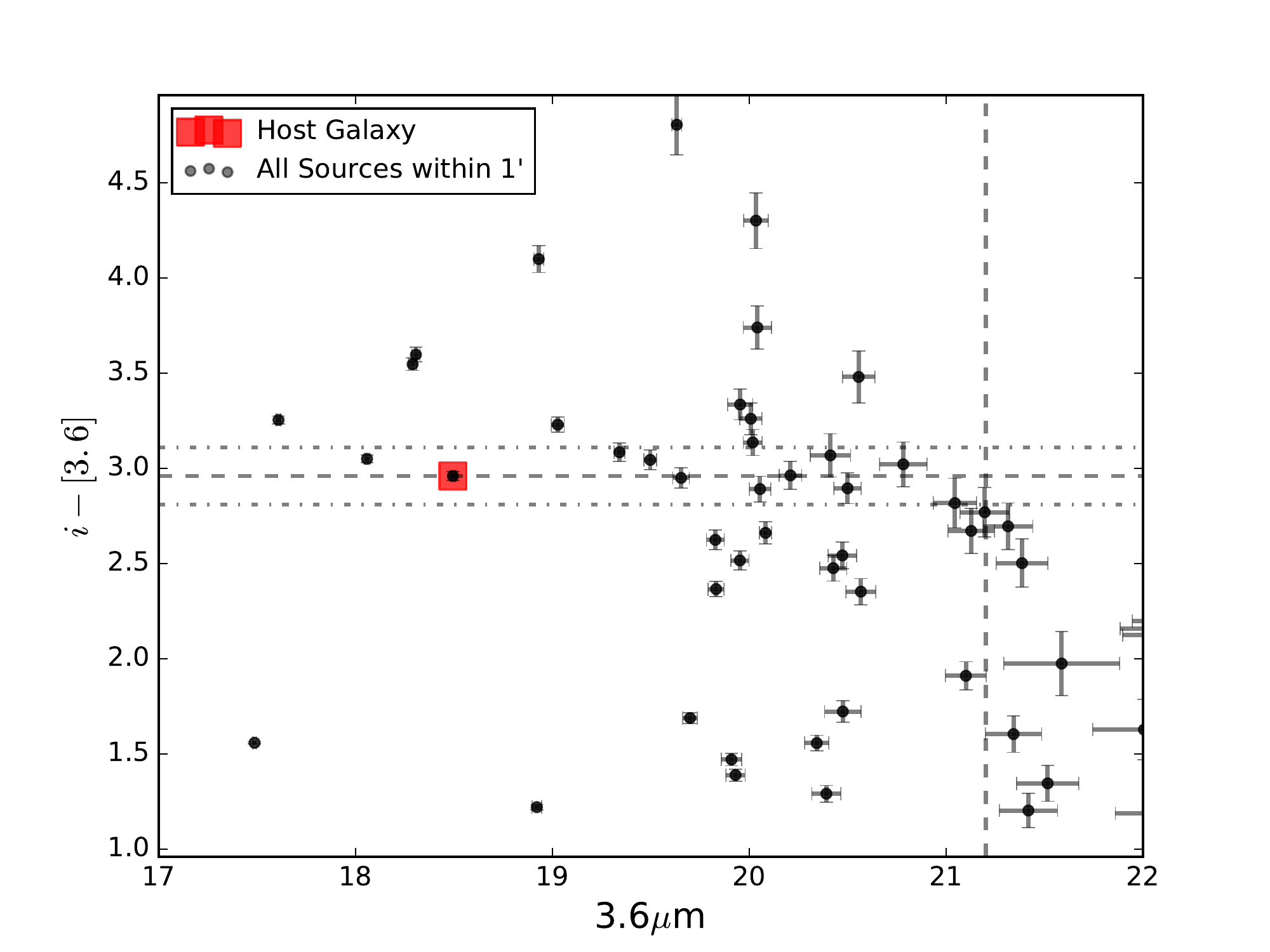}}

\caption{Histograms and their corresponding CMDs for COBRA005837.2+011326 and COBRA121712.2+241525.  Each plot shows the distributions of galaxy colors in $i - [3.6]$ for the galaxies within 1$\arcmin$ of either the radio source (shown in blue in the histogram) or the new red sequence center (shown in red in the histogram).  The background distribution is shown in grey.  All three of these distributions are normalized to one another.  On the CMDs, the host is shown in a red square, while all other galaxies detected within 1$\arcmin$ of the radio source are shown in black circles. The horizontal grey dashed line shows the color of the host galaxy.  The two grey dot-dashed lines show our estimate of the red sequence color range.  The value of the m*+1 magnitude limit is shown in the vertical dashed line.  For a full explanation of how the red sequence center is chosen, see \textsection{5}.  Both examples shown are red sequence cluster candidates.  COBRA005837.2+011326 is at $z$ $\approx$ 0.71 and the 3.6$\mu$m magnitude limit is 20.91 (m*+1) and the $i$-band magnitude limit is 23.50.  COBRA121712.2+241525 is at $z$ $\approx$ 0.90 and the 3.6$\mu$m magnitude limit is 21.20 (m*+1)  and the $i$-band magnitude limit is 24.50.  \label{Fig:colorhist}}
\end{figure*}

We measure the $i - [3.6]$ color of each galaxy in each field for which both $Spitzer$ 3.6\,$\mu$m and LMI $i$-band imaging are available, down to the m*+1\,mag sensitivity limit in 3.6\,$\mu$m whenever possible.  Because each cluster candidate lies at a different redshift, and some cluster candidates are too distant for us to achieve the m*+1 magnitude  limit given the sensitivity of our imaging, we divide our sample into two subsamples that reflect the different ways they have been analyzed.  We call the first the m*+1 sample, to indicate that our colors are reliable to this level, and we call the other the magnitude-limited sample.  The m*+1 sample and magnitude-limited cluster samples consist of 35 and 38 fields, respectively.  At 3.6\,$\mu$m, the m*+1 magnitude limit corresponds to $z$ $<$ 1.1 clusters, so all fields at $z$ $<$ 1.1 are in the m*+1 subsample by construction (the values of m*+1 in this sample range from 3.6\,$\mu$m 19.80\,mag to 21.35\,mag).  The higher-$z$ clusters are in the magnitude-limited subsample.  Within this subsample, our 3.6\,$\mu$m magnitude limit is 21.4\,mag for all fields, meaning we reach a different absolute magnitude limit for each field in the sample.  For reference, our 3.6\,$\mu$m magnitude limit of 21.4\,mag corresponds to an m* galaxy at $z$ = 2.1.  In order to match the most possible galaxies in each field, we do not use the m*+1 value in $i$-band as the magnitude limit.  Since some galaxies will be brighter than our m*+1 limit in 3.6\,$\mu$m, but fainter in $i$-band, we use the magnitude limit of our DCT observations for the $i$-band observations regardless of the cluster candidate redshift to best identify the most possible cluster members (this allows the colors of our red sequence galaxies to range from $\approx$ 0.6\,mag to 3.4\,mag depending on the redshift for our m*+1 subsample and $\approx$ 3.5\,mag to 5.0\,mag for our magnitude-limited sample).    

To examine the richness of red sequence galaxies relative to foreground and background interlopers, we plot histograms of galaxy color and CMDs in Figure\,\ref{Fig:colorhist}.  As shown in the left-hand side of Figure\,\ref{Fig:colorhist}, there is a spread in the peak of galaxies relative to the background around the color of the host galaxy, although we see strong evidence for an evolved red sequence in the right-hand side of Figure\,\ref{Fig:colorhist}.  To minimize background/foreground contamination, and best estimate red sequence galaxies, we adopt a red sequence width of $\pm$0.15\,mag, in agreement with literature \citep[e.g.,][]{Blakeslee2003,Mei2006,Mei2009,Snyder2012,Lemaux2012,Cerulo2016}, to estimate red sequence members.  Specifically, we are using the presence of a red sequence to inform our understanding of the environments of these bent radio AGN.  Our use of the $\pm$0.15\,mag width accounts for the $\pm$3$\sigma$ detection of the typical red sequence width of 0.05\,mag, which allows us to detect the full range of potential red sequence galaxies.  This is explicitly shown in the right-hand side of Figure\,\ref{Fig:colorhist}, where the red sequence color range is centered on the dashed horizontal line and bounded by the dot-dashed horizontal lines.  

To further verify our host galaxy redshift estimates, we compare these colors to the color distribution of galaxies within 1$\arcmin$ of the radio source to estimate the redshift of the cluster.  We find that 41 of 73 fields with redshift estimates show a strong overdensity at the host color/redshift in $i - [3.6]$.  Of the 41 fields, three are quasar fields with spectroscopic host redshifts.  The lack of a well-defined peak in color-space for some fields does not discredit those redshift estimates as our search for a peak at the color of the host is predicated on having a strong, evolved red sequence, which may not be a characteristic for all COBRA cluster candidates, especially those at higher redshifts \citep[e.g.,][]{Krick2009,Brodwin2013,Hennig2017}.  Furthermore, 19 of the fields are at redshifts ($z$ $>$ 1.4) where our observations are not sensitive enough to detect fainter red sequence galaxies (see \textsection{4.1.1} for a full description of these fields), meaning that the histograms identify only foreground structure.  Although not all 19 of these fields have 3.6\,$\mu$m and 4.5\,$\mu$m observations, we find evidence for a peak in the color distribution near our host color in six of these fields with our $[3.6] - [4.5]$ analysis (see \textsection{4.2} for a description of this analysis).  

In many of the fields where the host color doesn't lie at a peak in the color distribution in the field, the histograms do not show a strong peak at any color, which may mean that some of these bent radio sources are not in clusters.  However, in some of them, including the few fields with host galaxy photometric redshifts from SDSS in our sample, there is a peak at a bluer or redder color than our expected redshift range.  This could indicate a foreground/background cluster unassociated with the radio sources and/or an error in the host redshift (i.e., the redshift of the host galaxy is incorrect, or the host is incorrectly identified).  In future work, where we do a more rigorous red sequence fit, we may revisit some of these fields.

For fields without a host redshift estimate, we explore the galaxies surrounding the radio source to determine a redshift estimate.  We find strong evidence for the existence of a red sequence in four COBRA clusters at 0.6 $<$ $z$ $<$ 0.9 (marked in Table \ref{tb:4}).  This yields a total of 77 fields with redshift estimates, 39 in the m*+1 sample and 38 in the magnitude-limited sample.  

We quantify our red sequence measurements by estimating the overdensity of red sequence galaxies relative to the background.  For this analysis, we use $\pm$ 0.15\,mag as our red sequence width.  In choosing this width, we reflect a narrow range of red sequence colors needed to include most potential red sequence galaxies and mirror the uncertainty on our redshift estimates.  However, we use the host color to uniformly include all fields with a redshift estimate (for the four fields with redshift estimates from color histograms, we use the color of the peak and for the quasars, we use the EzGal modeled color associated with the spectroscopic redshift).  It is possible for fields where a host galaxy is intrinsically bluer or redder than other potential red sequence members that we will not identify all cluster galaxies using a set color range, especially given that the red sequence can have a non-zero slope \citep[e.g.,][]{Brodwin2006,Eisenhardt2007,Stott2009,Cooke2016}.  We address how to account for these potential cluster members in \textsection{4.1.2} using imaging from the ORELSE survey \citep{Lubin2009} to correct for populations of redder and bluer galaxies at our target redshift.

To better constrain our cluster candidate demographics, we need a background field with which to compare to.  We use the overlapping area between the SpUDS and UKIDSS Ultra Deep Survey (UDS; PI O. Almaini) fields as our $i - [3.6]$ background field.  Excluding regions with saturated stars, we construct 238 1$\arcmin$ radius regions and determine a unique $i - [3.6]$ background measurement for each field to mirror the magnitude limits of each field.  To create uniformity between our cluster and $i - [3.6]$ backgrounds measurements, we restrict the $i - [3.6]$ background counts to those sources with a color within 0.15\,mag of the host galaxy color.  

\begin{equation}\label{Eq:NEW}
    \sigma_{cluster} = \frac{N_{counts} - M_{Gauss}}{\sigma_{Gauss}}\footnote{For some fields, the mean number of galaxies in the Gaussian background is greater than the number of galaxies in the target region.  We report these fields as having negative overdensities and negative significances.}
\end{equation}

\begin{figure}
\centering
\figurenum{7}
\epsscale{1}
\includegraphics[scale=0.5,trim={0.45in 0.1in 0.0in 0.4in},clip=True]{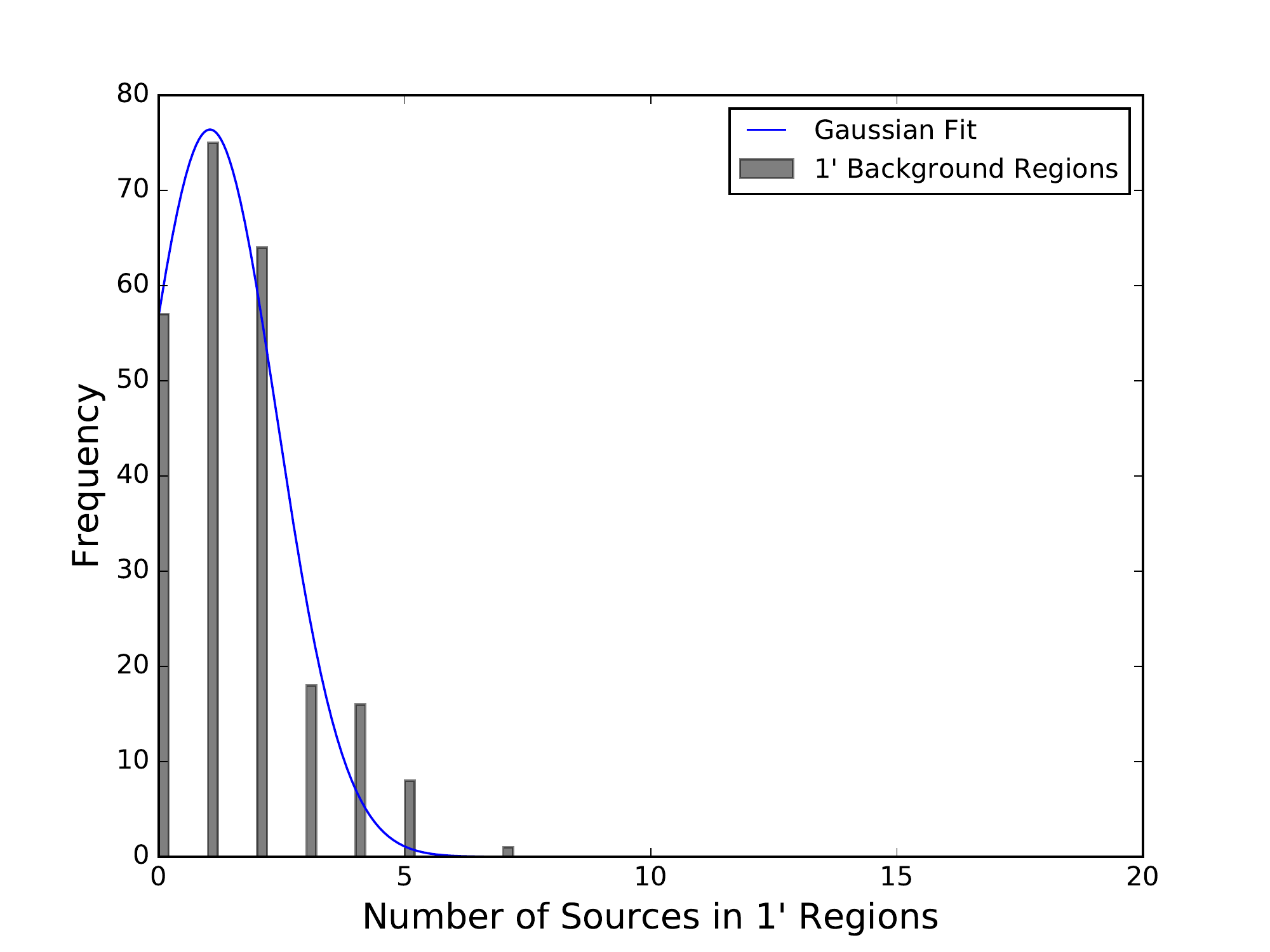}

\caption{Histogram showing the number of background counts in our 238 regions for the combined SpUDS and UDS fields that we use as our $i - [3.6]$ background region.  The blue line shows our fit to the data. This plot shows the $i - [3.6]$ background distribution for COBRA121712.2+241525 at $z$ $\approx$ 0.9.  We construct this background distribution using an m*+1 magnitude limit of 21.20\,mag in 3.6\,$\mu$m and a magnitude limit of 24.50\,mag in $i$-band.\label{Fig:Background}}
\end{figure}

As in \citet{Galametz2012} and \citet{Paterno-Mahler2017}, we find that each $i - [3.6]$ background distribution can be approximated by a Gaussian distribution, except for a high-density tail representing large scale structure in the field (see Figure\,\ref{Fig:Background}).  Because of this, we determine the mean and variance of our Gaussian fit by excluding this tail.  Like \cite{Paterno-Mahler2017}, we measure the significance of each overdensity using Equation\,\ref{Eq:NEW}, where $\sigma_{cluster}$ is the significance of each red sequence cluster measurement, $N_{counts}$ is the number of detected red sequence galaxies within 1$\arcmin$ of the radio source, $M_{Gauss}$ is the mean of the Gaussian fit to each background distribution, and $\sigma_{Gauss}$ is the standard deviation of the Gaussian fit to each background distribution.  $N_{counts}$, $\Delta$N (the net number of red sequence sources, N$_{counts}$ - M$_{Gauss}$), and $\sigma$ for all cluster candidates are included in Tables\,\ref{tb:ich1AGN}\,,\,\ref{tb:ch1ch2AGN},\,$\&$\,\ref{tb:riAGN} and for all fields in Table\,\ref{tb:2}.  Like \citet{Paterno-Mahler2017}, we use the 2$\sigma$ cut as our lower limit for red sequence cluster candidates.  

Additionally, since the mean of the Gaussian distribution for our $i - [3.6]$ background is less than 1.0 red sequence galaxy per 1$\arcmin$ region (given our magnitude limits and color range) in many cases, we require at least three galaxies to be identified to be a red sequence cluster candidate.  This removes fields where a two galaxy detection yields a significance exceeding 2$\sigma$, which occurs within our magnitude-limited sample.  Given the relatively bright limits required on the luminosity of galaxies that enter into our subsamples, specifically of the magnitude-limited subsample, it is possible that at the highest-redshifts, these overdensity measurements overestimate the true strength of the detection.  To compare the success of our red sequence cluster search to our random background, we measure how many of our 238 $i - [3.6]$ background regions are above our 2$\sigma$ cluster candidate threshold.  We find that $\approx$ 5$\%$ of these background regions fall above this 2$\sigma$ cluster candidate threshold in the m*+1 sample, while 5 - 10$\%$ of these background regions fall above this threshold in the magnitude-limited sample.

\subsubsection{$i - [3.6]$ Results}
This red sequence analysis is the first step toward determining which high-$z$ fields host galaxy clusters.  While previous works \citep[e.g.,][]{Wing2011,Wing2013} have analyzed bent sources in well studied photometric and spectroscopic samples at low-$z$, only COBRA113733.8+300010 is spectroscopically confirmed \citep{Blanton2003}, and only two of the other 89 fields host confirmed cluster candidates with multi-wavelength observations (see \textsection{6.3} for a discussion of previously identified clusters within our sample).  In the m*+1 sample (magnitude-limited sample), 20 of the 39 (7 of the 38) fields are red sequence cluster candidates (see Table\,\ref{tb:ich1AGN}).  For the complete list of overdensities, see Table\,\ref{tb:2}.  Of these 27 fields, two are cluster candidates not previously identified in \citet{Paterno-Mahler2017} (see Table\,\ref{tb:ich1AGN}).  Although our strongest red sequence cluster candidate has 11 red sequence galaxies when centered on the AGN, we do find, that like \citet{Cerulo2016}, the expected number of $i - [3.6]$ galaxies based on our SpUDS-UDS $i - [3.6]$ background that have colors similar to those of our target red sequence is low, with most regions expecting fewer than two galaxies, based on our magnitude limits.  This mean Gaussian background value is well below that reported in \citet{Paterno-Mahler2017}, and only serves to strengthen the efficacy of the red sequence color cut at identifying real structures.  

\begin{deluxetable*}{lccccccccccccc}
    \tablenum{4}
    \tablecolumns{14}
    \tabletypesize{\small}
    \tablecaption{COBRA $i - [3.6]$ Red Sequence Cluster Candidates\label{tb:ich1AGN}}
    \tablewidth{0pt}
    \tabletypesize{\footnotesize}

    \setlength{\tabcolsep}{0.05in}
    \tablehead{
    \colhead{Field}&
    \colhead{Redshift ($z$)\tablenotemark{l}}&
    \multicolumn{6}{c}{Red Sequence Overdensity}&
    \multicolumn{6}{c}{Combined Overdensity}
    \cr
    \colhead{}&
    \colhead{}&
    \multicolumn{3}{c}{AGN Center}&
    \multicolumn{3}{c}{RS Center}&
    \multicolumn{3}{c}{AGN Center}&
    \multicolumn{3}{c}{RS Center}
    \cr
    \colhead{}&
    \colhead{}&
    \colhead{N\tablenotemark{a}}& 
    \colhead{$\Delta$N\tablenotemark{b}}&  
    \colhead{$\sigma$\tablenotemark{c}} &
    \colhead{N\tablenotemark{a}}& 
    \colhead{$\Delta$N\tablenotemark{b}}&
    \colhead{$\sigma$\tablenotemark{c}} &
    \colhead{N$_{total}$\tablenotemark{d}}& 
    \colhead{$\Delta$N$_{total}$\tablenotemark{e}}&
    \colhead{$\sigma_{total}$\tablenotemark{f}}&
    \colhead{N$_{total}$\tablenotemark{d}}& 
    \colhead{$\Delta$N$_{total}$\tablenotemark{e}}&
    \colhead{$\sigma_{total}$\tablenotemark{f}}
    \cr
    }

\startdata
COBRA005837.2+011326\tablenotemark{g} & 0.71 & 6 & 5.5 & 3.6 & 7 & 6.5 & 4.3 & 28 & 4.4 & 4.0 & 27 & 4.9 & 4.4\\
COBRA012058.9+002140\tablenotemark{g} & 0.75 & 5 & 4.2 & 2.5 & 4 & 3.2 & 1.9 & 36 & 4.4 & 3.8 & 33 & 3.6 & 3.1\\
COBRA014741.6$-$004706\tablenotemark{g}\tablenotemark{i} & 0.60 & 5 & 4.0 & 2.0 & 5 & 4.0 & 2.0 & 23 & 3.5 & 2.8 & 23 & 3.6 & 2.9\\
COBRA015313.0$-$001018\tablenotemark{g}\tablenotemark{j} & 0.44 & 1 & 1.0 & 0.7 & 3 & 3.0 & 2.1 & 6 & 0.1 & 0.1 & 12 & 1.6 & 2.3\\
COBRA074025.5+485124\tablenotemark{h} & 1.10 & 4 & 4.0 & 3.6 & 3 & 3.0 & 2.7 & 33 & 2.6 & 3.9 & 29 & 1.9 & 2.9\\
COBRA074410.0+274011\tablenotemark{h} & 1.30 & 4 & 3.7 & 4.0 & 7 & 6.7 & 7.3 & 41 & 2.6 & 4.3 & 45 & 4.0 & 6.7\\
COBRA075516.6+171457\tablenotemark{g} & 0.64 & 2 & 1.8 & 1.1 & 6 & 5.8 & 3.4 & 29 & 2.7 & 2.4 & 39 & 6.4 & 5.6\\
COBRA100745.5+580713\tablenotemark{g} & 0.656 & 6 & 6.0 & 3.1 & 5 & 5.0 & 2.6 & 24 & 3.2 & 2.6 & 25 & 2.9 & 4.3\\
COBRA100841.7+372513\tablenotemark{h} & 1.20/1.35 & 5 & 4.4 & 5.1 & 5 & 4.4 & 5.1 & 35 & 3.7 & 5.0 & 34 & 3.4 & 4.5\\
COBRA103434.2+310352\tablenotemark{h} & 1.20 & 5 & 4.6 & 4.8 & 5 & 4.6 & 4.8 & 40 & 4.0 & 5.8 & 42 & 4.2 & 6.1\\
COBRA104254.8+290719\tablenotemark{g} & 1.35/1.05 & 3 & 3.0 & 3.3 & 5 & 5.0 & 5.4 & 32 & 1.7 & 3.4 & 36 & 2.4 & 4.9\\
COBRA113733.8+300010\tablenotemark{g} & 0.96 & 8 & 7.4 & 4.6 & 7 & 6.4 & 4.0 & 36 & 4.2 & 4.1 & 36 & 4.0 & 3.9\\
COBRA120654.6+290742\tablenotemark{g}\tablenotemark{i}\tablenotemark{k} & 0.85 & 2 & 1.7 & 0.9 & 3 & 2.7 & 1.4 & 30 & \nodata & \nodata & 30 & 3.1 & 2.4\\
COBRA121712.2+241525\tablenotemark{g} & 0.90 & 11 & 10.0 & 7.3 & 12 & 11.0 & 8.1 & 46 & 7.2 & 6.2 & 45 & 7.0 & 6.1\\
COBRA123940.7+280828\tablenotemark{g} & 0.92 & 7 & 6.0 & 3.6 & 7 & 6.0 & 3.6 & 37 & 3.5 & 2.9 & 38 & 3.8 & 3.1\\
COBRA125047.4+142355\tablenotemark{g} & 0.90 & 5 & 4.2 & 2.5 & 5 & 4.2 & 2.5 & 32 & 3.8 & 2.9 & 36 & 4.6 & 3.5\\
COBRA130729.2+274659\tablenotemark{h}\tablenotemark{j} & 1.144 & 1 & 1.0 & 0.9 & 6 & 6.0 & 5.5 & 19 & 0.4 & 0.7 & 25 & 2.9 & 4.4\\
COBRA133507.1+132329\tablenotemark{h}\tablenotemark{j} & 1.25 & 2 & \nodata & \nodata & 3 & 3.0 & 4.5 & 21 & 0.9 & 1.7 & 22 & 1.3 & 2.5\\
COBRA134104.4+055841\tablenotemark{g} & 0.90 & 3 & 2.6 & 2.2 & 3 & 2.6 & 2.2 & 19 & 2.3 & 2.2 & 22 & 2.7 & 2.6\\
COBRA135136.2+543955\tablenotemark{g} & 0.55 & 10 & 9.9 & 6.4 & 11 & 10.9 & 7.0 & 31 & 6.4 & 6.7 & 32 & 7.0 & 7.3\\
COBRA135838.1+384722\tablenotemark{g} & 0.81 & 1 & 0.3 & 0.2 & 4 & 3.3 & 2.4 & 21 & 1.1 & 1.1 & 28 & 2.7 & 2.6\\ 
COBRA142238.1+251433\tablenotemark{g} & 1.00 & 5 & 4.3 & 2.6 & 9 & 8.3 & 5.1 & 37 & 3.7 & 3.7 & 38 & 4.8 & 4.7\\
COBRA145023.3+340123\tablenotemark{h} & 1.20 & 2 & \nodata & \nodata & 3 & 3.0 & 3.9 & 29 & \nodata & \nodata & 38 & 2.6 & 5.2\\
COBRA150238.1+170146\tablenotemark{h} & 1.10 & 5 & 4.2 & 2.9 & 5 & 4.2 & 2.9 & 27 & 1.7 & 2.2 & 26 & 1.6 & 2.1\\
COBRA151458.0$-$011749\tablenotemark{g} & 0.80 & 6 & 5.0 & 3.0 & 6 & 5.0 & 3.0 & 36 & 4.8 & 4.5 & 36 & 4.8 & 4.5\\
COBRA152647.5+554859\tablenotemark{h} & 1.10 & 3 & 3.0 & 4.1 & 3 & 3.0 & 4.1 & 29 & 2.1 & 3.3 & 33 & 2.5 & 4.0\\
COBRA154638.3+364420\tablenotemark{g} & 0.939 & 3 & 1.9 & 1.4 & 6 & 4.9 & 3.5 & 23 & 1.2 & 1.1 & 23 & 2.0 & 1.9\\
COBRA162955.5+451607\tablenotemark{g} & 0.78 & 5 & 3.8 & 2.2 & 5 & 3.8 & 2.2 & 31 & 3.4 & 3.0 & 39 & 4.8 & 4.2\\
COBRA164551.2+153230\tablenotemark{g}\tablenotemark{k} & 0.65 & 2 & 1.0 & 0.7 & 3 & 2.0 & 1.4 & 17 & 1.5 & 1.4 & 18 & 2.1 & 2.0\\
COBRA164611.2+512915\tablenotemark{g} & 0.351 & 9 & 8.7 & 8.7 & 8 & 7.7 & 7.7 & 26 & 5.3 & 8.5 & 22 & 4.5 & 7.3\\
COBRA164951.6+310818\tablenotemark{g} & 0.52 & 4 & 4.0 & 2.3 & 4 & 4.0 & 2.3 & 18 & 2.5 & 2.8 & 17 & 2.0 & 2.3\\
COBRA170105.4+360958\tablenotemark{g} & 0.80 & 7 & 6.0 & 4.2 & 7 & 6.0 & 4.2 & 42 & 6.4 & 5.8 & 39 & 5.9 & 5.3\\
COBRA170614.5+243707\tablenotemark{g} & 0.71 & 8 & 7.2 & 5.2 & 9 & 8.2 & 5.9 & 41 & 6.1 & 5.5 & 39 & 5.9 & 5.3\\
COBRA171330.9+423502\tablenotemark{g}\tablenotemark{i} & 0.698 & 4 & 4.0 & 2.0 & 2 & 2.0 & 1.0 & 19 & 1.9 & 1.5 & 17 & 1.0 & 0.8\\
COBRA172248.2+542400\tablenotemark{h} & 1.45/1.25 & 3 & 3.0 & 4.7 & 3 & 3.0 & 4.7 & 29 & 1.5 & 2.2 & 30 & 1.6 & 2.4\\
COBRA221605.1$-$081335\tablenotemark{g} & 0.70 & 5 & 4.4 & 3.2 & 4 & 3.4 & 2.4 & 23 & 2.7 & 2.1 & 21 & 2.2 & 1.9\\
COBRA232345.9+002925\tablenotemark{g}\tablenotemark{i}\tablenotemark{k} & 0.73 & 3 & 2.2 & 1.7 & 3 & 2.2 & 1.7 & 22 & 2.2 & 2.1 & 26 & 2.4 & 2.3\\
\enddata
\tablenotetext{a}{N = The total number of red sequence members in the 1$\arcmin$ region.}
\tablenotetext{b}{$\Delta$N = The excess of counts in the 1$\arcmin$ region above the background.}
\tablenotetext{c}{$\sigma$ = The significance calculated using Equation\,\ref{Eq:NEW}.}
\tablenotetext{d}{N$_{total}$ = The total number of galaxies (red sequence, redder, and bluer) in the 1$\arcmin$ region.}
\tablenotetext{e}{$\Delta$N$_{total}$ = The excess of galaxies in the 1$\arcmin$ region above the combined background adjusted for the red sequence completeness fraction.}
\tablenotetext{f}{$\sigma_{total}$ = The significance calculated using Equation\,\ref{Eq:NEWover}.}
\tablenotetext{g}{Fields in the m*+1 sample.}
\tablenotetext{h}{Fields in the magnitude-limited sample.}
\tablenotetext{i}{New cluster candidates (not in \citet{Paterno-Mahler2017}) with AGN Center.}
\tablenotetext{j}{New cluster candidates (not in \citet{Paterno-Mahler2017}) with just the Red Sequence Center.}
\tablenotetext{k}{Fields that are only combined cluster candidates, not red sequence cluster candidates.}
\tablenotetext{l}{Fields with mutliple redshift estimates are due to disagreements in our EzGal photo-$z$ estimates for the different colors used.}
\end{deluxetable*}

We present only the number of detected sources in Table\,\ref{tb:2} for the 19 fields at $z$ $>$ 1.4 because our sensitivity in the $i - [3.6]$ color is insufficient for a statistical analysis.  We do likewise for the five fields which appear to be red sequence cluster candidates, but which have fewer than three red sequence galaxies.  Setting aside such fields, we find 27 of the 53 remaining fields have strong red sequences at or above the 2$\sigma$ threshold.  Several of the remaining fields have red sequence significances between 1$\sigma$ and 2$\sigma$ (see Table\,\ref{tb:ich1AGNstat}), including 13 cluster candidates in \citet{Paterno-Mahler2017} that are not red sequence cluster candidates despite being in our statistically analyzed sample.  It is possible that these are either less massive galaxy groups, poorer clusters, or clusters with larger star forming populations. 

In examining our CMDs, we find that the vast majority of our AGN host galaxies are among the brightest red sequence galaxies (see the right-hand side of Figure\,\ref{Fig:colorhist}).  Although most host galaxies are at least 1.0\,mag brighter than an m* galaxy, many do not appear to be proto-BCGs.  Because centering our search region on the AGN implicitly requires the AGN to be at the cluster center, and thus a proto-BCG, that we see host galaxies that do not appear to be the BCG likely means that these sources are not always at the cluster center, even under the ideal assumption that the BCG is situated in the center of of its parent cluster.

\begin{deluxetable*}{cccccc}
    \tablenum{5}
    \tablecolumns{5}
    \tabletypesize{\small}
    \tablecaption{COBRA Galaxy Detection Significance \label{tb:ich1AGNstat}}
    \tablewidth{0pt}
    \tabletypesize{\footnotesize}

    \setlength{\tabcolsep}{0.05in}
    \tablehead{
    \colhead{Sample} &
    \colhead{Cluster Candidates}&
    \multicolumn{3}{c}{Non-Cluster Candidates}
    \cr
    \colhead{}&
    \colhead{$\sigma$ $>$ 2}\vline&
    \colhead{$\sigma$ $<$ 0}\vline&
    \colhead{0 $<$ $\sigma$ $<$ 1}\vline&
    \colhead{1 $<$ $\sigma$ $<$ 2}\vline 
    }
    
\startdata
$i - [3.6]$ m*+1 & 20 & 1 & 12 & 6 \\
$i - [3.6]$ magnitude-limited\tablenotemark{a} & 7 & 3 & 2 & 2 \\
$[3.6] - [4.5]$ subsample & 4 & 6 & 6 & 4 \\
$r - i $ subsample & 8 & 0 & 5 & 1 \\
\enddata
\tablenotetext{a}{This excludes the 19 fields in the magnitude-limited sample for which we are unable to accurately measure a background value.  We also exclude the five fields for which we are not confident in the red sequence overdensity based on having a measurement above 2$\sigma$ for fewer than three red sequence members.}
\end{deluxetable*}

\subsubsection{$i - [3.6]$ Statistical Analysis \& Combined Overdensity}
Since the current COBRA data set lacks spectroscopic verification and our photometric redshift estimates are made solely based on red sequence colors, it is possible that we are identifying interloping galaxies that have the same photometric color, but are not at the redshift we estimate via EzGal.  Additionally, some fraction of galaxies that do not fall within our red sequence range are at our target redshift.  To correct our measurements to better account for all potential cluster galaxies, we statistically analyze our sample and our color cuts to estimate what fraction of the galaxies that we identify are within our expected redshift range ($\pm$0.10) of the host galaxy's redshift.  To do this, we take advantage of data from the Observations of Redshift Evolution in Large Scale Environments (ORELSE) survey.   ORELSE is a spectroscopic and photometric survey designed to examine the effects of galaxy environment at high-$z$ (0.5 $<$ $z$ $<$ 1.4) by examining the structure out to 10$h^{-1}_{70}$\,Mpc around 20 known clusters at $z$ $>$ 0.6 \citep{Lubin2009,Lemaux2018,Hung2019}.  The high quality spectroscopic data ($\approx$ 100 - 500 confirmed members per structure) and the deep photometric observations have been used to identify high-$z$ structures beyond those it was intended to study \citep[e.g.,][]{Gal2008,Lemaux2018}.

Although ORELSE spans $\approx$ 5 square degrees, for our analysis, we use the data presented in \citet{Hung2019} for four ORELSE fields; SC1604, SC0910, SC0849, and CL1137 (which overlaps with COBRA113733.8+300010, the spectroscopically confirmed cluster from \citet{Blanton2003}).  These fields were chosen because they have confirmed structures at both the low- and high-$z$ ends of the COBRA redshift range and are among the fields with the most comprehensive spectroscopic samples in ORELSE.  One of the strengths of the ORELSE photometric data is the extensive number of bands with deep observations.  Additionally, the ORELSE photometric redshifts are based on multi-band optical to mid-IR SED fitting as opposed to our single or multi-color estimates \citep{Hung2019}.

To compare to the COBRA sample, we take advantage of the similar wavebands covered by both surveys.  All ORELSE fields were observed with $Spitzer$ IRAC 3.6\,$\mu$m and 4.5\,$\mu$m, and each has fainter magnitude limits than our COBRA fields.  However, while every field has coverage with some variation on the $i$-band filter, not every field contains sufficient observations with an SDSS-like $i$-band.  Additionally, for those fields that do have an SDSS-like $i$-band, they are generally much shallower and narrower than these observations in other $i$-band variations \citep{Hung2019}.  To combat this problem and thus allow for us to compare our sample to the largest possible statistical sample, we convert the ORELSE $I^{+}$ and $I_{c}$ bands into an SDSS-like $i$-band using a transformation similar to that discussed in \citet{Gal2008} and shown below (Equation\,\ref{Eq:Shift}). 

\begin{equation}\label{Eq:Shift}
    \sl{i} = A\,\times\,I_{C} + B\,\times\,(R_{C} - I_{C}) + C
\end{equation}{}

\begin{figure}
\centering
\figurenum{8}
\epsscale{1}
\includegraphics[scale=0.5,trim={0.4in 0.1in 0.0in 0.3in},clip=True]{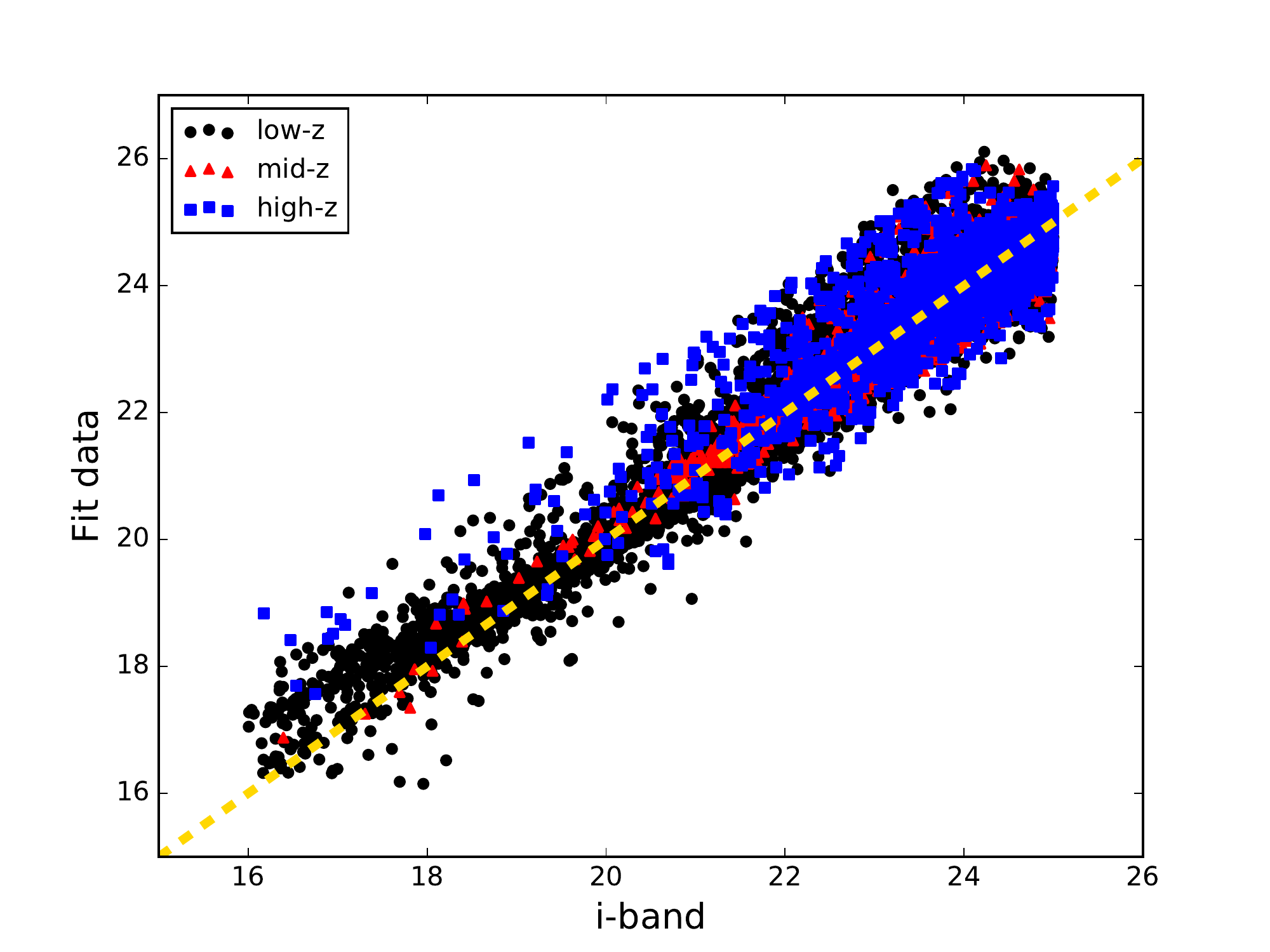}
\caption{A comparison of the data converted to an SDSS-like $i$-band from I$_{c}$ (labelled as Fit data) to the SDSS-like $i$-band observations.  We divide the three samples by redshift, with the low-$z$ sample falling at 0 $<$ $z$ $\leq$ 0.85 (black circles), the mid-$z$ sample falling at 0.85 $<$ $z$ $\leq$ 1.2 (red triangles), and the high-$z$ sample falling at $z$ $>$ 1.2 (blue squares).  The normalized median standard deviation for each sample is 0.18\,mag, 0.16\,mag, and 0.20\,mag respectively.  We plot a one-to-one trend dashed gold line to show how well fit the converted data is to the observed data.  \label{Fig:ColorshiftIci}}
\end{figure}

To do the conversion, we treat A, B, and C as free parameters that we fit to our data using a $\chi^{2}$ minimization.  We first convert I$^{+}$ to I$_{c}$ using a similar version of Equation\,\ref{Eq:Shift} and then convert the data to an SDSS-like $i$-band.  To reduce the scatter in the measurements and our fitting, we divide our sample into three fiducially estimated redshift bins, 0 $<$ $z$ $\leq$ 0.85, 0.85 $<$ $z$ $\leq$ 1.2, and $z$ $>$ 1.2 (See Figure\,\ref{Fig:ColorshiftIci}).  This reduces the noticeable amount of scatter at fainter magnitudes.  We find that a simple magnitude cut did not impact the scatter.  To measure the statistical dispersion of this transformation, we measure the normalized median, absolute deviation and find rather small values for each sample set (0.18\,mag for the low-$z$ sample, 0.16\,mag for the mid-$z$ sample, and 0.20\,mag for the high-$z$ sample).  We applied this same correction to all fields of comparison.

\begin{figure*}
\figurenum{9}
\subfigure{\includegraphics[scale=0.5,trim={0.3in 0.1in 0.6in 0.5in},clip=True]{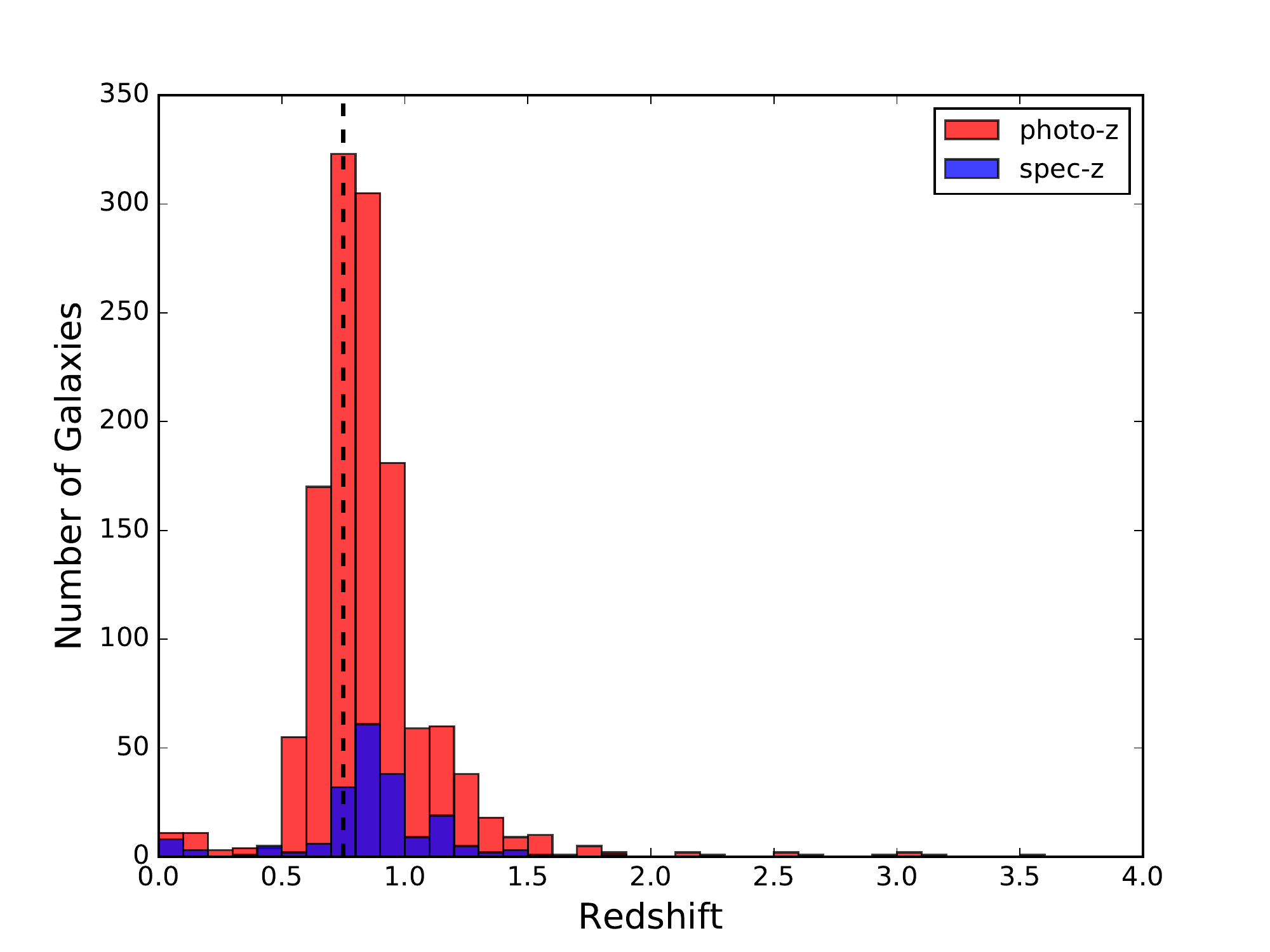}}
\subfigure{\includegraphics[scale=0.5,trim={0.3in 0.1in 0.6in 0.5in},clip=True]{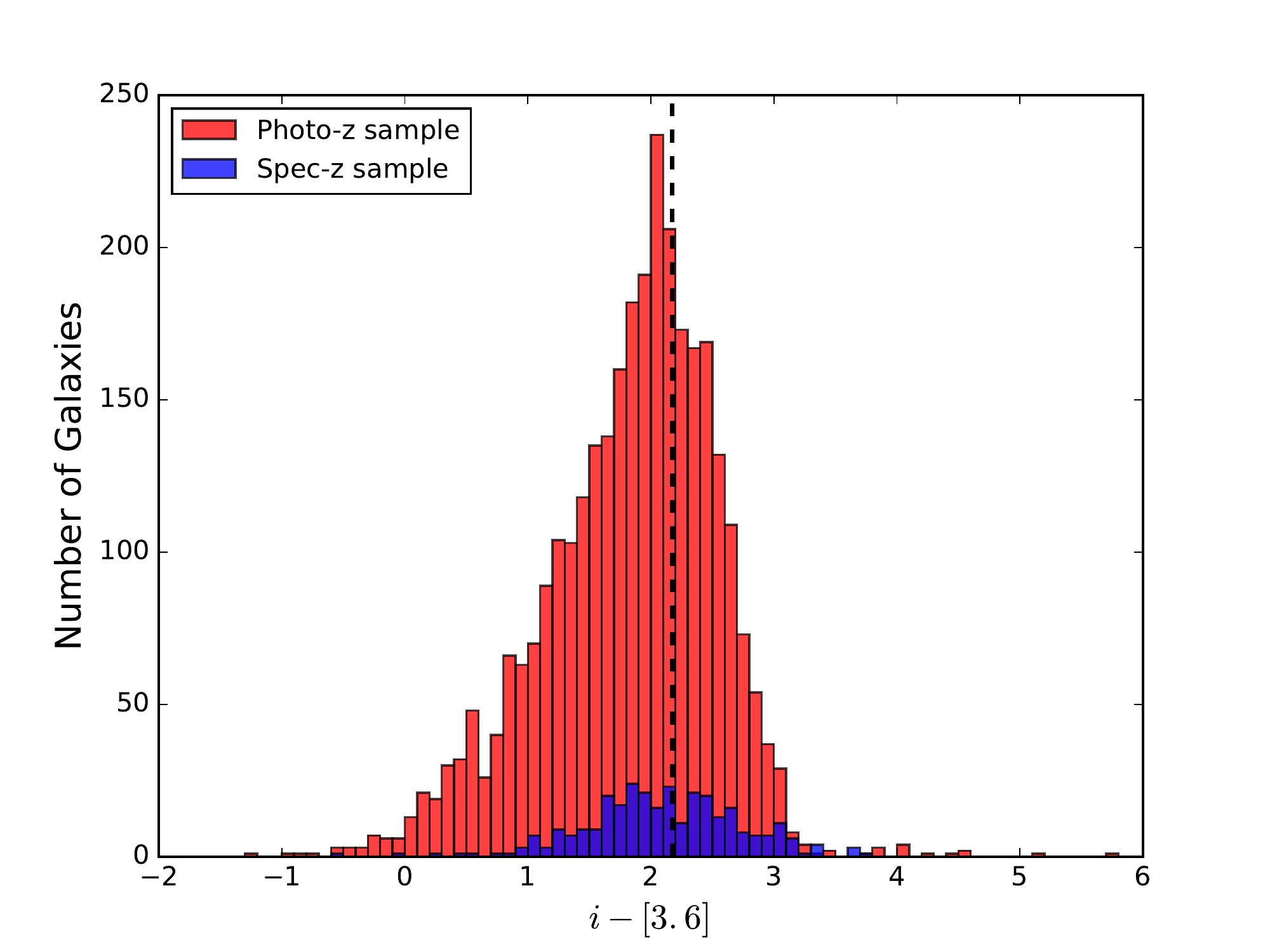}}

\caption{Two different portrayals of the effectiveness of our red sequence cuts using ORELSE data with the detection limits for COBRA012058.9+002140.  On the left hand side, we plot the distribution of redshifts for the expected red sequence color range ($\pm$0.15 about the host color).  On the right-hand side, we plot the distribution of colors for the expected redshift range ($\pm$0.1 about the host redshift estimate).  In each, the photo-$z$ sample is shown in red, while the spec-$z$ sample is shown in blue.  The vertical dashed lines show the host galaxy's redshift (0.75) and color respectively ($i - [3.6]$ = 2.18\,mag).}  \label{Fig:ORELSEfig1}
\end{figure*}

With the transformed ORELSE data, we can estimate the fraction of red sequence, bluer, and redder galaxies that are in a given redshift range to determine correction factors for our measurements.  We apply each individual COBRA magnitude limit to create a unique ORELSE data set for each COBRA field.  To estimate the fraction of red sequence galaxies that have our expected redshift, we remove sources without photometric or spectroscopic redshifts to create 77 unique samples with all the galaxies within the red sequence color range for each field.  To properly sample the ORELSE data, we randomly sample the data set 1,000 times and measure the fraction of fields that fall within the $\pm$0.1 of our redshift estimate.  Although we see general agreement between the ORELSE photometric and spectroscopic samples, because of the large difference in sample size, we follow the values derived from the ORELSE photometric sample.  We find that between 35$\%$ - 55$\%$ of red ORELSE galaxies have the redshift we expect from our EzGal models  (See the left-hand panel of Figure\,\ref{Fig:ORELSEfig1}).  The complete list of expected red sequence fractions is found in Table\,\ref{tb:completeness}. 

To verify that we are looking in the correct color range for a given redshift, we plot the color distribution of all ORELSE galaxies within $\pm$ 0.1 of our redshift estimate and find that out to $z$ $\approx$ 1.2 the host color is at or near the peak of the color distribution (see the right hand panel of Figure\,\ref{Fig:ORELSEfig1}).  When we examine ORELSE galaxies at $z$ $>$ 1.3, we find that the peak is generally bluer than our host color, which follows typical cluster galaxy evolution because red sequence early-type galaxies are not always the dominant population at high-$z$.  It should be noted that this methodology treats each ORELSE galaxy individually.  It is possible that if we accounted for the spatial clustering of galaxies of a similar color, these values may be higher.

For all of our fields, ORELSE galaxies exceed both the red and blue limits of the red sequence color selection range (an example is shown in the right-hand side of Figure\,\ref{Fig:ORELSEfig1}.  While significantly bluer galaxies may be AGN or actively forming stars, the width, specifically slightly blueward of the host galaxy color likely results from the natural slope in the red sequence leading to a population of slightly bluer galaxies still at our target redshift.  Similar to how we estimate the fraction of red sequence galaxies at the target redshift, we also measure what fraction of galaxies at each target redshift are bluer or redder than the COBRA red sequence range.  The bluer and redder subsets are made by using the same COBRA magnitude cuts for each field and identifying ORELSE galaxies bluer than our minimum red sequence color and redder than our maximum red sequence color.  From these bluer and redder samples of ORELSE galaxies, we find that 15$\%$ - 20$\%$ of the bluer ORELSE galaxies and 10$\%$ - 15$\%$ of the redder ORELSE galaxies typically lie within our redshift range, with the fraction of bluer galaxies decreasing with redshift (see Table\,\ref{tb:completeness} for the complete list of fractions).  

The goal of using the ORELSE data is to correct our overdensity measurements to account for the fraction of red sequence galaxies that are at our target redshift and account for redder and bluer galaxies that are at the target redshift.  In doing this, we can account for some of the slightly bluer galaxies we miss by assuming the red sequence has no slope.  We quantify our new overdensity measurement by determining a combined overdensity that factors in the number of red sequence galaxies, redder, and bluer galaxies above our background distribution (again measured using unique versions of the combined SpUDS and UDS fields for each COBRA field).  The combined overdensity measurement (Equation\,\ref{Eq:NEWover}) is an expansion of the original overdensity measurement (Equation\,\ref{Eq:NEW}), and fully accounts for the completeness fraction of the COBRA red sequence measurements, as well as the populations of redder and bluer galaxies measured using the ORELSE data (see Table\,\ref{tb:combo} for the complete list of values).

\begin{equation}\label{Eq:NEWover}\footnotesize
    \sigma_{combined} = \frac{(N_{RS} - M_{RS})f_{RS} + (N_{B} - M_{B})f_{B} + (N_{R} - M_{R})f_{R}}{((\sigma_{RS}f_{RS})^2 + (\sigma_{B}f_{B})^2 + (\sigma_{R}f_{R})^2)^{0.5}}\footnote{As in Equation\,\ref{Eq:NEW}, for some fields, the weighted mean number of galaxies in the Gaussian background regions (for each of the three color ranges) is greater than the number of galaxies in the target region.  We report these fields as having negative overdensities and negative significances.}
\end{equation}

In the combined overdensity, N$_{RS}$ is the number of red sequence galaxies in the 1$\arcmin$ region, M$_{RS}$ is the mean of the Gaussian distribution of the red sequence background, and $\sigma_{RS}$ is the standard deviation of the Gaussian fit of the background red sequence distribution.  These three values are identical to those in Equation\,\ref{Eq:NEW}.  N$_{B}$ is the number of bluer galaxies, M$_{B}$ is the mean of the Gaussian distribution of bluer galaxies, and $\sigma_{B}$ is the standard deviation of the Gaussian fit for the background bluer galaxies.  The versions with the ``R" subscript represent the same components, but for the redder galaxies.  We use the same physical background regionss for the bluer and redder galaxies our red sequence backgrounds.  f$_{RS}$, f$_{B}$, and f$_{R}$ are the fraction of red sequence, bluer, and redder galaxies that fall within $\pm$0.1 of the cluster redshift estimate (see Table\,\ref{tb:completeness} for the complete list of values).  To verify that our redder and bluer backgrounds also follow a Gaussian distribution, we checked the values of the median, 16th, and 84th percentile and find strong agreement with our Gaussian fit.   Like our previous red sequence overdensities, we require at least three red sequence galaxies for a given field to be a combined overdensity cluster candidate.

\begin{figure}
\centering
\figurenum{10}
\epsscale{1}
\includegraphics[scale=0.4,trim={0.75in 0.5in 0.0in 0.95in},clip=True]{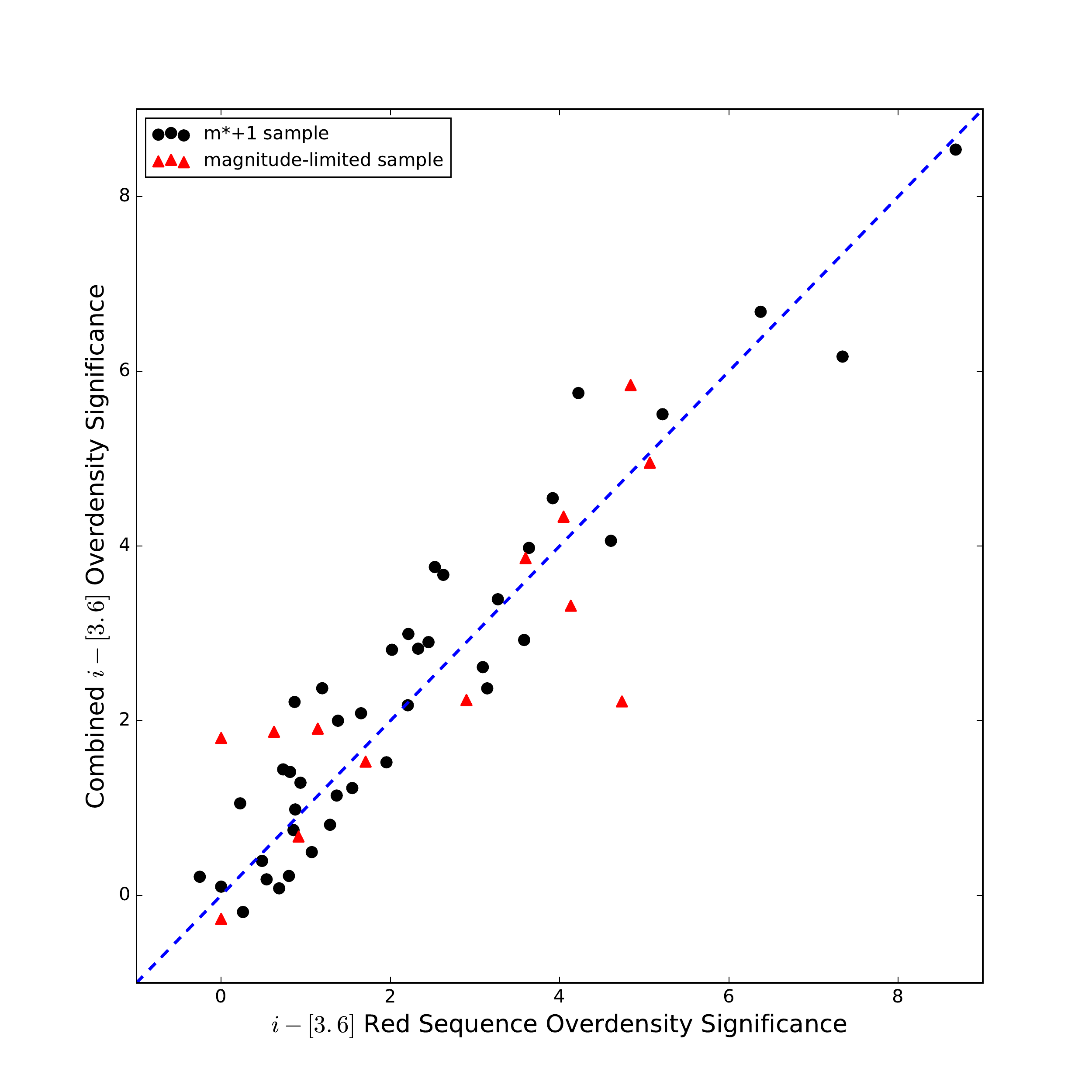}
\caption{A comparison of the $i - [3.6]$ red sequence overdensity to the combined $i - [3.6]$ overdensity.  The m*+1 sample is shown in black circles, while the magnitude-limited sample is shown in red triangles.  A one-to-one dashed line is shown in blue.  \label{Fig:Combinedredsequencecheck}}
\end{figure}

As shown in Figure\,\ref{Fig:Combinedredsequencecheck}, the two overdensity measurements have a linear relationship, with the combined overdensity giving a similar value to the red sequence overdensity.  The nearly one-to-one trend between the red sequence signficance and combined overdensity significance strengthens our confidence in the detection of our COBRA red sequence clusters because clusters should also be overdense, even if just slightly, in the number of non-red galaxies relative to the field.  

Since the initial COBRA overdensity measurements from \citet{Paterno-Mahler2017} did not account for galaxy color, these combined overdensities allow us to further characterize the fractions of bluer and redder cluster galaxies and better estimate which fields are the strongest cluster candidates.  In total, we identify 20 combined cluster candidates in the m*+1 sample and 7 in the magnitude-limited sample.  Of these fields, one is not a red sequence cluster candidate, although it has a 1.7$\sigma$ overdensity in just the red sequence measurement.  However, one red sequence cluster candidate does fall below the 2$\sigma$ threshold for the combined overdensity.  Overall, the similarity between these two measurements further strengthens our confidence that our red sequence detections are, in fact, real cluster candidates.

\subsection{$[3.6] - [4.5]$ Analysis $\&$ Results}
A $[3.6] - [4.5]$ $>$ $-$0.15\,mag color cut can efficiently identify high-$z$ ($z$ $>$ 1.2) galaxies, including red sequence and star forming galaxies \citep[e.g.,][]{Papovich2008,Wylezalek2014,Cooke2015}.  Due to the effectiveness of this cut, and the bimodal nature of this color-redshift distribution at low-$z$, we apply this color cut to the 20 fields at $z$ $>$ 1.2 within our sample with these observations.  Although this cut is an excellent identifier of high-$z$ galaxies, our measurements are hindered by the limiting magnitude of our images, 21.4\,mag in both bands, which corresponds to an EzGal modeled m* galaxy at $z$ = 2.1 in 3.6\,$\mu$m and an m* galaxy at $z$ = 2.25 in 4.5\,$\mu$m.  

With the $[3.6] - [4.5]$ color cut, and the SpUDS 3.6\,$\mu$m and 4.5\,$\mu$m combined images as our $[3.6] - [4.5]$ background, we identify four of the highest-redshift cluster candidates in the COBRA survey above the 2$\sigma$ significance (see Table\,\ref{tb:ch1ch2AGN}; for the overdensities of the remaining fields, see Table\,\ref{tb:2}).  Two of the cluster candidates are not identified in \textsection{4.1.1} or Table\,\ref{tb:ich1AGN} as they are at $z$ $>$ 1.4.  The other two $[3.6] - [4.5]$ cluster candidates are identified by our $i - [3.6]$ red sequence analysis.  However, each cluster candidate is identified in \citet{Paterno-Mahler2017}.

\begin{deluxetable*}{lccccccccccccccc}
    \tablenum{8}
    \tablecolumns{14}
    \tabletypesize{\small}
    \tablecaption{COBRA $[3.6] - [4.5]$ Cluster Candidates\label{tb:ch1ch2AGN}}
    \tablewidth{0pt}
    \tabletypesize{\footnotesize}

    \setlength{\tabcolsep}{0.05in}
    \tablehead{
    \colhead{Field}&
    \colhead{Redshift ($z$)\tablenotemark{i}}&
    \multicolumn{6}{c}{Red Sequence Overdensity}&
    \multicolumn{6}{c}{Combined Overdensity}
    \cr
    \colhead{}&
    \colhead{}&
    \multicolumn{3}{c}{AGN Center}&
    \multicolumn{3}{c}{RS Center}&
    \multicolumn{3}{c}{AGN Center}&
    \multicolumn{3}{c}{RS Center}
    \cr
    \colhead{}&
    \colhead{}&
    \colhead{N\tablenotemark{a}}& 
    \colhead{$\Delta$N\tablenotemark{b}}&
    \colhead{$\sigma$\tablenotemark{c}} &
    \colhead{N\tablenotemark{a}}& 
    \colhead{$\Delta$N\tablenotemark{b}}&
    \colhead{$\sigma$\tablenotemark{c}} &
    \colhead{N$_{total}$\tablenotemark{d}}& 
    \colhead{$\Delta$N$_{total}$\tablenotemark{e}}&
    \colhead{$\sigma_{total}$\tablenotemark{f}} &
    \colhead{N$_{total}$\tablenotemark{d}}& 
    \colhead{$\Delta$N$_{total}$\tablenotemark{e}}&
    \colhead{$\sigma_{total}$\tablenotemark{f}}
    \cr
    }
    
\startdata
COBRA072805.2+312857\tablenotemark{g} & 1.75 & 22 & 12.2 & 3.3 & 19 & 9.2 & 2.5 & 47 & 7.9 & 3.6 & 46 & 6.2 & 2.9\\
COBRA100841.7+372513 & 1.20/1.35 & 21 & 11.2 & 3.0 & 25 & 15.2 & 4.1 & 39 & 6.8 & 3.1 & 40 & 9.0 & 4.1\\
COBRA103256.8+262335\tablenotemark{g} & 2.18 & 18 & 8.2 & 2.2 & 17 & 7.2 & 1.9 & 34 & 5.0 & 2.3 & 31 & 4.3 & 2.0\\
COBRA104254.8+290719 & 1.35/1.05 & 21 & 11.2 & 3.0 & 22 & 12.2 & 3.3 & 41 & 7.0 & 3.2 & 44 & 7.7 & 3.5\\
COBRA121128.5+505253\tablenotemark{h} & 1.364 & 15 & 5.2 & 1.4 & 20 & 10.2 & 2.7 & 41 & 3.9 & 1.7 & 45 & 6.7 & 3.1\\
COBRA141155.2+341510\tablenotemark{h} & 1.818 & 12 & 2.2 & 0.6 & 18 & 8.2 & 2.2 & 39 & 2.2 & 1.0 & 37 & 5.2 & 2.4\\
COBRA222729.1+000522\tablenotemark{h} & 1.513 & 11 & 1.2 & 0.3 & 20 & 10.2 & 2.7 & 34 & 1.3 & 0.6 & 45 & 6.7 & 3.1\\
\enddata
\tablenotetext{a}{N = The total number of red sequence members in the 1$\arcmin$ region.}
\tablenotetext{b}{$\Delta$N = The excess of counts in the 1$\arcmin$ region above the background.}
\tablenotetext{c}{$\sigma$ = The significance calculated using Equation\,\ref{Eq:NEW}.}
\tablenotetext{d}{N$_{total}$ = The total number of galaxies (red sequence and bluer) in the 1$\arcmin$ region.}
\tablenotetext{e}{$\Delta$N$_{total}$ = The excess of galaxies in the 1$\arcmin$ region above the combined background adjusted for the red sequence completeness fraction.}
\tablenotetext{f}{$\sigma_{total}$ = The significance calculated using a slightly altered version of Equation\,\ref{Eq:NEWover} (without the redder term).}
\tablenotetext{g}{Fields that are cluster candidates with the AGN center, but not identified in Table\,\ref{tb:ich1AGN}.}
\tablenotetext{h}{Fields that are cluster candidates with just the high-$z$ galaxy center, but not the AGN center or in Table\,\ref{tb:ich1AGN}.}
\tablenotetext{i}{Fields with mutliple redshift estimates are due to disagreements in our EzGal photo-$z$ estimates for the different colors used.}
\end{deluxetable*}

\subsubsection{$[3.6] - [4.5]$ Statistical Analysis \& Combined Overdensity}
\begin{figure}
\centering
\figurenum{11}
\epsscale{1}
\includegraphics[scale=0.4,trim={0.6in 0.5in 0.0in 0.9in},clip=True]{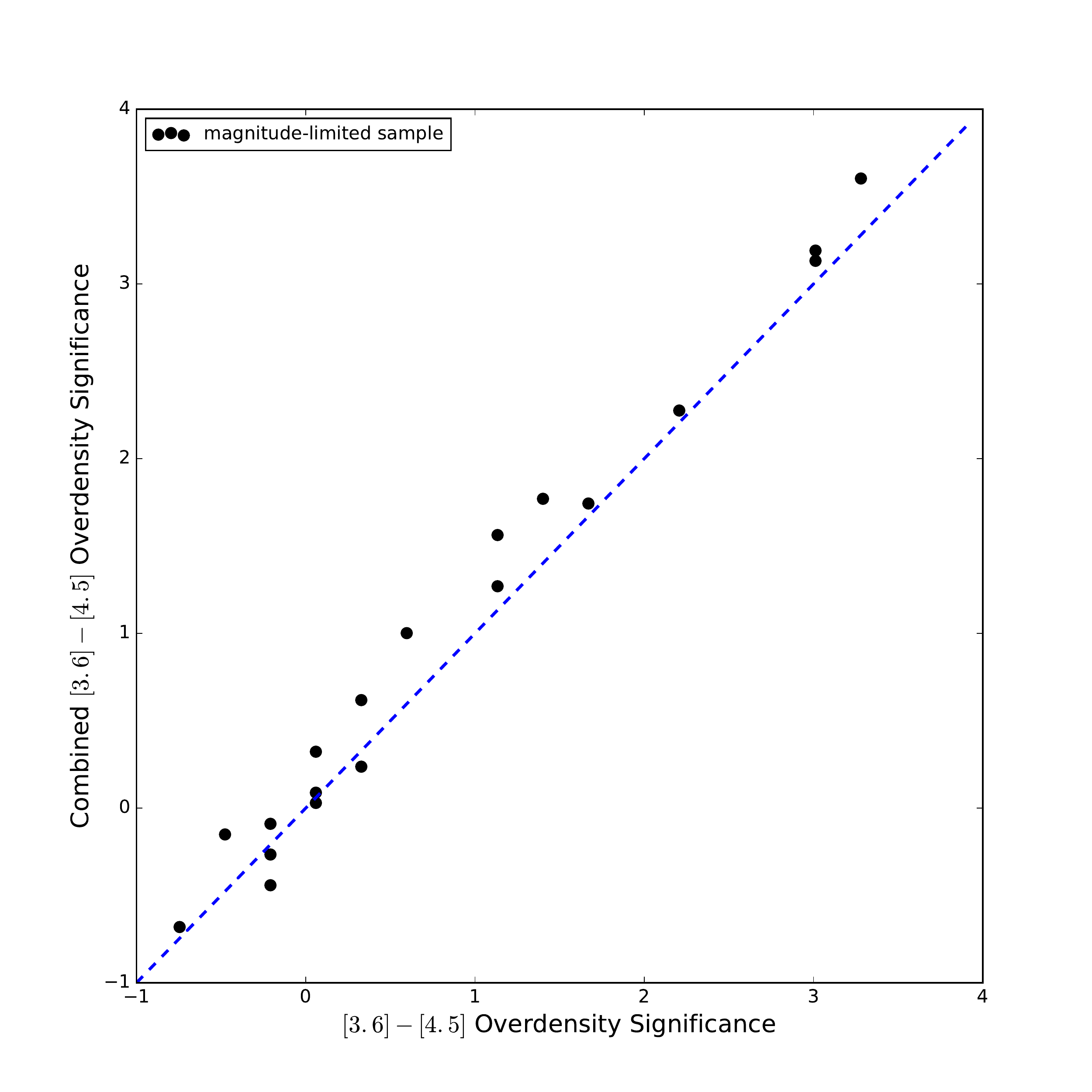}
\caption{A comparison of the $[3.6] - [4.5]$ $>$ -0.15 overdensity to the combined $[3.6] - [4.5]$ overdensity.  A one-to-one dashed line is shown in blue.   \label{Fig:Combinedch1ch2}}
\end{figure}

Like our analysis in \textsection{4.1.2}, we again use data from the ORELSE sample to estimate what fraction of ORELSE galaxies that fall within our given color range are within our redshift range.  Although ORELSE is designed to go out to $z$ $\approx$ 1.4 \citep{Lubin2009}, the number of galaxies with photometric redshifts is roughly flat out to $z$ $\approx$ 1.8 (with $\approx$ 20,000 galaxies per redshift bin with $\Delta$z = 0.1), thus allowing us to estimate the fraction of high-$z$ ORELSE galaxies that are within our color range ($[3.6] - [4.5]$ $>$ $-$0.15) for the majority of the COBRA sample.  Given that number of high-$z$ ORELSE galaxies decreases at redshifts similar to where we approach the COBRA magnitude limit, we treat the ORELSE sample as a representative sample.  Since the ORELSE data set contains both $Spitzer$ bands, no transformations are necessary.  Using a similar methodology to the $i - [3.6]$ analysis, we measure the fraction of ORELSE galaxies with $[3.6] - [4.5]$ $>$ -0.15 at $z$ $>$ 1.2, as well as the fraction of bluer galaxies at $z$ $>$ 1.2.  Given the well-studied nature of this color cut, it is not surprising that we find that 58.1$\%$ of galaxies in this range are at $z$ $>$ 1.2, while only 6.4$\%$ of bluer galaxies are at this redshift range, further reinforcing the effectiveness of this color cut.  Like the $i - [3.6]$ red sequence analysis, we see a tight correlation between the two measurements (see Figure\,\ref{Fig:Combinedch1ch2} and Table\,\ref{tb:combo} for the complete list of combined overdensities).  This trend is likely due to the small fraction of bluer high-$z$ galaxies, with both methodologies (red sequence and combined overdensity) identifying the same four high-$z$ cluster candidates.

\subsection{$r$ $-$ $i$ Analysis $\&$ Results}
\begin{figure*}
\figurenum{12}
\subfigure{\includegraphics[scale=0.5,trim={0.4in 0.1in 0.6in 0.5in},clip=True]{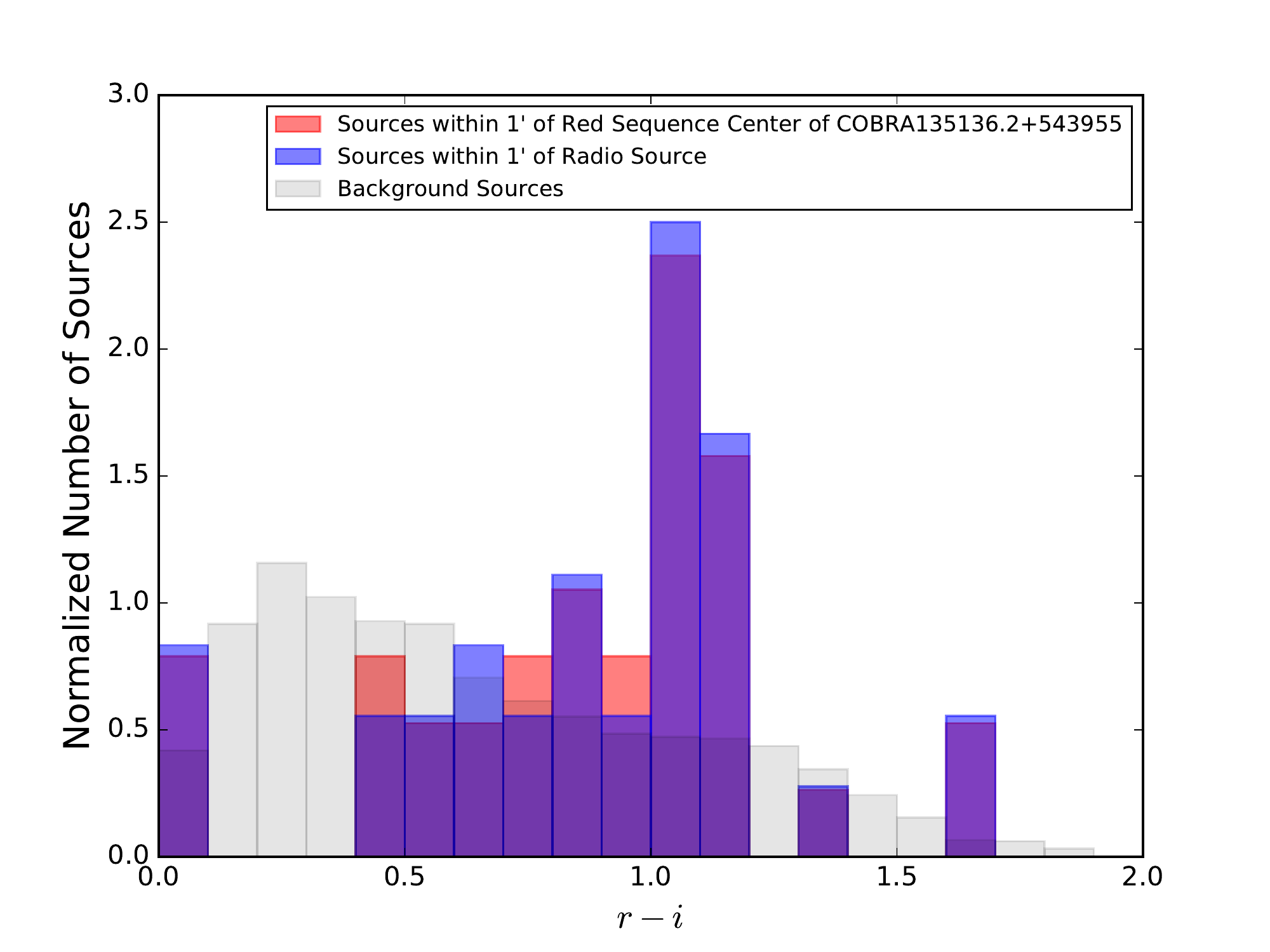}}
\subfigure{\includegraphics[scale=0.5,trim={0.3in 0.1in 0.6in 0.5in},clip=True]{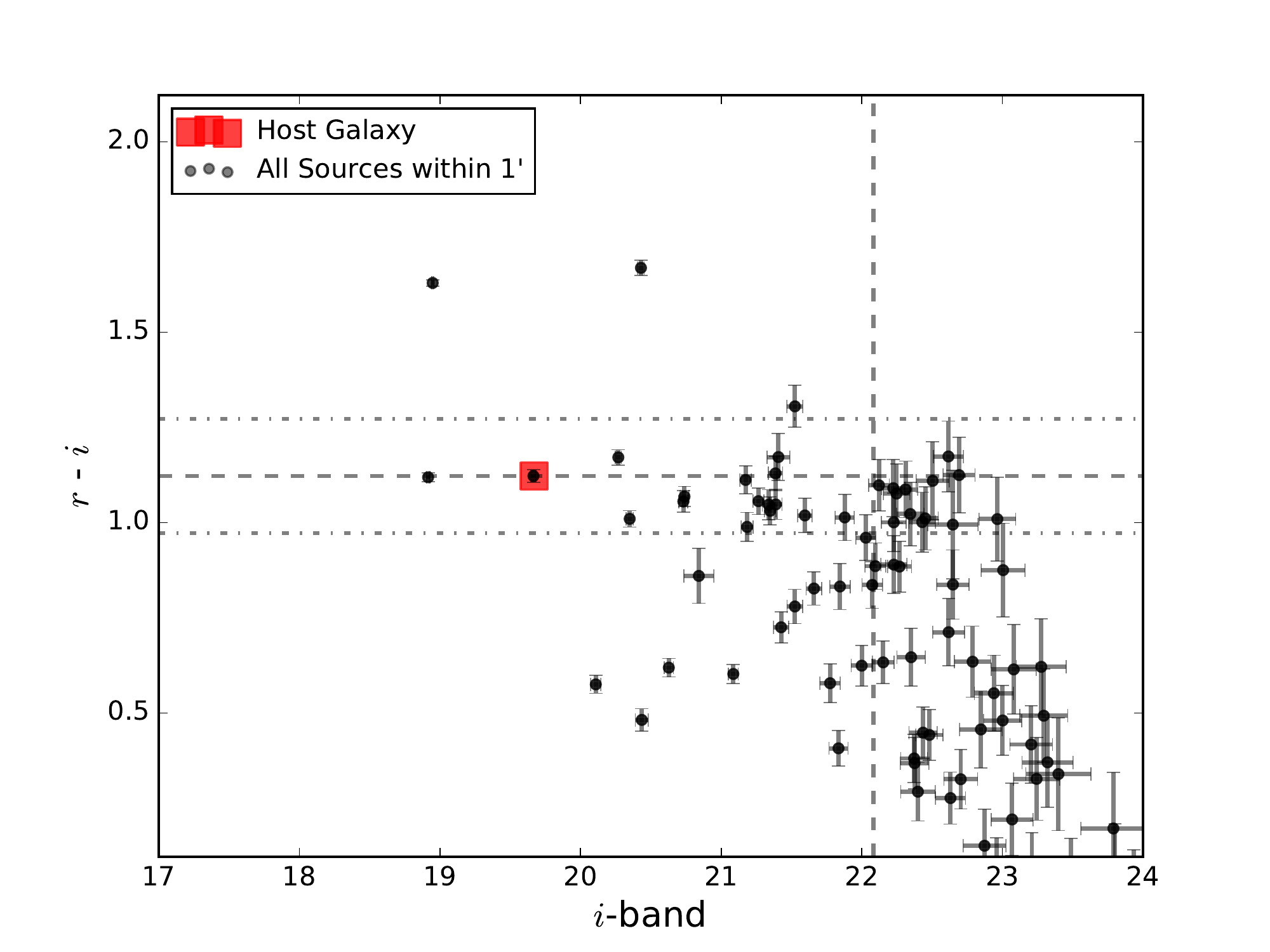}}
\subfigure{\includegraphics[scale=0.5,trim={0.4in 0.1in 0.6in 0.5in},clip=True]{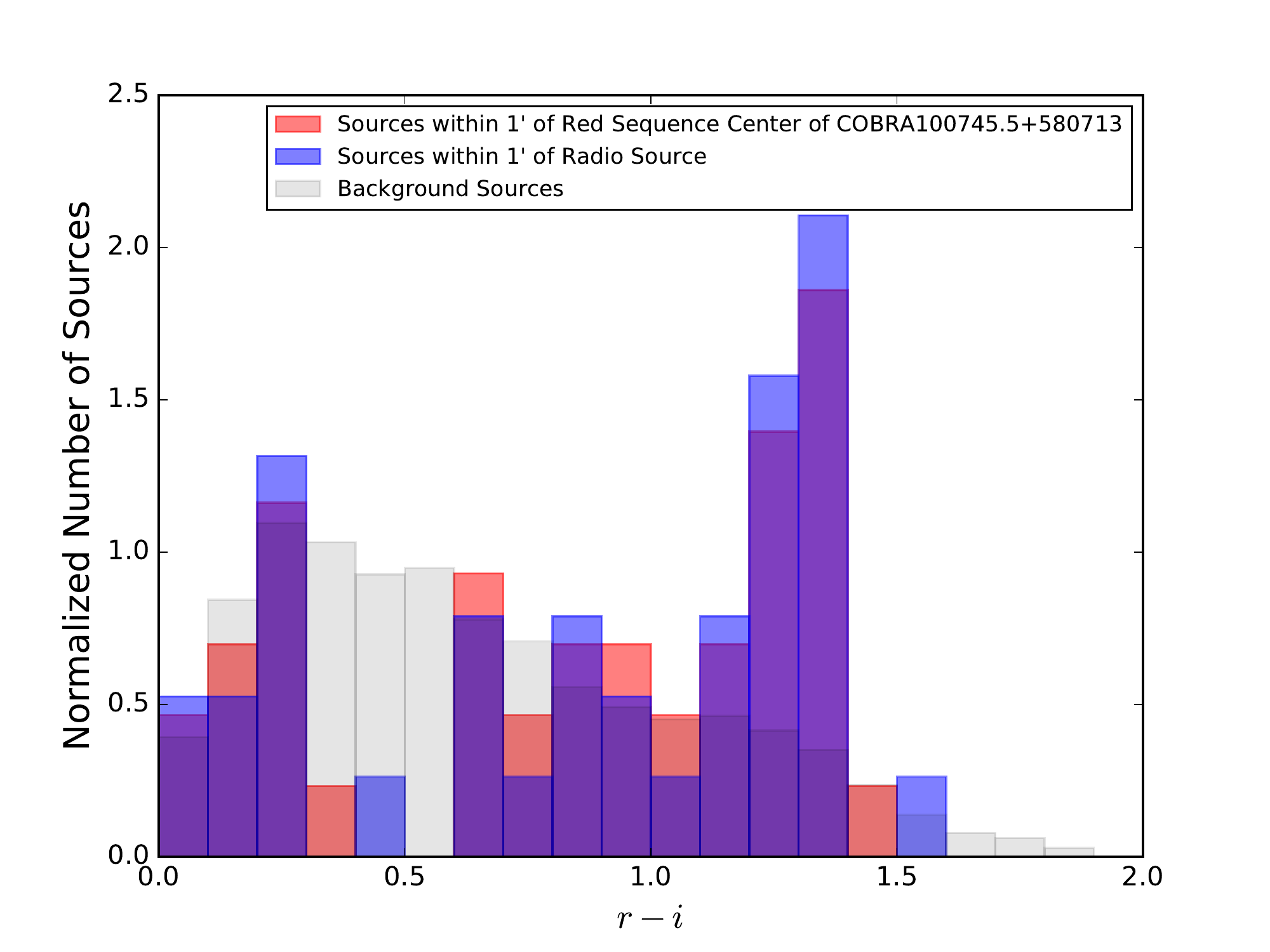}}
\subfigure{\includegraphics[scale=0.5,trim={0.3in 0.1in 0.6in 0.5in},clip=True]{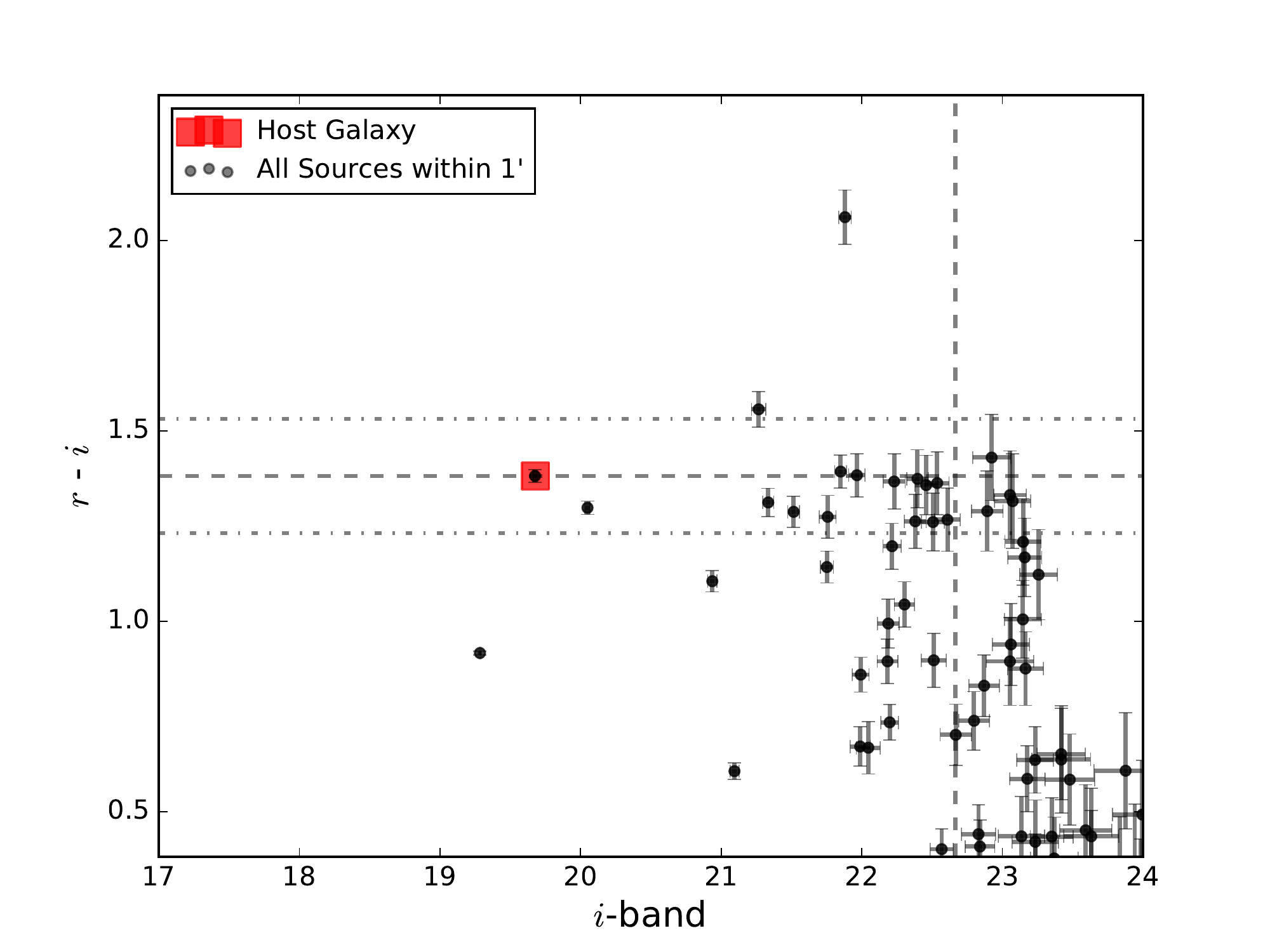}}

\caption{Histograms and their corresponding CMDs for COBRA135136.2+543955 and COBRA100745.5+580713.  Each plot shows the distributions of galaxy color in $r - i$ for the galaxies within 1$\arcmin$ of either the radio source (shown in blue) or the new red sequence center (shown in red).  The background distribution is shown in grey.  All three of these distributions are normalized to one another.  For a full explanation of how the red sequence center is chosen, see \textsection{5}.  For the CMDs, the host galaxies are shown in red squares, while the other galaxies detected within 1$\arcmin$ of the radio source are shown in black circles.  The horizontal grey dashed line shows the color of the host galaxy.  The two grey dot-dashed lines show our estimate of the red sequence color range.  The vertical dashed line shows the value of the m*+1 magnitude limit at the cluster candidates redshift.  Both examples shown are red sequence cluster candidates.  COBRA135136.2+543955 is at $z$ $\approx$ 0.55 and the $i$-band magnitude limit is 22.08 (m*+1) and the $r$-band magnitude limit is 24.50.  COBRA100745.5 is at $z$ = 0.656 and the $i$-band magnitude limit is 22.66 (m*+1) and the $r$-band magnitude limit is 25.00. \label{Fig:ricolorhist}  }
\end{figure*}

To further explore the m*+1 sample, we analyze the 14 fields with $r$- and $i$-band observations at $z$ $<$ 1.0, the m*+1 magnitude limit for our $i$-band images.  Although we have $r$- and $i$-band observations for fields at higher-redshifts, the m*+1 limit and the redshifted wavelength of the 4000\,\AA\,\,break falling outside of this color range at $z$ $\approx$ 0.91 lead to our focus on the m*+1 sample.  To again evaluate the spread in our red sequence galaxies, we plot $r - i$ histograms and CMDs (see Figure\,\ref{Fig:ricolorhist}).  Unlike the $i - [3.6]$ histograms, we see a large bluer distribution at $r - i$ $>$ 0.9\,mag due to foreground stars and galaxies.  Based on this blue peak and the nature of the $r$ $-$ $i$ color-redshift trend (see Figure\,\ref{Fig:4}) and in agreement with the literature for high-$z$ cluster red sequences  \citep[e.g.,][]{Blakeslee2003,Mei2006,Mei2009,Snyder2012,Lemaux2012,Cerulo2016}, we estimate the red sequence width as $\pm$ 0.15\,mag, identical to that of our $i - [3.6]$ analysis.  

Since there is no UDS $r$-band image, we take advantage of the large FOV on our DCT observations taken with the LMI to create a composite $r - i$ background measurement.  From our 21 fields with observations in both bands and an average seeing less than 1$\farcs$2, we create a composite background $r - i$ region for each field made up of 252 1$\arcmin$ regions.  Like in our $i - [3.6]$ analysis, we create a unique $r - i$ background for each of the 14 fields from the same regions.  

We find that 8 of our 14 fields are red sequence cluster candidates (see Table\,\ref{tb:riAGN}; for the complete list of overdensities, see Table\,\ref{tb:2}).  Of these eight fields, seven were identified as cluster candidates with the $i - [3.6]$ color, which further confirms our prior findings.  The only addition is COBRA075516.6+171457, which is one of the strongest candidates in $r$ $-$ $i$, but has a 1.1$\sigma$ overdensity measurement in $i - [3.6]$.  This discrepancy may be due to the differing angular resolution of the LMI and IRAC.  In our DCT image, we see one galaxy surrounded by smaller satellites, while in our $Spitzer$ image, the BCG is blended with these satellites. This removes potential red sequence galaxies from the $i - [3.6]$ analysis and artificially lowers the overdensity (and impacts the host color).

\begin{deluxetable*}{lccccccccccccc}
    \tablenum{9}
    \tablecolumns{14}
    \tabletypesize{\small}
    \tablecaption{COBRA $r - i$ Red Sequence Cluster Candidates\label{tb:riAGN}}
    \tablewidth{0pt}
    \tabletypesize{\footnotesize}

    \setlength{\tabcolsep}{0.05in}
    \tablehead{
    \colhead{Field}&
    \colhead{Redshift ($z$)}&
    \multicolumn{6}{c}{Red Sequence Overdensity}&
    \multicolumn{6}{c}{Combined Overdensity}
    \cr
    \colhead{}&
    \colhead{}&
    \multicolumn{3}{c}{AGN Center}&
    \multicolumn{3}{c}{RS Center}&
    \multicolumn{3}{c}{AGN Center}&
    \multicolumn{3}{c}{AGN Center}
    \cr
    \colhead{}&
    \colhead{}&
    \colhead{N\tablenotemark{a}}& 
    \colhead{$\Delta$N\tablenotemark{b}}&
    \colhead{$\sigma$\tablenotemark{c}} &
    \colhead{N\tablenotemark{a}}& 
    \colhead{$\Delta$N\tablenotemark{b}}&
    \colhead{$\sigma$\tablenotemark{c}} &    
    \colhead{N$_{total}$\tablenotemark{d}}& 
    \colhead{$\Delta$N$_{total}$\tablenotemark{e}}&
    \colhead{$\sigma_{total}$\tablenotemark{f}}&
    \colhead{N$_{total}$\tablenotemark{d}}& 
    \colhead{$\Delta$N$_{total}$\tablenotemark{e}}&
    \colhead{$\sigma_{total}$\tablenotemark{f}}
    \cr
    } 
\startdata
COBRA005837.2+011326 & 0.71 & 9 & 7.2 & 3.9 & 10 & 8.2 & 4.5 & 39 & 3.1 & 1.9 & 39 & 3.3 & 2.0\\
COBRA012058.9+002140 & 0.80 & 19 & 15.3 & 5.9 & 17 & 13.3 & 5.1 & 68 & 9.6 & 5.0 & 68 & 8.9 & 4.6\\
COBRA075516.6+171457\tablenotemark{g} & 0.64 & 12 & 9.3 & 5.0 & 14 & 11.3 & 6.0 & 46 & 5.4 & 4.4 & 52 & 6.3 & 5.1\\
COBRA100745.5+580713 & 0.656 & 14 & 12.2 & 6.5 & 14 & 12.2 & 6.5 & 40 & 4.4 & 3.4 & 41 & 4.6 & 3.6\\
COBRA113733.8+300010 & 0.96 & 13 & 9.1 & 3.8 & 10 & 6.1 & 2.5 & 55 & 0.1 & 0.0 & 59 & 0.4 & 0.2\\
COBRA121712.2+241525 & 0.90 & 28 & 13.4 & 5.5 & 17 & 12.4 & 5.1 & 78 & 6.8 & 3.2 & 79 & 6.8 & 3.1\\
COBRA135136.2+543955 & 0.55 & 16 & 13.9 & 9.0 & 17 & 14.9 & 9.6 & 36 & 3.4 & 3.0 & 38 & 3.8 & 3.3\\
COBRA135838.2+384722\tablenotemark{g} & 0.81 & 4 & $-$0.2 & $-$0.1 & 10 & 5.8 & 2.3 & 44 & $-$0.0 & $-$0.0 & 68 & 5.4 & 3.1\\
COBRA164611.2+512915 & 0.351 & 18 & 16.3 & 8.8 & 16 & 14.3 & 7.7 & 26 & 5.5 & 6.6 & 23 & 4.7 & 5.6\\
\enddata
\tablenotetext{a}{N = The total number of red sequence members in the 1$\arcmin$ region.}
\tablenotetext{b}{$\Delta$N = The excess of counts in the 1$\arcmin$ region above the background.}
\tablenotetext{c}{$\sigma$ = The significance calculated using Equation\,\ref{Eq:NEW}.}
\tablenotetext{d}{N$_{total}$ = The total number of galaxies (red sequence, redder, and bluer) in the 1$\arcmin$ region.}
\tablenotetext{e}{$\Delta$N$_{total}$ = The excess of galaxies in the 1$\arcmin$ region above the combined background adjusted for the red sequence completeness fraction.}
\tablenotetext{f}{$\sigma_{total}$ = The significance calculated using Equation\,\ref{Eq:NEWover}.}
\tablenotetext{g}{Fields that are not identified as cluster candidates in Table\,\ref{tb:ich1AGN} with the radio source center.}
\end{deluxetable*}

\subsubsection{$r - i$ Statistical Analysis \& Combined Overdensity}
Following our analysis of the fraction of ORELSE red sequence galaxies that lie within our target redshift range for the $i - [3.6]$ analysis, we perform a similar analysis using the ORELSE data for our $r - i$ analysis.  Like with the SDSS-like $i$-band, not all ORELSE fields have available or expansive SDSS-like $r$-band observations.  Thus, we convert the R$_{C}$ to an SDSS-like $r$-band using identical methodology as I$_{C}$ to $i$-band (see Equation\,\ref{Eq:Shift}).  We again test the statistical dispersion of this sample using the normalized median, absolute deviation and find small values, showing the strength of the fit (0.15\,mag for the low-$z$ sample, 0.14\,mag for the mid-$z$ sample, and 0.17\,mag for the high-$z$ sample).  Using the shifted ORELSE $r$- and $i$-band data, we measure the fraction of ORELSE galaxies that are red sequence, redder, or bluer galaxies and use this information to measure a combined overdensity (Equation\,\ref{Eq:NEWover}).  These completeness fraction values are generally slightly below the $i - [3.6]$ values (see Table\,\ref{tb:completeness} for the complete list).  

\begin{figure}
\centering
\figurenum{13}
\epsscale{1}
\includegraphics[scale=0.4,trim={0.6in 0.5in 0.0in 0.9in},clip=True]{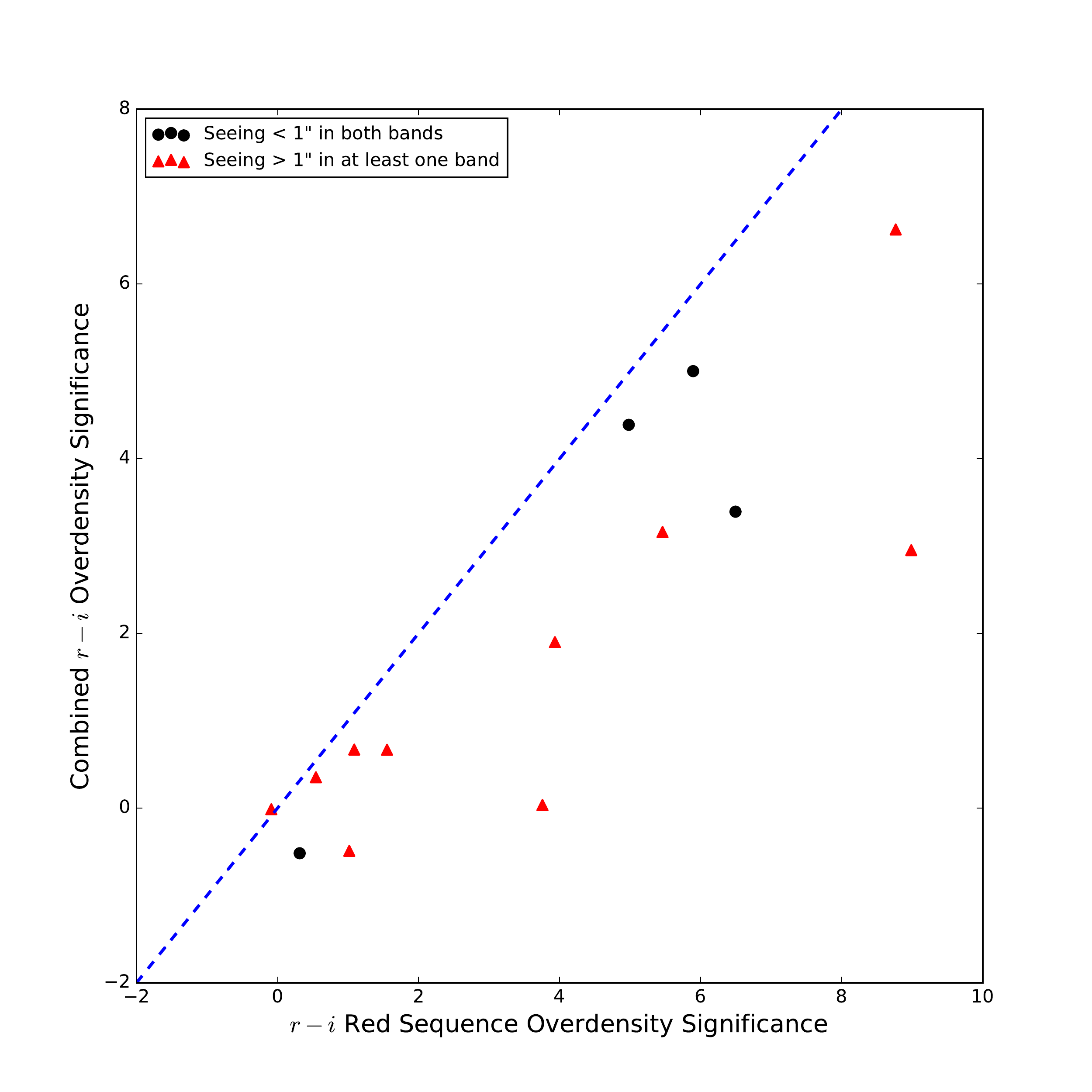}
\caption{A comparison of the $r - i$ red sequence overdensity to the combined overdensity for the 14 m*+1 fields with $r$- and $i$-band observations.  Fields with seeing greater than 1$\arcsec$ in at least one band are shown in red triangles, while fields with seeing less than 1$\arcsec$ in both bands are shown in black circles.  A one-to-one dashed line is shown in blue.   \label{Fig:Combinedri}}
\end{figure}

Unlike the strong one-to-one trend between the $i - [3.6]$ red sequence overdensity significance and the combined $i - [3.6]$ overdensity significance (see Figure\,\ref{Fig:Combinedredsequencecheck} and Table\,\ref{tb:combo} for the complete table of combined overdensities), the $r - i$ red sequence overdensity significance appears to be significantly offset from it's combined counterpart (see Figure\,\ref{Fig:Combinedri}).  In examining the overdensity values and the CMDs, the major discrepancy appears to arise in lack of bluer galaxies measured relative to what is expected from our background.  By dividing the subsample by the seeing, we highlight that fields with better seeing are somewhat closer to the one-to-one trend, especially at higher overdensities.  For the fields where this discrepancy is the largest, we visually inspected the DCT images and our histograms of the individual level of completeness for each field.  We find that our $r$-band images generally have worse sky conditions and slightly worse seeing than our best fields used to create the background.  Additionally, the magnitude limit of these $r$-band images is closer to the peak in completeness than for our fields taken with worse seeing.  As a result, the number of faint blue galaxies relative to the expected background is greatly diminished for these fields.  This demonstrates that the variability in sky conditions impacts our confidence in the $r - i$ analysis and further emphasizes why we treat it as a supplemental measurement relative to the $i - [3.6]$ red sequence analysis.  An alternative explanation is that at lower redshifts, specifically where the the $r - i$ color is most effective, cluster populations are far more distinct from the field populations, with significantly fewer bluer galaxies at bright magnitudes.  Since we don't factor in a specific cluster population to our ORELSE analysis, we may be weighting our measurement too heavily toward the bluer galaxies in this color, making our statistics less valid.

\section{Determining the Cluster Center and Re-Evaluating Cluster Candidates}

As discussed in \citet{Sakelliou2000}, bent, double-lobed radio sources need not reside in the center of galaxy clusters.  As evidenced by our CMDs (see Figure\,\ref{Fig:colorhist} and Figure\,\ref{Fig:ricolorhist}), not all of our host galaxies are BCGs, allowing for the possibility that some are fast-moving galaxies on cluster outskirts.  Since clusters hosting bent, double-lobed radio sources are found in merging and relaxed clusters, bent sources may be farther offset from the cluster center.  Thus, unlike the previous overdensity measurements (discussed in \textsection{4} and in \citet{Paterno-Mahler2017}), we should not automatically assume the bent AGN is located at the cluster center.  To more accurately locate each COBRA cluster candidate, we evaluate the 2D spatial distribution of all surrounding galaxies in color space, specifically looking for galaxies at similar colors to the host galaxy.  By identifying a cluster center based on red sequence galaxies, we should be able to better trace the true cluster center for relaxed clusters because this center has the strongest correlation with the X-ray cluster center \citep[e.g.,][]{Rumbaugh2018}.  

In this section, we measure the surface density of red sequence sources to better estimate cluster centers.  From these new centers, we re-measure both the red sequence overdensity and the combined overdensity to identify additional red sequence COBRA cluster candidates.  We measure the surface density within our $Spitzer$ and DCT images by counting all sources of the target color within a 10$\arcsec$ radius of each point in a regular grid of 10$\arcsec$ spacing imposed on the image.  At this grid spacing, there is double-counting of sources due to the overlap of search radii. However, the overall shape of the density distribution is unaffected, especially after the image is smoothed, as described below.    
 
To determine a new cluster center, we smooth the surface density images using a Gaussian kernel.  Then, we determine where the peak overdensity is relative to the radio source, assigning higher weight to sources closer to the radio source to avoid identifying other structures.  Thus, each new cluster center is the location of the peak overdensity of our red sequence galaxies in a given field.  Since our composite images are $\approx$ 5$\arcmin$ $\times\,$ 5$\arcmin$, it is possible that the highest density peak could be offset by as much as 3$\arcmin$ from the radio source ($\approx$ 1.4\,Mpc at $z$ = 1.0).  We follow \citet{Sakelliou2000}, who found that the majority of their wide-angle tail radio sources are offset from the cluster center by $\leq$ 300\,kpc, although one radio source is offset by $\approx$ 1.6\,Mpc.  Given that the average redshift of COBRA clusters is $z$ $\approx$ 1.0, 300\,kpc is $\approx$ 0$\farcm$6.  For this reason, in the few cases with multiple density peaks and no obvious strongest peak, we select the closer peak as the cluster center.  We note that we in no way limit our offset to 0$\farcm$6, but use the result from \citet{Sakelliou2000} to keep our offsets on reasonable galaxy cluster scales.

Although some of the fields change their center minimally, a new ``central" location was chosen for each field based on the surface density of red sequence galaxies.  These positions are noted in Table\,\ref{tb:2}.  However, we stress that although this technique allows us to identify regions of peak density, most density structures do not encompass the entire 1$\arcmin$ radius circular region.  Because of this, for some fields, the overall overdensity is slightly lower when centered on the most overdense region because we lose a galaxy or two located on the edge of the 1$\arcmin$ search region.

\subsection{$i - [3.6]$ Analysis $\&$ Results}
\begin{figure*}
\begin{center}
\figurenum{14}
\subfigure{\includegraphics[scale=0.75,trim={1.85in 3.in 2.19in 3.55in},clip=true]{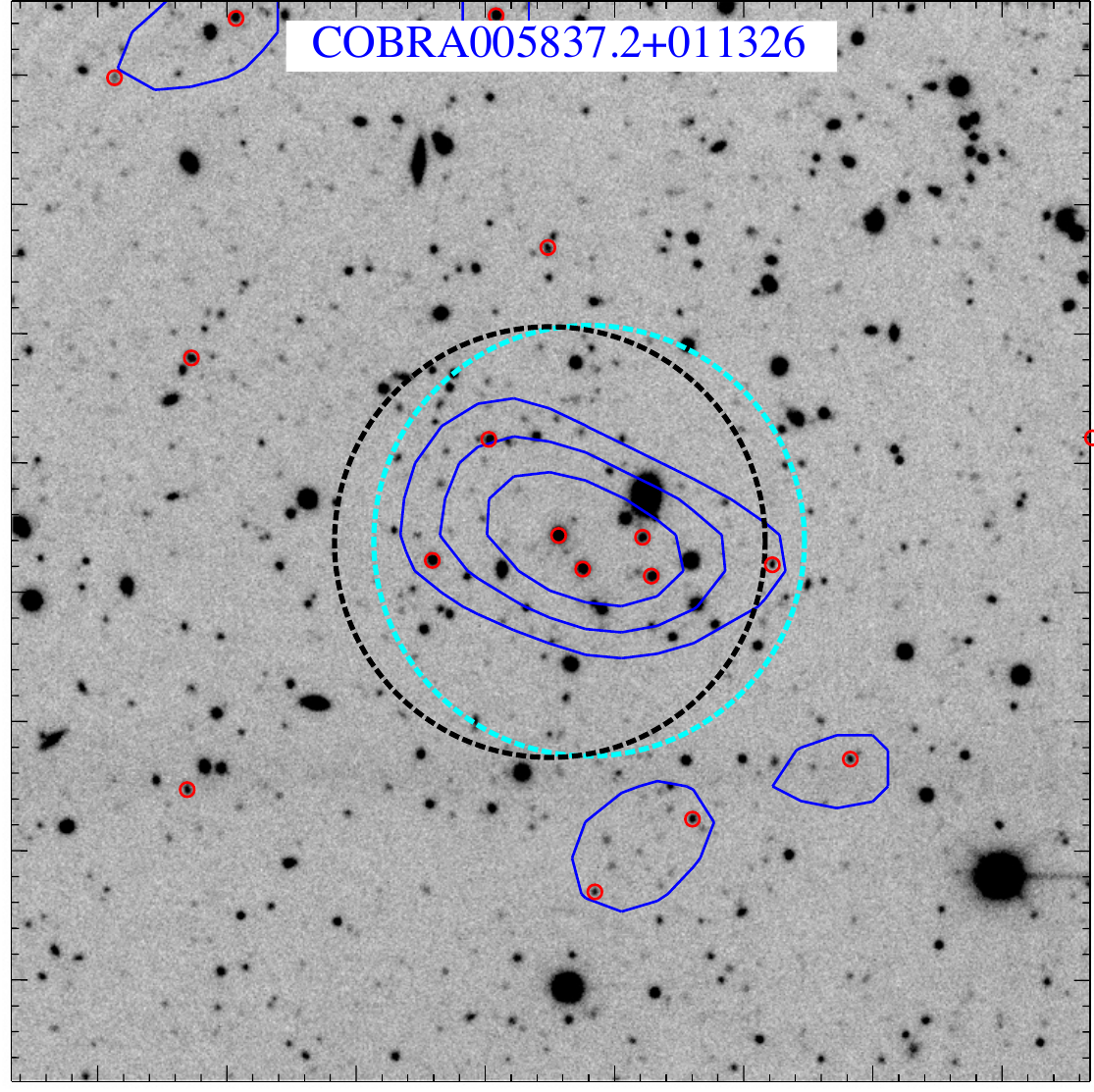}}
\subfigure{\includegraphics[scale=0.75,trim={1.85in 3.in 2.18in 3.55in},clip=true]{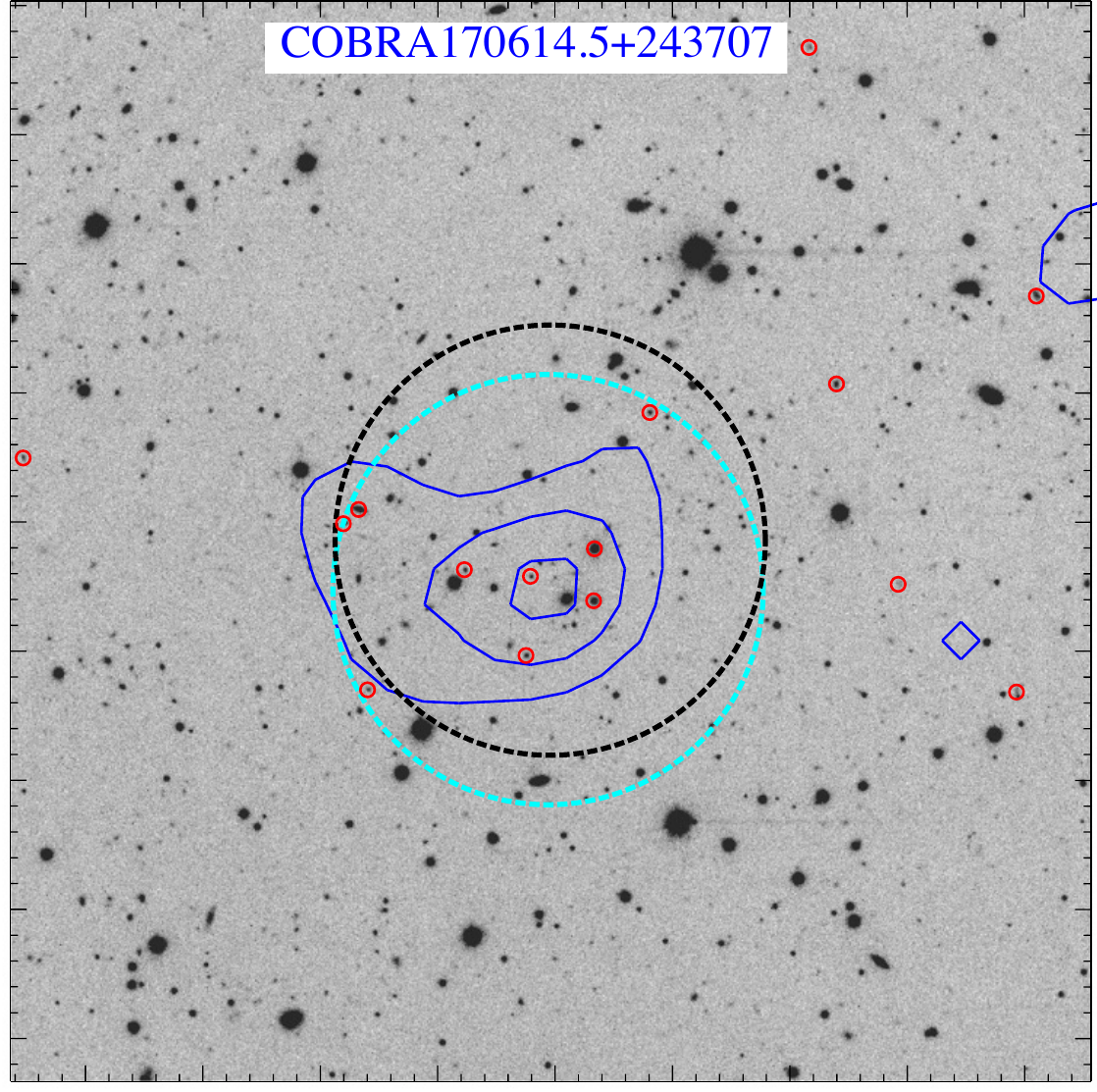}}
\subfigure{\includegraphics[scale=0.75,trim={1.85in 3.in 2.18in 3.53in},clip=true]{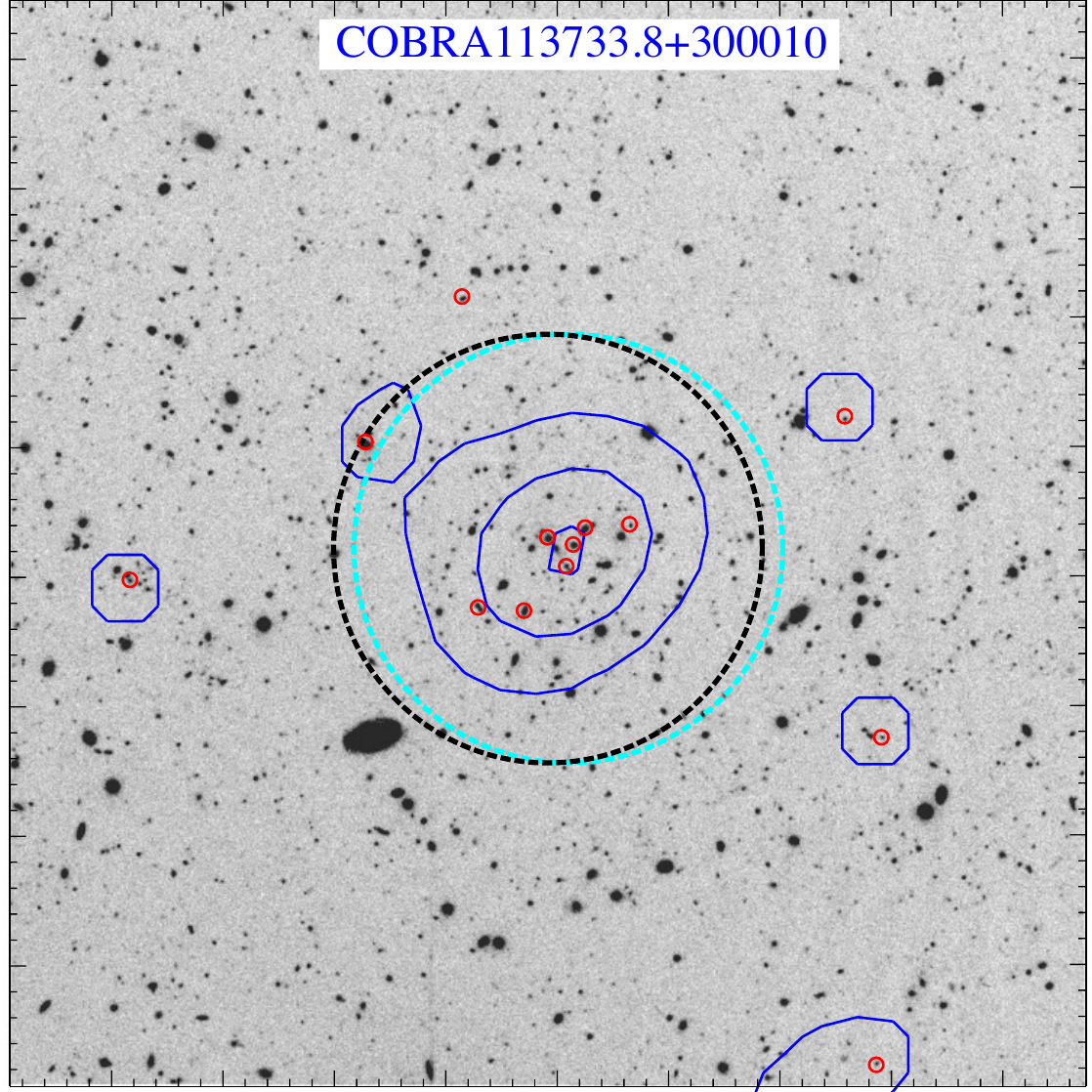}}
\subfigure{\includegraphics[scale=0.75,trim={1.85in 3.in 2.18in 3.55in},clip=true]{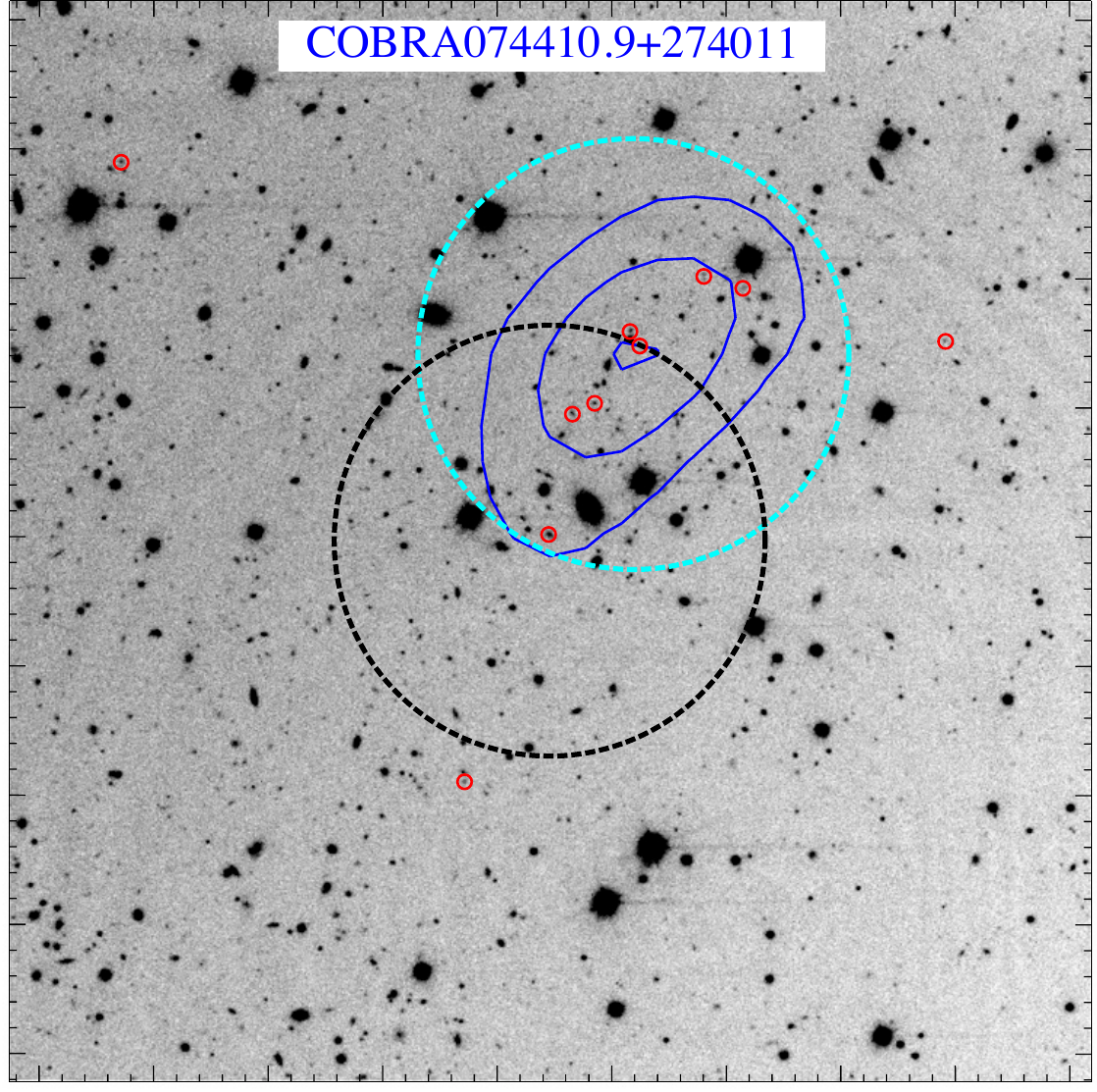}}
\caption{Examples of 5$\arcmin$ $\times$ 5$\arcmin$ cutouts of $i$-band images taken on the DCT of COBRA fields.  The red circular regions show red sequence members (within $\pm$ 0.15\,mag of the host's $i - [3.6]$ color).  The blue contours represent the red sequence surface density contours for each field.  The 1$\arcmin$ radius black dashed circle is centered on the radio source (with the host galaxy at the center), while the 1$\arcmin$ radius cyan dashed circle is centered on the distribution of red sequence galaxies.  COBRA005837.2+011326 ($z$ $\approx$ 0.7) is a 3.6$\sigma$ red sequence detection when centered on the AGN, while it is a 4.3$\sigma$ red sequence detection when centered on the distribution of red sources.  COBRA170614.5+243707 ($z$ $\approx$ 0.71) is a 5.2$\sigma$ red sequence overdensity when centered on the AGN, while it is a 5.9$\sigma$ red sequence detection when centered on the distribution of red sources.   COBRA113733.8+300010 ($z$ = 0.96) is a 4.6$\sigma$ red sequence detection when centered on the AGN, while it is a 4.0$\sigma$ red sequence detection when centered on the distribution of red sources.  COBRA074410.9+274011 ($z$ $\approx$ 1.3) is a 4.0$\sigma$ red sequence detection when centered on the AGN, while it is a 7.3$\sigma$ detection when centered on the distribution of red sources (the strongest red sequence overdensity in the magnitude-limited sample).  The overdensities and new cluster centers for all COBRA fields are given in Table\,\ref{tb:2}.}
\label{Fig:CONTOURS1}
\end{center}
\end{figure*}

We measure the surface density of red sequence cluster galaxies in $i - [3.6]$ for the 39 m*+1 fields and the 38 magnitude-limited fields (see Figure\,\ref{Fig:CONTOURS1} for examples of the red sequence surface density measurements).  From our newly determined cluster centers, we estimate the red sequence overdensity using identical background statistics to \textsection{4.1}.    We identify 22 fields in the m*+1 sample and 10 fields in the magnitude-limited sample that are red sequence cluster candidates (see Table\,\ref{tb:ich1AGN}; for the complete list of overdensities and new cluster centers, see Table\,\ref{tb:2}).  

Of the 32 red sequence cluster candidates we identify with the new cluster centers, seven are not identified in the $i - [3.6]$ analysis when centered on the radio source, and three are newly identified cluster candidates increasing the total number of COBRA cluster candidates to 195 (See Table\,\ref{tb:ich1AGN}).  Since we had previously only done single-band overdensity measurements, the identification of new red sequence cluster candidates that are not overdense at the 2$\sigma$ level based on the $Spitzer$ observations may be indicative of a population of smaller, red galaxy groups that may be underdense and potentially offset from the bent radio source.  Each of these newly identified red sequence cluster candidates is also a cluster candidate via the combined overdensity, further emphasizing the importance of accounting for galaxy color.  

Of the newly detected cluster candidates using the new red sequence center, two are fields with two red sequence galaxies detected when centered on the AGN.  Two other new red sequence cluster candidates, COBRA135838.1+384722 and COBRA154638.3+364420, highlight the necessity of determining an offset between the bent, double-lobed radio source and the galaxy cluster center because each is a cluster candidate in \citet{Paterno-Mahler2017} based on $\approx$ 3.0$\sigma$ overdensities in the 2$\arcmin$ search region centered on the radio source.  By moving to our new cluster centers, these fields become red sequence cluster candidates.  Another new $i - [3.6]$ red sequence cluster candidate is COBRA075516.6+171457, identified in $r$ $-$ $i$, which we note in \textsection{4.3} has a host galaxy that is blended due to the resolution of $Spitzer$.  Because of this, the cluster center identified in Table\,\ref{tb:2} may be inaccurate, although the galaxies identified are real.       

Despite the increase in the number of cluster candidates, five m*+1 fields and two magnitude-limited fields show a slight decrease in the number of detected sources with this method (see Table\,\ref{tb:2} for the fields which show a slight decrease in overdensity), including two AGN centered red sequence cluster candidates.  Since the average difference is 1.1 sources, this is likely due to non-central red galaxies falling outside of the 1$\arcmin$ circular region with our new center or asymmetric distributions of red sequence cluster galaxies.

\begin{figure}
\centering
\figurenum{15}
\epsscale{1}
\includegraphics[scale=0.4,trim={0.7in 0.5in 0.0in 0.95in},clip=True]{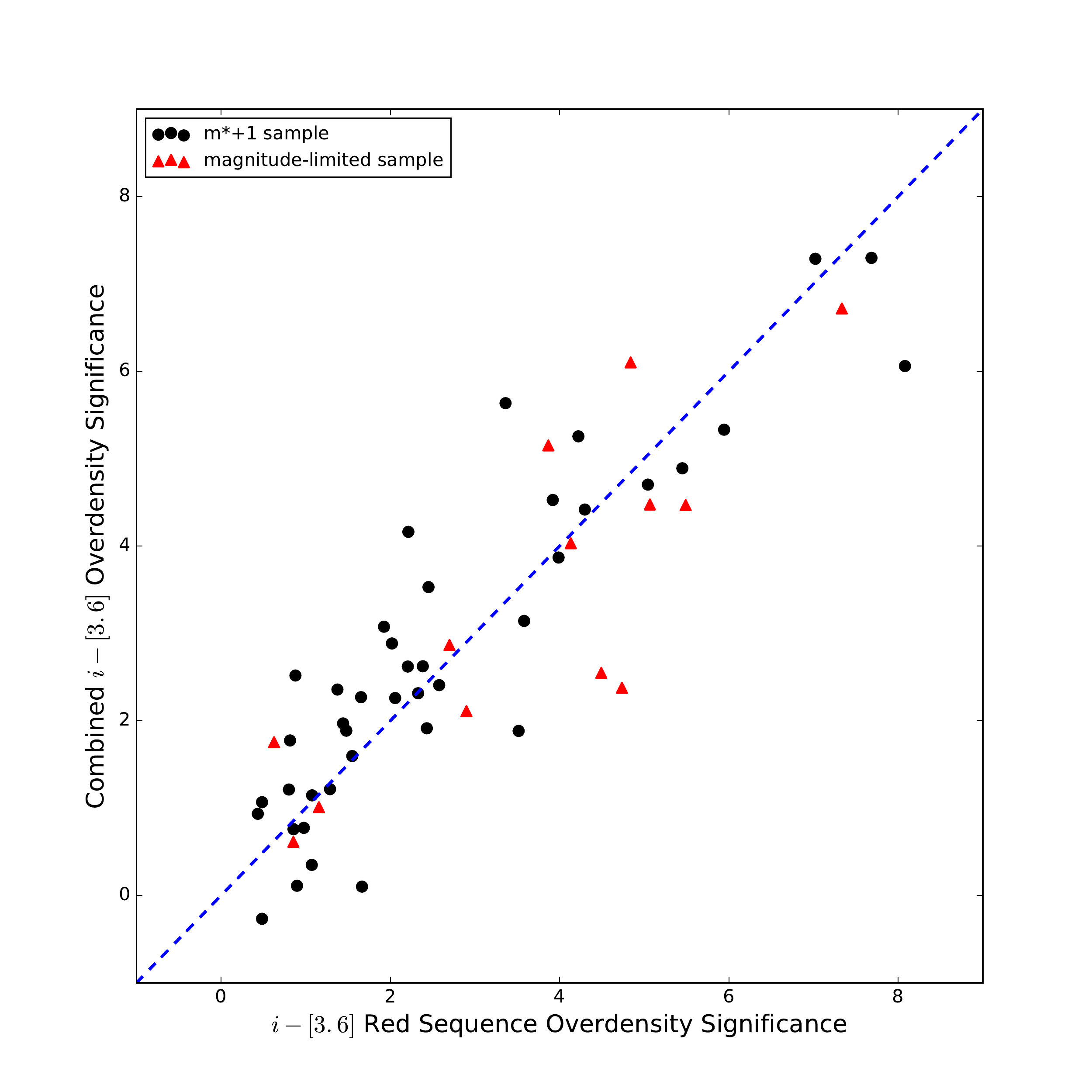}
\caption{A comparison of the $i - [3.6]$ red sequence overdensity to the combined $i - [3.6]$ overdensity for our shifted overdensities centered on the distribution of red sequence galaxies.  The m*+1 sample is shown in black circles, while the magnitude-limited sample is shown in red triangles.  A one-to-one dashed line is shown in blue.  \label{Fig:CombinedredsequencecheckShift}}
\end{figure}

Like in \textsection{4.1.2}, we also re-measure the combined overdensity using identical background statistics.  We identify 24 combined overdensity cluster candidates in the m*+1 sample and ten combined overdensity cluster candidates in the magnitude-limited sample.  This sample includes both the newly identified red sequence cluster candidates as well as the one new field (COBRA120654.6+290742 at $z$ $\approx$ 0.85).  When we compare the new red sequence overdensity to the combined overdensity (see Figure\,\ref{Fig:CombinedredsequencecheckShift}), we find a bit more scatter than when centered on the AGN, but still an agreement between the two measurements.  The differences likely factor from our new cluster center being chosen based specifically on the overdensity of red sequence galaxies, and not a weighted distribution of redder and bluer galaxies.  The strong similarity between the two measurements still emphasizes the similar rate at which each measurement identifies cluster candidate fields.

\subsection{$[3.6] - [4.5]$ Analysis $\&$ Results}
For the 20 fields at $z$ $>$ 1.2 with $[3.6] - [4.5]$ colors, six show evidence for cluster candidacy based on our color surface density analysis (see Table\,\ref{tb:ch1ch2AGN}; for the complete list of overdensities and new cluster centers, see Table\,\ref{tb:2}).  Of these six, three are new cluster candidates (one AGN centered cluster candidate falls just below the 2$\sigma$ threshold using this method).  All three new color cluster candidates (1.3 $<$ $z$ $<$ 1.85) are cluster candidates identified in \citet{Paterno-Mahler2017} and are quasars.  Interestingly, one of these fields, COBRA121128.5+505253, has just two $i - [3.6]$ red sequence sources and a significance above 2$\sigma$ that we choose not to report because of our three red sequence galaxy criterion.  This overlap lends some validity to the strength of that measurement.  We measure a slight decrease in seven fields ($\approx$ 1.3 sources on average) due to sources at the edge of our search region.  A discussion of the cross-correlation between central positions for fields in $i - [3.6]$ and $[3.6] - [4.5]$ can be found in \textsection{6}. 

\begin{figure}
\centering
\figurenum{16}
\epsscale{1}
\includegraphics[scale=0.4,trim={0.6in 0.5in 0.0in 0.9in},clip=True]{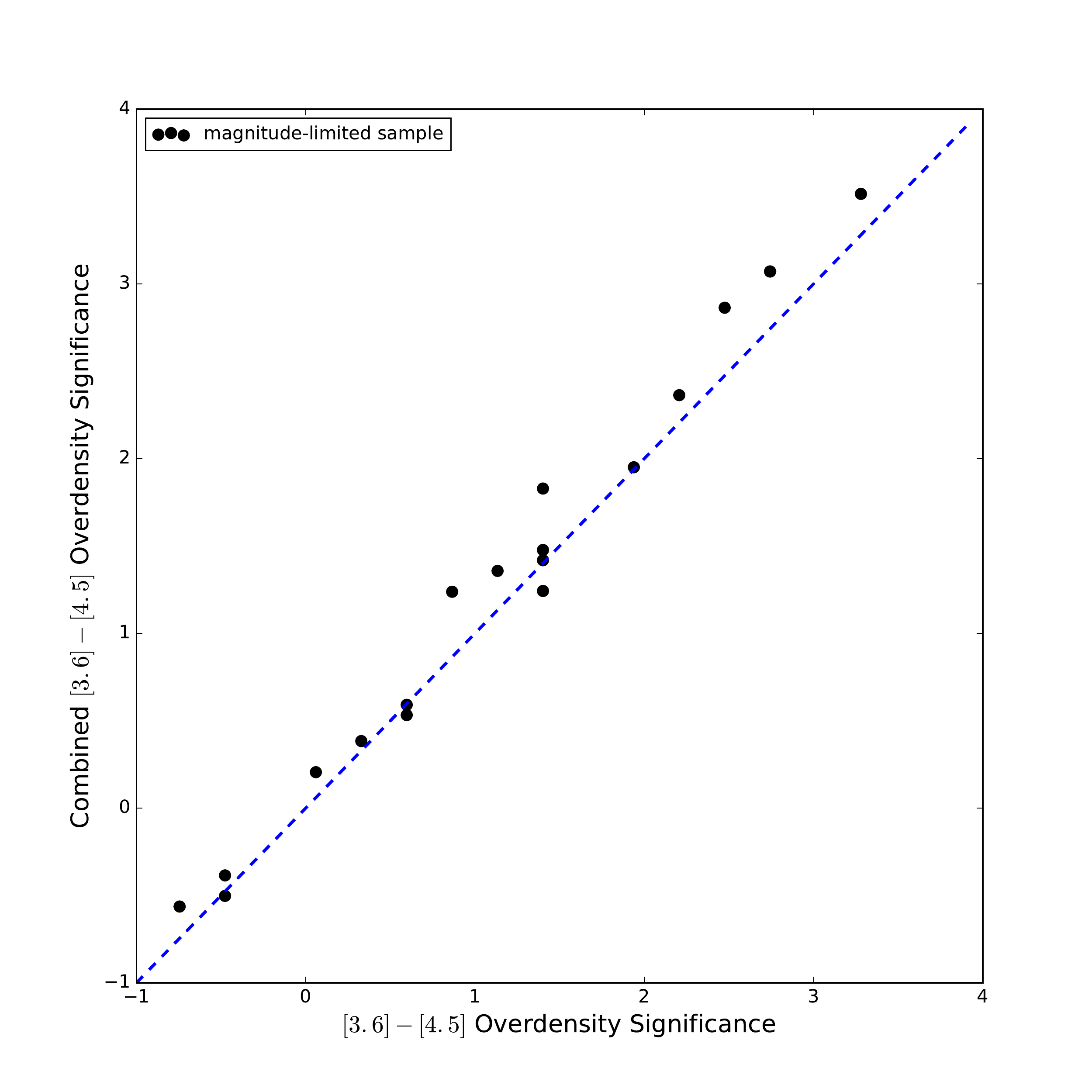}
\caption{A comparison of the $[3.6] - [4.5]$ $>$ -0.15 overdensity to the combined $[3.6] - [4.5]$ overdensity for the analysis centered on the distribution of high-$z$ galaxies.  A one-to-one dashed line is shown in blue.   \label{Fig:Combinedch1ch2shift}}
\end{figure}

When we measure the combined overdensity, we again find a strong one-to-one relation between the two overdensity measurements (see Figure\,\ref{Fig:Combinedch1ch2shift}).  All six shifted red sequence cluster candidates are identified as combined overdensity cluster candidates, as is the field that dips just below the cluster candidate threshold.   

\subsection{$r$ $-$ $i$ Analysis $\&$ Results} 
\begin{figure}
\centering
\figurenum{17}
\epsscale{1}
\includegraphics[scale=0.4,trim={0.6in 0.5in 0.0in 0.9in},clip=True]{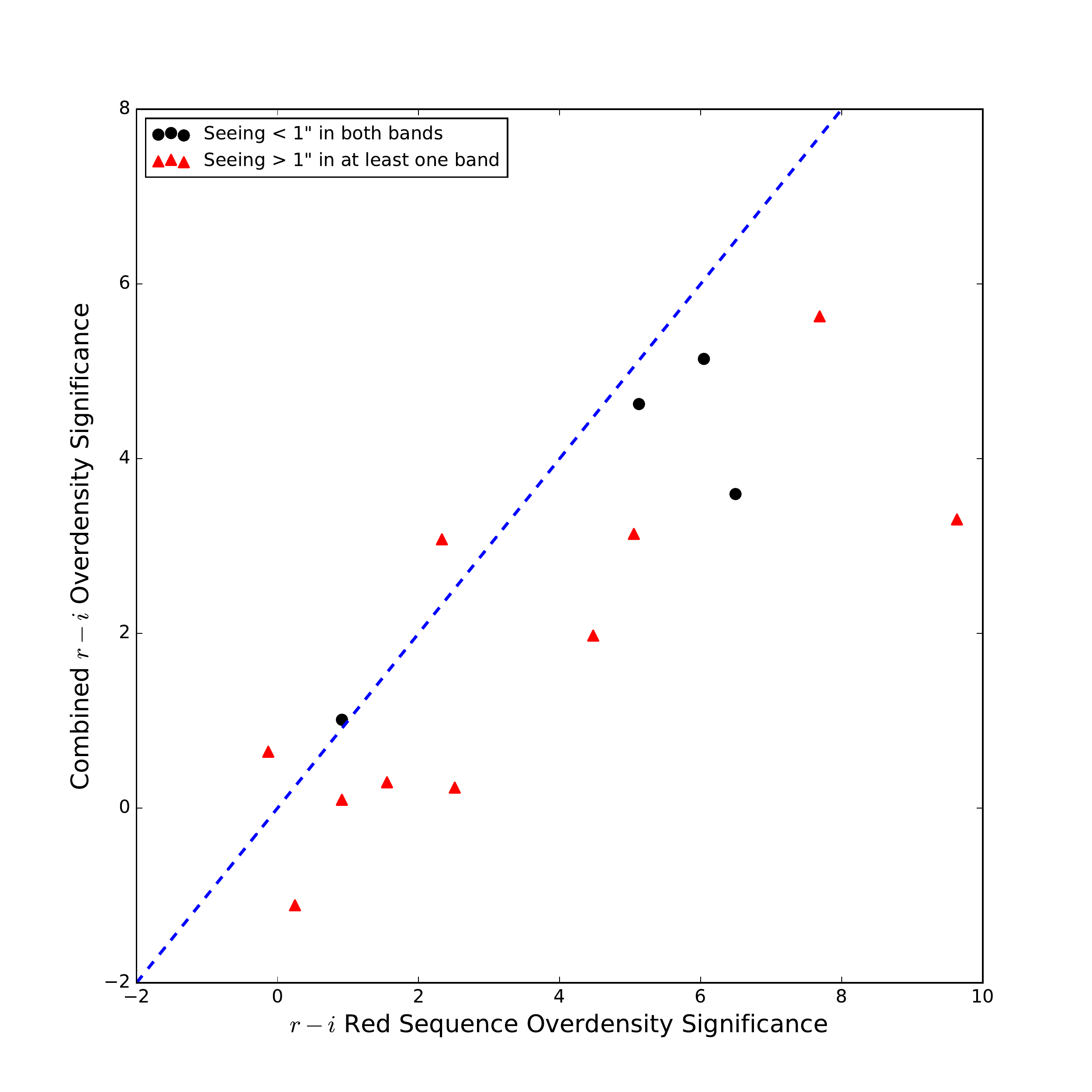}
\caption{A comparison of the $r - i$ red sequence overdensity to the combined overdensity for the 14 m*+1 fields with $r$- and $i$-band observations when centered on the distribution of red sequence galaxies.  Fields with seeing greater than 1$\arcsec$ in at least one band are shown in red triangles, while fields with seeing less than 1$\arcsec$ in both bands are shown in black circles.  A one-to-one dashed line is shown in blue.   \label{Fig:Combinedrishift}}
\end{figure}

For the 14 fields at $z$ $<$ 1.0 with $r - i$ analysis (see \textsection{4.3} for an explanation of the fields and field parameters chosen), we find nine fields are cluster candidates based on our red sequence surface density analysis (see Table\,\ref{tb:riAGN}; for the complete list of overdensities and new cluster centers, see Table\,\ref{tb:2}).  Five fields show a slight decrease in the number of detected sources ($\approx$ 2.0 sources on average) due to the change of the edge region.  Since all fields in the $r - i$ sample are in the $i - [3.6]$ sample, we discuss the correlation of cluster center and overdensity in \textsection{6}.  When we measure the combined overdensity using these new $r - i$ red sequence cluster centers, we find, that like in Figure\,\ref{Fig:Combinedri}, there exists a similar amount of scatter and a large offset from the one-to-one linear relations (see Figure\,\ref{Fig:Combinedrishift}), again reinforcing that the differences in the two values for our $r - i$ analysis are likely symptomatic of the image quality or our underlying assumptions overvaluing blue galaxies.    

\section{Discussion}

\begin{figure}
\figurenum{18}
\includegraphics[scale=0.5,trim={0.4in 0.1in 0in 0.55in},clip=True]{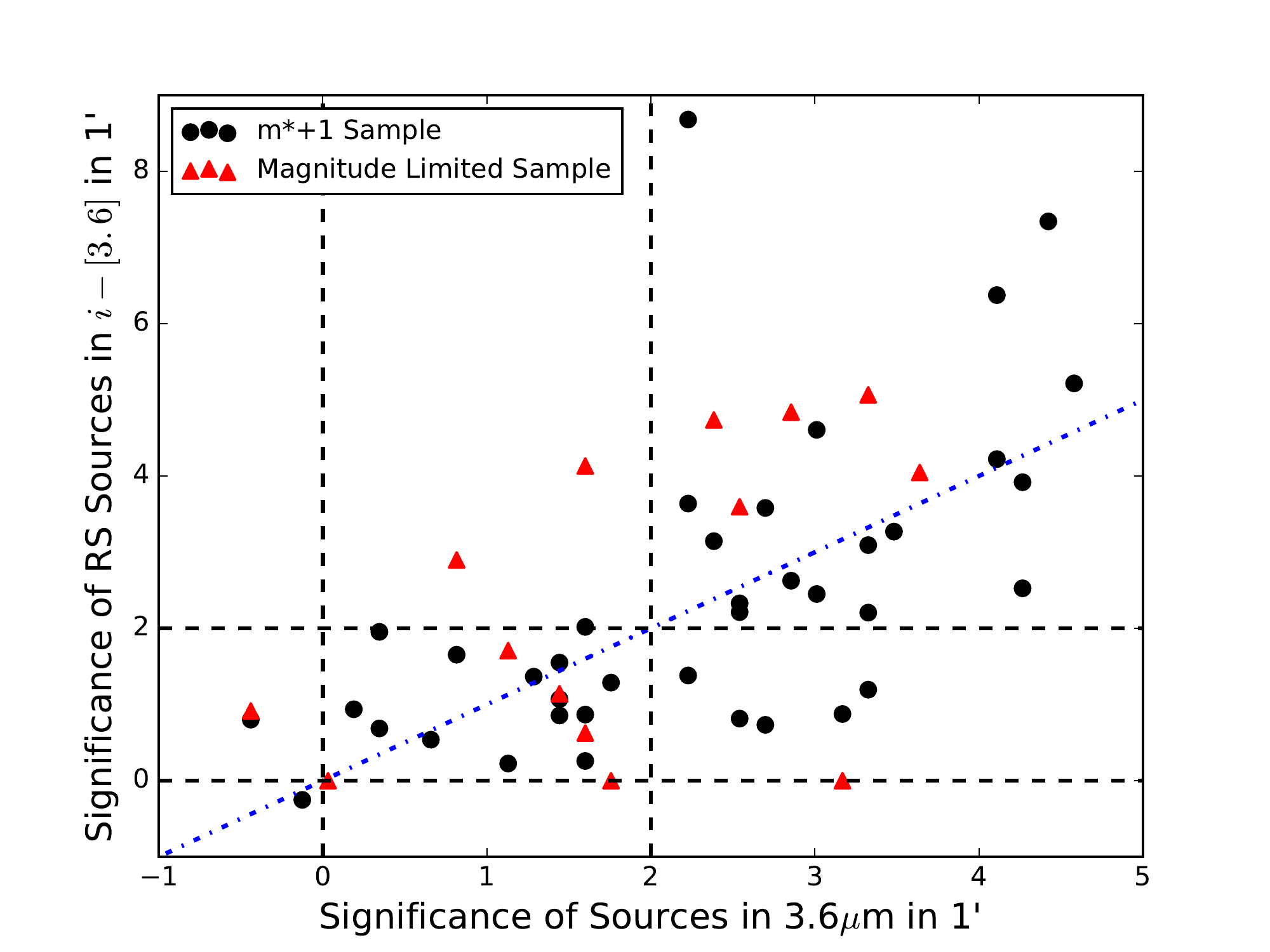}
\caption{A comparison of the significance of the \citet{Paterno-Mahler2017} 3.6\,$\mu$m-based overdensity measurements to those derived from our AGN centered $i - [3.6]$ red sequence analysis.  We plot the m*+1 sample in black circles and the magnitude-limited sample in red triangles.  A line of unity is plotted in the dashed-dot blue line and the lines denoting 0$\sigma$ and 2$\sigma$ are shown in the dashed black lines. \label{Fig:ich1-COMP} }  
\end{figure}

\begin{figure*}
\begin{center}
\figurenum{19}
\includegraphics[scale=0.45,trim={0.35in 0.1in 0.5in 0.3in},clip=True]{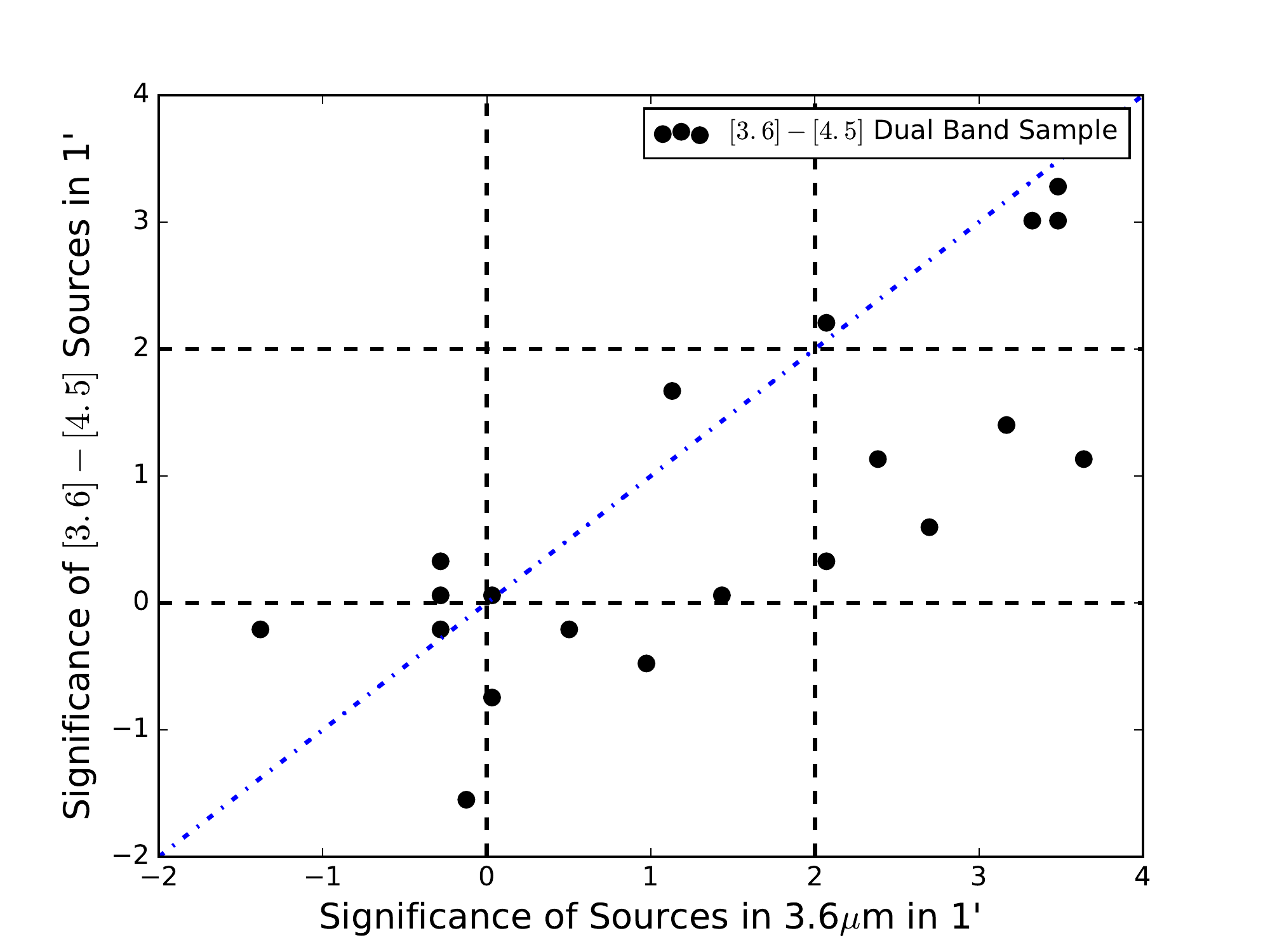}
\includegraphics[scale=0.45,trim={0.35in 0.05in 0.6in 0.25in},clip=True]{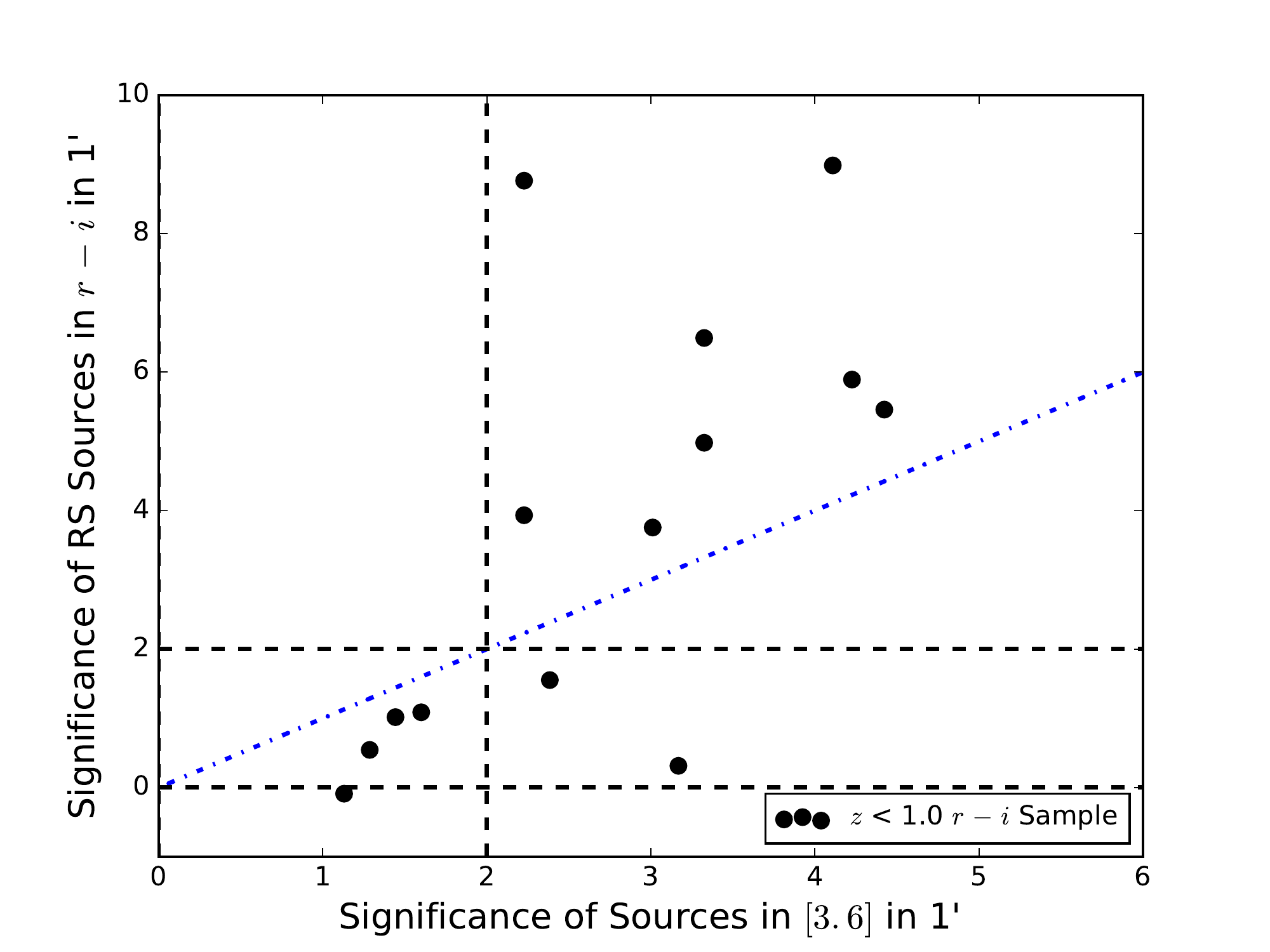}
\caption{A comparison of the AGN centered significance of our fields with $[3.6] - [4.5]$ to their 3.6\,$\mu$m overdensity significance (left) and $r$ $-$ $i$ with their 3.6\,$\mu$m overdensity significance (right).  Both plots have lines of unity plotted in a blue dashed-dot line, while the lines showing a 0$\sigma$ and 2$\sigma$ measurement are shown in black dashed lines. \label{Fig:ch1ch2-comp}}  
\end{center}
\end{figure*}

In \textsection{6.1}, we compare our color-based cluster significance estimates to those of \citet{Paterno-Mahler2017}.  We base this comparison on our analysis centered on the COBRA AGNs to ensure identical search regions to those of \citet{Paterno-Mahler2017}.  In \textsection{6.2}, we examine the impact of using cluster centers derived from the galaxy distributions as described in \textsection{5}.  We discuss the likelihood of cluster mis-identification in \textsection{6.3}.  Finally, we compare COBRA to other high-$z$ cluster searches in \textsection{6.4}. 

\subsection{Comparing Overdensity measurements}
\subsubsection{Comparisons to 3.6\,$\mu$m overdensities}
The original COBRA overdensity measurements are single-band IRAC 3.6\,$\mu$m measurements, with no color component.  The additional optical photometry allows us to determine which galaxies are potential red sequence members and thus improve our understanding of the systems that host bent, double-lobed radio sources.  Figure\,\ref{Fig:ich1-COMP} compares the 3.6\,$\mu$m overdensity estimates from \citet{Paterno-Mahler2017} to our $i - [3.6]$ color-selected AGN centered overdensities for COBRA fields common to both works.  Although the two estimates generally agree, Figure\,\ref{Fig:ich1-COMP} highlights that for clusters detected at $>$ 3$\sigma$ in \citet{Paterno-Mahler2017}, the $i - [3.6]$ color selection is more effective at identifying the strongest cluster candidates than the single-band measurement.  Although not shown, a similar trend exists between our combined overdensities and our $Spitzer$ overdensities.  

The red sequence color cut effectively removes foreground and background contaminants, better highlighting the cluster candidates.  This can be seen particularly well in the magnitude-limited sample, where six of the seven cluster candidates exceed their $Spitzer$ counterparts, likely due to the removal of the numerous lower-redshift galaxies that lie along the line of sight.  

Although there is a strong one-to-one trend, some fields that are not red sequence clusters are cluster candidates in \citet{Paterno-Mahler2017}.  Some of these fields are cluster candidates based on the 2$\arcmin$ search region, while others lack a redshift estimate.  However, four of these fields have between a 1$\sigma$ and 2$\sigma$ red sequence overdensity in $i - [3.6]$.  As noted before, the lack of a 2$\sigma$ red sequence measurement does not remove a cluster candidate detection.  There are many clusters lacking large populations of red sequence members, and as such, the fields that are cluster candidates based on the 3.6\,$\mu$m observations are also included as cluster candidates.  Additionally, some of these fields may have inaccurate redshift estimates caused by bluer or redder host galaxies than our EzGal models predict, especially if the host galaxy's colors are affected by an AGN component.  

Since the goal of this analysis is to uniformly identify red sequence populations in these fields, we treat the red sequence as having no intrinsic slope.  Doing so allows us to measure a red sequence population for fields that might not host cluster candidates, but also may remove some potential red sequence galaxies that are slightly bluer than our color range, which could diminish some of the red sequence overdensities, especially if the slope is coupled with a redder host galaxy.  The population of fainter, slightly bluer galaxies, can be seen if we examine the CMDs of the strongest red sequence cluster candidates.  In future work, we plan to thoroughly study the red sequence populations of our strongest cluster candidates to account for this.     

Figure\,\ref{Fig:ch1ch2-comp} compares the $[3.6] - [4.5]$ and $r - i$ color-derived overdensity estimates to those of \citet{Paterno-Mahler2017}.   For $[3.6] - [4.5]$ sample, on average, the cluster measurements are more significant when using the 3.6\,$\mu$m galaxy density as the sole criterion than the $[3.6] - [4.5]$ measurement (Figure\,\ref{Fig:ch1ch2-comp}, left panel).  Because the $[3.6] - [4.5]$ color cut is a well-studied and effective selection criteria, that our 3.6\,$\mu$m selection generally yields a greater overdensity for these fields is surprising.  It is possible that the high-$z$ nature of these fields and the IR bands used in these images combine yielding a higher fraction of galaxies in our $[3.6] - [4.5]$ background being at high-$z$ as opposed to our $i - [3.6]$ and $r - i$ color cuts which isolate only a small fraction of the total galaxies detected.  Additionally, the relative bright magnitude limits required on the luminosity of galaxies relative to an m* galaxy at high-$z$ means that we can only detect a small fraction of the highest-redshift galaxies.  

By contrast, in the right panel of Figure\,\ref{Fig:ch1ch2-comp}, the $r - i$ color cut has a comparable efficiency to the 3.6\,$\mu$m overdensity.  Like our $i - [3.6]$ color cut, above 3$\sigma$, the $r - i$ color cut is far more effective than the single-band measurement.  The similarity between the effectiveness of these two red sequence color cuts likely stems from the similar selection methods in terms of isolating red sequence galaxies.  That we see a similar trend in both the $i - [3.6]$ and $r - i$ red sequence analyses emphasizes that our strongest $Spitzer$ overdensities ($>$ 3$\sigma$) generally indicate strong red sequence cluster candidates, which can help further prioritize fields to observe as we continue the optical campaign (See Figure\,\ref{Fig:Allsources}).  

\subsubsection{Comparing our Color Selections to One Another}
\begin{figure*}
\begin{center}

\figurenum{20}
\includegraphics[scale=0.45,trim={0.35in 0.1in 0.5in 0.3in},clip=True]{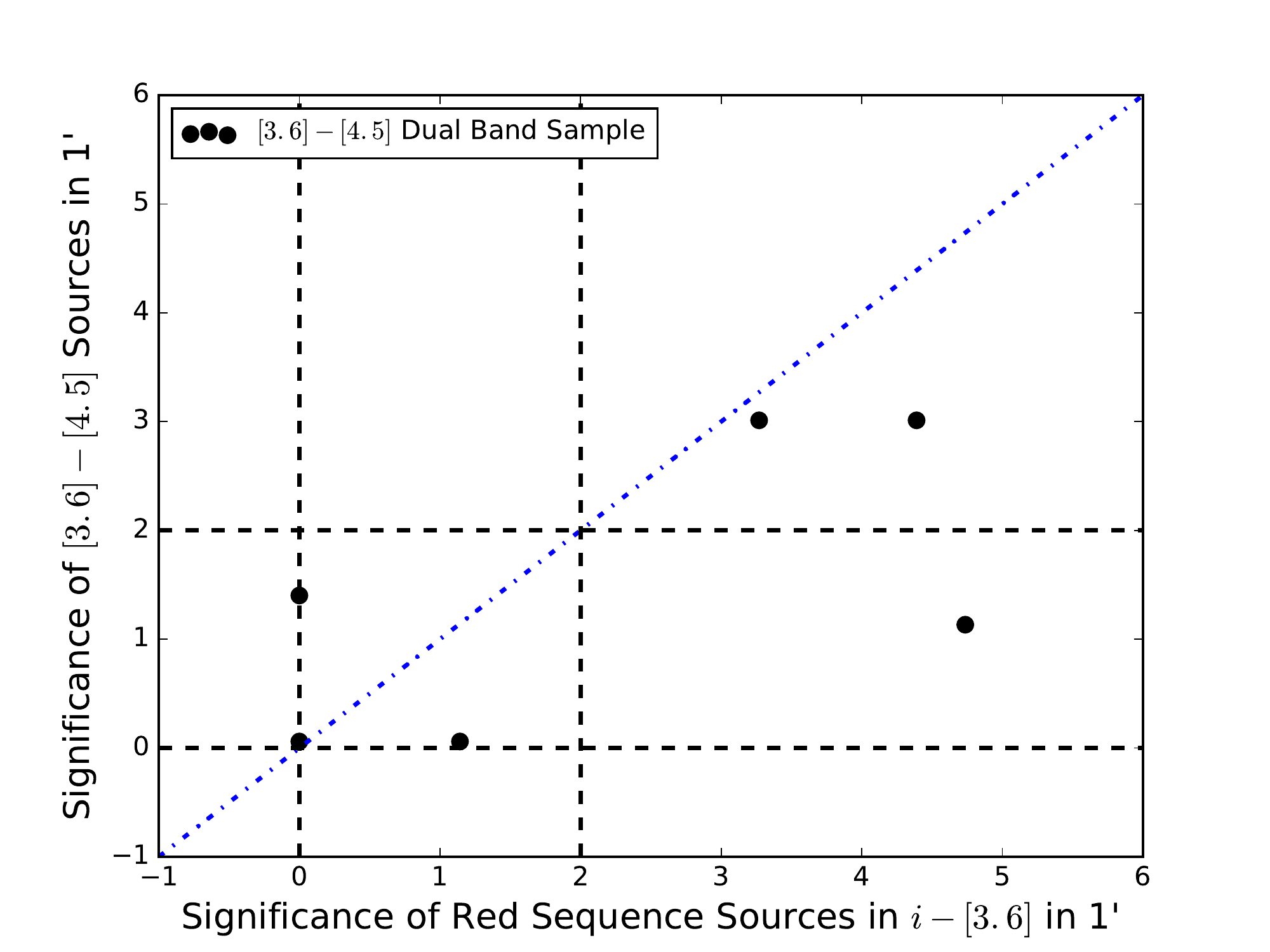}
\includegraphics[scale=0.45,trim={0.35in 0.05in 0.6in 0.25in},clip=True]{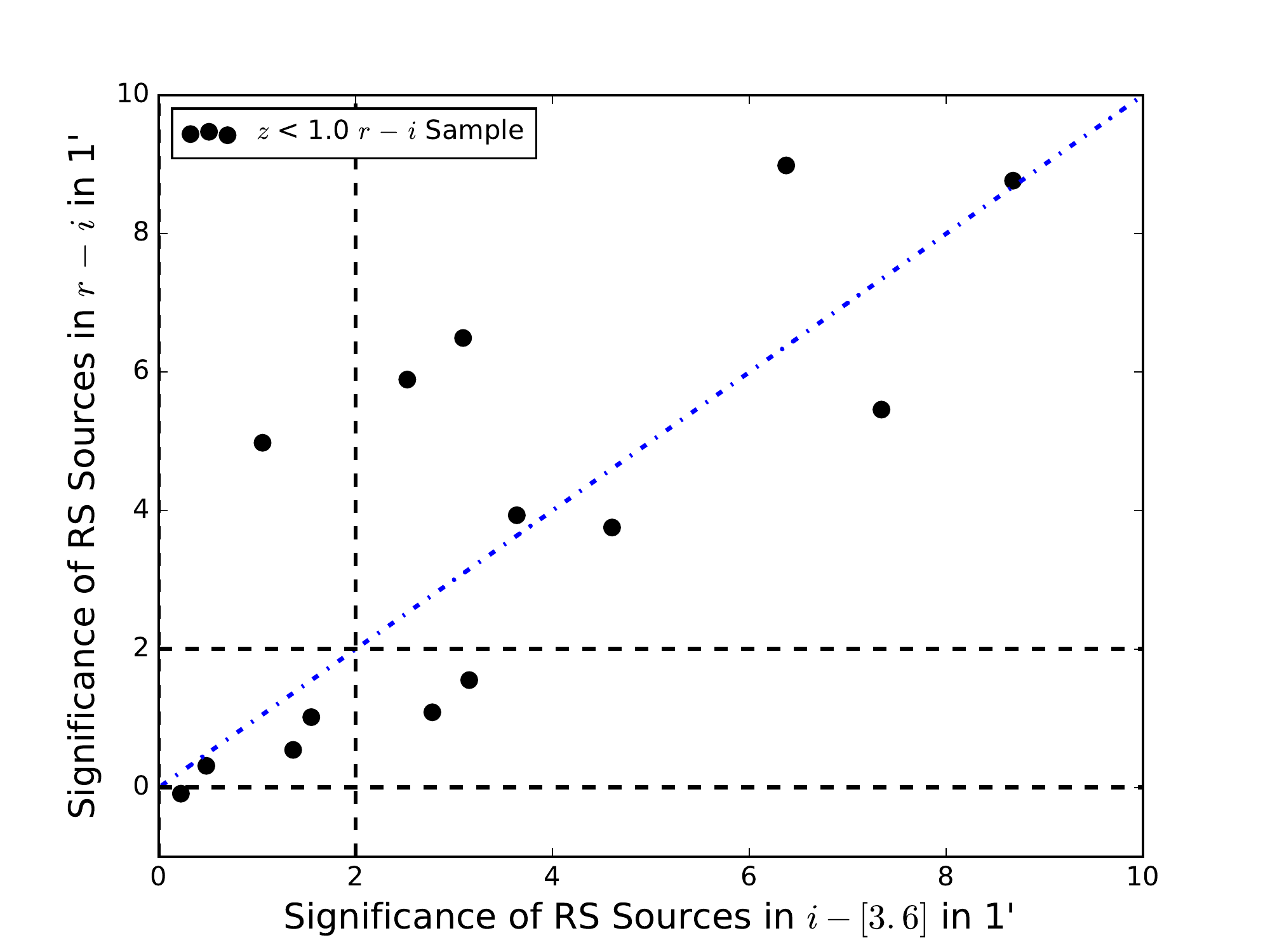}
\caption{A comparison of the significance of our $[3.6] - [4.5]$ fields with our $i - [3.6]$ (left) and $r - i$ with our $i - [3.6]$ analysis (right).  Both plots have lines of unity plotted in a blue dashed-dot line, while the lines showing a 0$\sigma$ and 2$\sigma$ measurement are shown in black dashed lines.  The left hand plot shows the six fields with $[3.6] - [4.5]$ and $i$ $-$ $[3.6]$ overdensity measurements, while the right hand plot contains the subset of 14 m*+1 fields (see Table\,\ref{tb:1} and Table\,\ref{tb:2} for the list of fields). \label{Fig:ri-comp}}  
\end{center}
\end{figure*}

We compare our color-based overdensity measurements to each other to estimate the reliability of each color cut.  Since we only have six overlapping fields with $i - [3.6]$ and $[3.6] - [4.5]$, we cannot make any definitive statements on trends between these two colors.  We note that two of the three red sequence cluster candidates in the overlapping sample are cluster candidates based on our $i - [3.6]$ and $[3.6] - [4.5]$ analysis (see Figure\,\ref{Fig:ri-comp}).  This may imply similar detection levels of galaxies with these two color cuts.  The other cluster candidate lies below the line of unity and is one of the magnitude-limited sample where the mean background value is nearly zero and thus the three galaxies detected may yield a slightly inflated red sequence overdensity significance.  

For the 14 fields with $i - [3.6]$ and $r - i$ shown in the right side of Figure\,\ref{Fig:ri-comp}, we see a roughly linear trend with a large degree of scatter.  Although one point is due to the blended host in COBRA075516.6+171457 (see \textsection{4.3}), the remaining scatter may be due to poor seeing for some of our $r$-band observations (as discussed in \textsection{4.3.1}), since 6 of the 14 $r$-band images have seeing worse than 1$\farcs$2.  Additionally, based on the measured fractions of red sequence galaxies at the expected redshifts, the level of red sequence contamination is higher for the $r - i$ color, thus resulting in greater foreground/background contamination, which may impact these measurements.  Although not shown, we see a similar amount of scatter when we compare the combined overdensities, especially for the fields with poorer seeing.  Overall, these results show how similarly each color identifies high-$z$ cluster candidates, specifically within the redshift ranges of our analysis.  This strengthens our confidence in combining our color analysis to show that 39 of the 77 fields with redshift estimates are red sequence cluster candidates.

We measure a total of 42 combined overdensity and red sequence cluster candidates with redshift estimates.  The additional fields from the combined overdensities generally have three or four red sequence members, with less than a 2.5$\sigma$ measurement.  While these still might be cluster candidates, the lack of red sequence 2$\sigma$ overdensities and the relatively small number of red sequence members makes us less confident in these additional fields.  In our strongest red sequence overdensities, we see extended populations of red galaxies brighter than m* with populations of fainter bluer and redder galaxies (as seen in our CMDs; see Figure\,\ref{Fig:colorhist} and Figure\,\ref{Fig:ricolorhist}), similar to what is expected of a cluster or group, which we do not see in all of these fields.

\subsection{Comparing Cluster Center Offsets and Color Surface Density Profiles}
\subsubsection{Measuring the Offset from the Cluster Center}
\begin{figure*}
\begin{center}
\figurenum{21}
\includegraphics[scale=0.5,trim={0.5in 0.15in 0.6in 0.5in},clip=True]{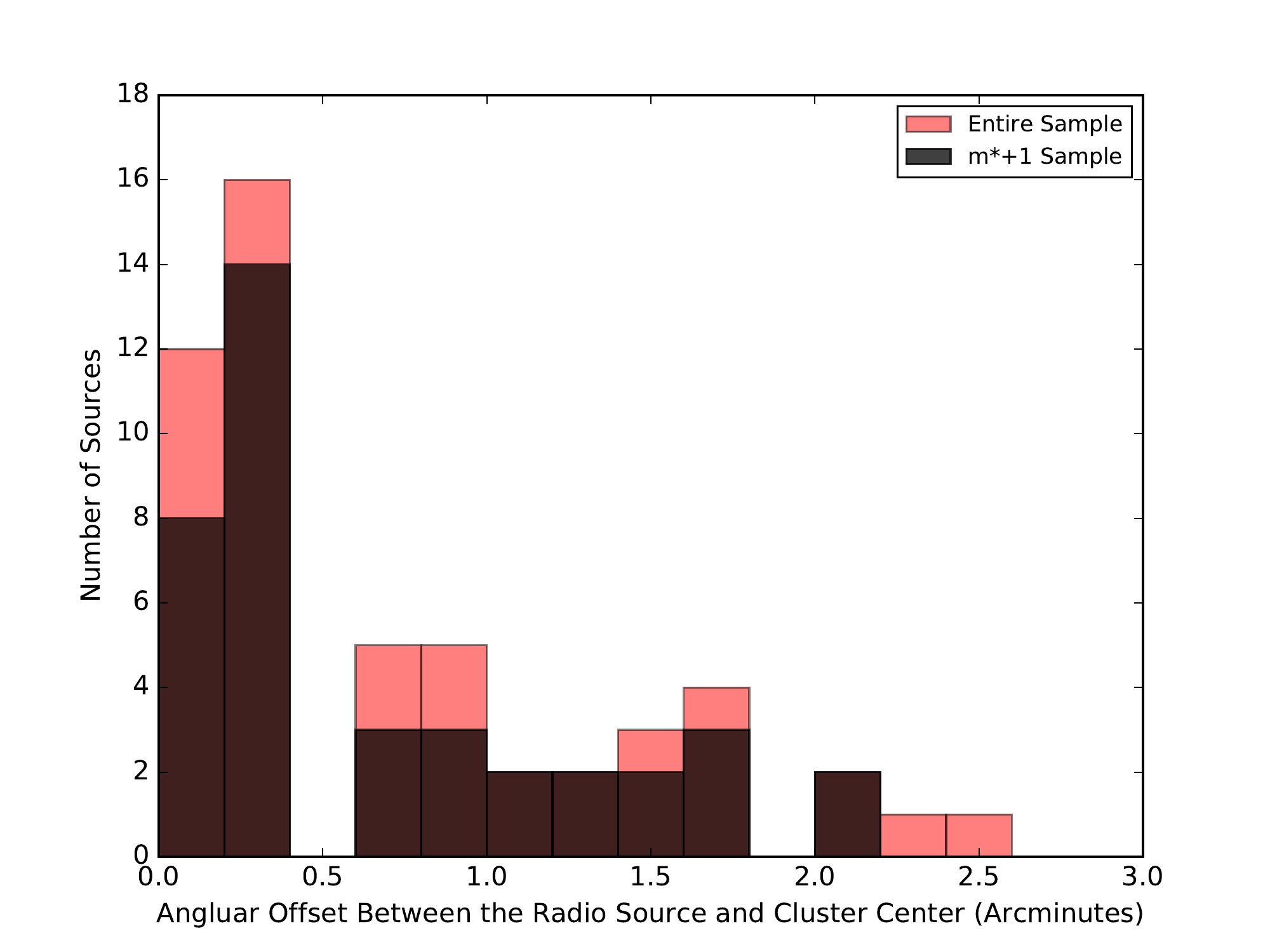}
\includegraphics[scale=0.5,trim={0.5in 0.15in 0.6in 0.5in},clip=True]{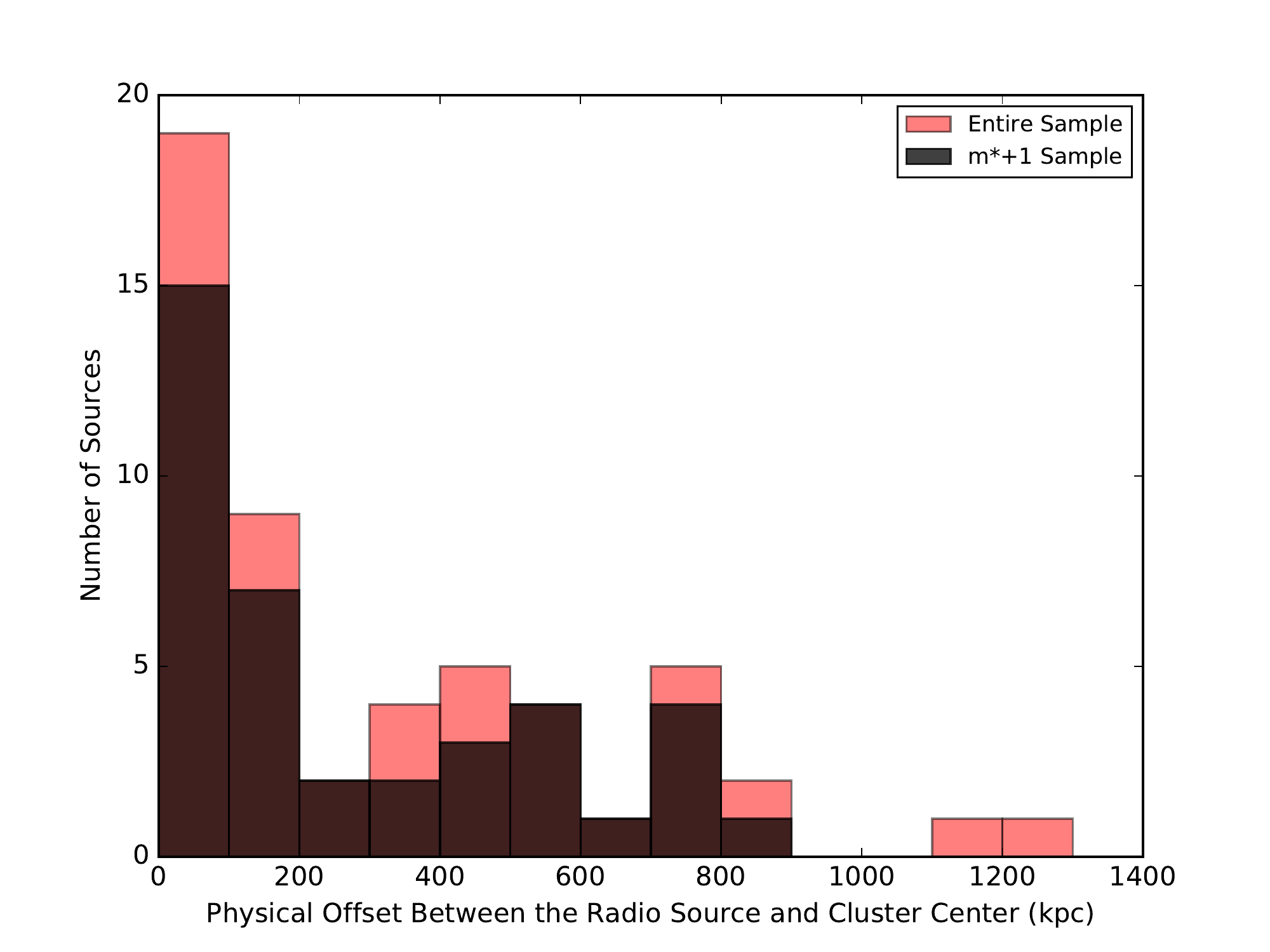}
\caption{Distribution of radio sources relative to the cluster center.  The left histogram shows the angular offset between the radio source and the cluster center estimate based on the $i - [3.6]$ color.  The right histogram shows the physical offset determined for each field based on our redshift estimate (see Table\,\ref{tb:4}).  The black distribution shows the m*+1 sample, while the red distribution is the entire distribution of COBRA clusters with redshift estimates.  We separate the two distribution because we have greater confidence in our m*+1 sample based on the measured magnitude limits.   \label{Fig:ich1-offset}}
\end{center}
\end{figure*}

\citet{Sakelliou2000} find that the majority of their wide-angle tail radio sources are offset by less than $\approx$ 300\,kpc  from the cluster center (with a few sources at greater offsets, and one source offset by $\approx$ 1.6\,Mpc).  We find a similar distribution for our sources.  In the $i$ $-$ $[3.6]$ analysis shown in Figure\,\ref{Fig:ich1-offset}, 24 of 39 m*+1 fields and 6 of 14 magnitude-limited fields that we can determine a significance of are offset by less than 300\,kpc from our cluster centers determined by the surface density of red sequence galaxies.  Given this measured offset and the correlation between red sequence cluster center and true cluster center in relaxed clusters \citep[e.g.,][]{Rumbaugh2018}, we find that many of our bent, double-lobed radio sources have either fallen through the cluster center or are infalling near the cluster center.  Additionally, when we examine our $i - [3.6]$ CMDs, we find that our host galaxies are generally brighter than m* $-$ 1.0, although they are not always BCGs.  When these host magnitudes are coupled with the measured offset, it allows for the possibility that our bent AGN need not be found in the center of clusters.  

When we measure the offset using our $r$ $-$ $i$ analysis, we find a similar trend; 10 of 14 are fields offset by less than 300\,kpc.  However, this trend is not as strong in $[3.6] - [4.5]$, where 11 of 20 sources are offset by greater than 300\,kpc.  As the $[3.6] - [4.5]$ subsample represents some of the highest-redshift cluster candidates in COBRA, and the fraction of unrelaxed clusters increases with increasing redshift, this offset may be because the red sequence cluster galaxies do not trace the true cluster center as accurately in unrelaxed clusters \citep{Rumbaugh2018}.  Physically, this larger offset could also be the result of these high-$z$ bent sources infalling into the cluster center for the first time.  However, this offset may result from our single color cut as opposed to a specific color range, since the color range aims to identify red sequence members, as opposed to any high-$z$ galaxies ($z$ $>$ 1.2).  Between the color cut and the high-$z$ nature of each field in the $[3.6] - [4.5]$ subsample, our cluster centers may be subject to more contamination and uncertainty using the dual-band $Spitzer$ analysis. 

\subsubsection{Comparing Color Cluster Centers and Surface Density Profiles}
\begin{figure*}
\begin{center}
\figurenum{22}
\subfigure{\includegraphics[scale=0.76,trim={1.85in 3.03in 2.18in 3.51in},clip=true]{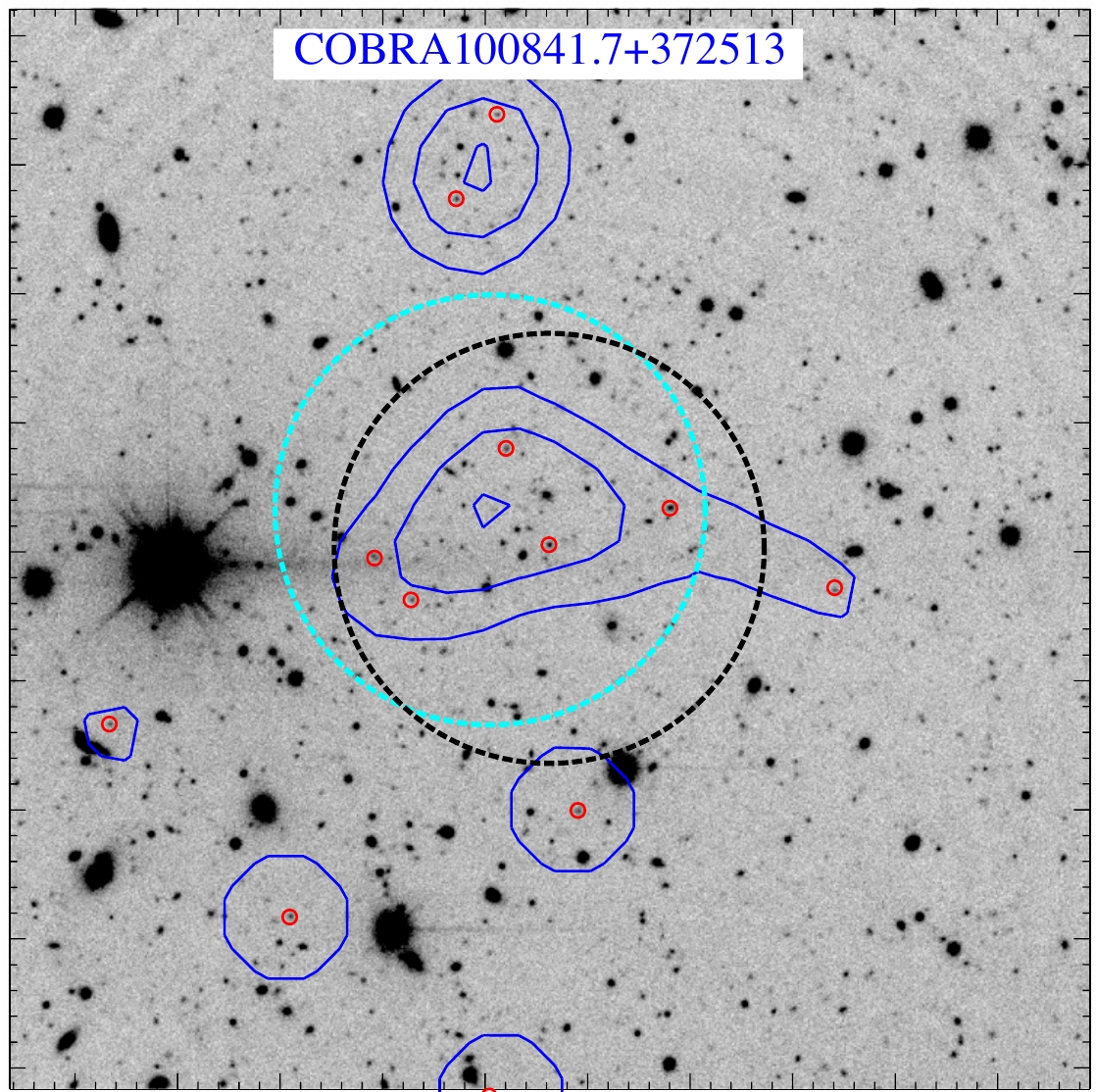}}
\subfigure{\includegraphics[scale=0.7,trim={1.68in 2.85in 1.98in 3.31in},clip=true]{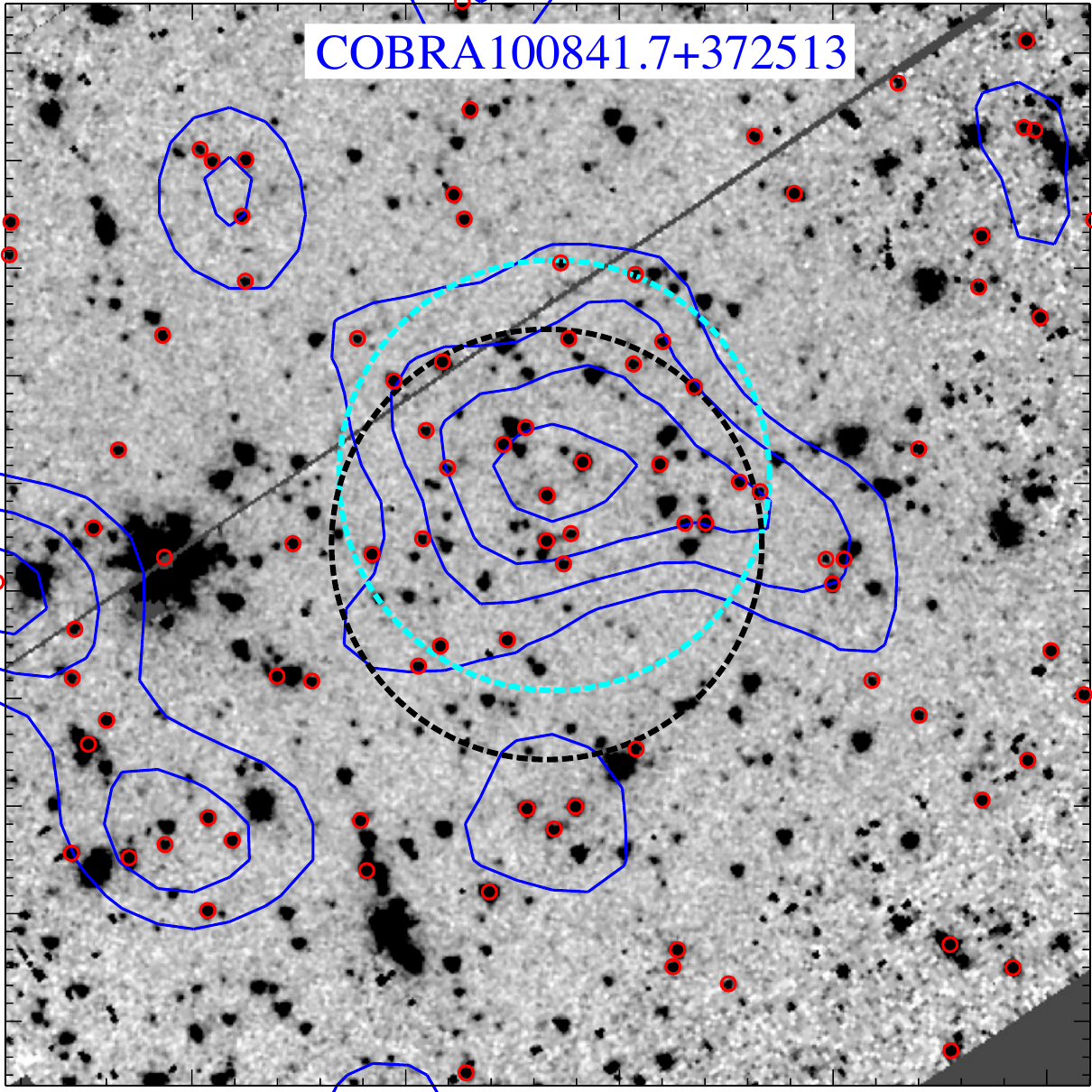}}

\caption{Examples of 5$\arcmin$ $\times$ 5$\arcmin$ cutouts of the $i - [3.6]$ and $[3.6] - [4.5]$ analysis for COBRA100841.7+372513 ($z$ $\approx$ 1.2/1.35).  The left image shows an $i$-band observation and the right image shows a 3.6\,$\mu$m $Spitzer$ observation.  In each, the red circles identify red sequence members (within $\pm$0.15\,mag of the host color for $i - [3.6]$ analysis and with a color greater than -0.15\,mag for the $[3.6] - [4.5]$ analysis).  The blue contours represent the red sequence surface density contours for each field.  The 1$\arcmin$ radius black dashed circle is centered on the radio source (with the host galaxy at the center), while the 1$\arcmin$ radius cyan dashed circle is centered on the center of the distribution of red sequence galaxies.  COBRA100841.7+372513 has a 5.1$\sigma$ detection when centered on the radio source and on the distribution of red sequence sources in $i - [3.6]$.  In the $[3.6] - [4.5]$ analysis, the cluster candidate shows a 3.0$\sigma$ detection centered on the radio source and a 4.1$\sigma$ detection when centered on the red sources.}
\label{Fig:CONTOURS2}
\end{center}
\end{figure*}

To compare the effectiveness of our independent color cuts at identifying similar galaxies in a given cluster, and thus determining similar cluster centers, we compare our limited overlapping sample.  For the six fields with $i - [3.6]$ and $[3.6] - [4.5]$ analyses, we find that four of the fields are offset from one another by less than 0$\farcm$5 ($\approx$ 220\,kpc).  The fields with larger cluster center differences may result from the low number of detected sources in the $i - [3.6]$ analysis; each field has less than three detected red sequence galaxies in $i - [3.6]$, leading to more uncertainty.  Additionally, some of the differences may result from the $i - [3.6]$ color cut identifying red sequence galaxies and the $[3.6] - [4.5]$ color cut identifying all high-$z$ galaxies.  Since clusters are not limited to our 1$\arcmin$ radial region, the larger offset does not preclude these color cuts from identifying the same structures.  Additionally, as highlighted in the combined analysis and the statistical analysis with the ORELSE data, a larger fraction of the redder and bluer galaxies in $i - [3.6]$ are actually at the target redshift, than the bluer galaxies in $[3.6] - [4.5]$.  Since the $[3.6] - [4.5]$ color cut does a better job of identifying high-$z$ galaxies and we expect larger populations of bluer galaxies at high-$z$, this may further explain the offsets and differences in structure.  As we continue to study COBRA fields, we will further address how these different color cuts compare to one another.

An example of a comparison between the $i - [3.6]$ and $[3.6] - [4.5]$ analysis is shown in Figure\,\ref{Fig:CONTOURS2}, which shows COBRA100841.7+372513, one of the strongest red sequence cluster candidates in $i - [3.6]$ and $[3.6] - [4.5]$ in the magnitude-limited sample.  Although we see a slight offset between the cluster centers ($\approx$ 170\,kpc), these methods identify three of the same galaxies in the inner 1$\arcmin$ region of the $i - [3.6]$ central region, as well as four other sources in the FOV.  Of the sources not detected in both color cuts, many result from either being below our magnitude limit in $i$-band, as is evident by comparing the two images, or fall outside of the color range in the other color.  Despite using different color cuts, we identify shared high-$z$ cluster galaxies between the two colors.

We find a similar level of agreement in our new cluster centers between our $r - i$ and $i - [3.6]$ analyses, with eight of the cluster centers being offset by less that 0$\farcm$6 ($<$ 300\,kpc) and an additional low-$z$ cluster candidate offset by $<$ 400\,kpc (though it is offset by $>$ 1$\arcmin$).    This likely results from both colors identifying red sequence galaxies.  Of the fields with offsets greater that 1$\arcmin$,  two have $r$ $-$ $i$ significances $<$ 1$\sigma$, which increases the degree of uncertainty in the central position of the distribution of galaxies.  Additionally, two other fields are more offset; one is at $z$ $\approx$ 0.351 (COBRA164611.2+512915, which is the field offset by $<$ 400\,kpc), where large angular sizes yield much smaller distances, and the other has a known blending issue with the BCG, thus increasing the error in a color estimated center (COBRA075516.6+171457).  It is also worth noting that the large separation between potential centers for COBRA135838.1+384722 ($>$ 1\,Mpc), likely implies two distinct structures associated with the radio source.  

\begin{figure*}
\begin{center}
\figurenum{23}
\subfigure{\includegraphics[scale=0.75,trim={1.85in 3.in 2.19in 3.55in},clip=true]{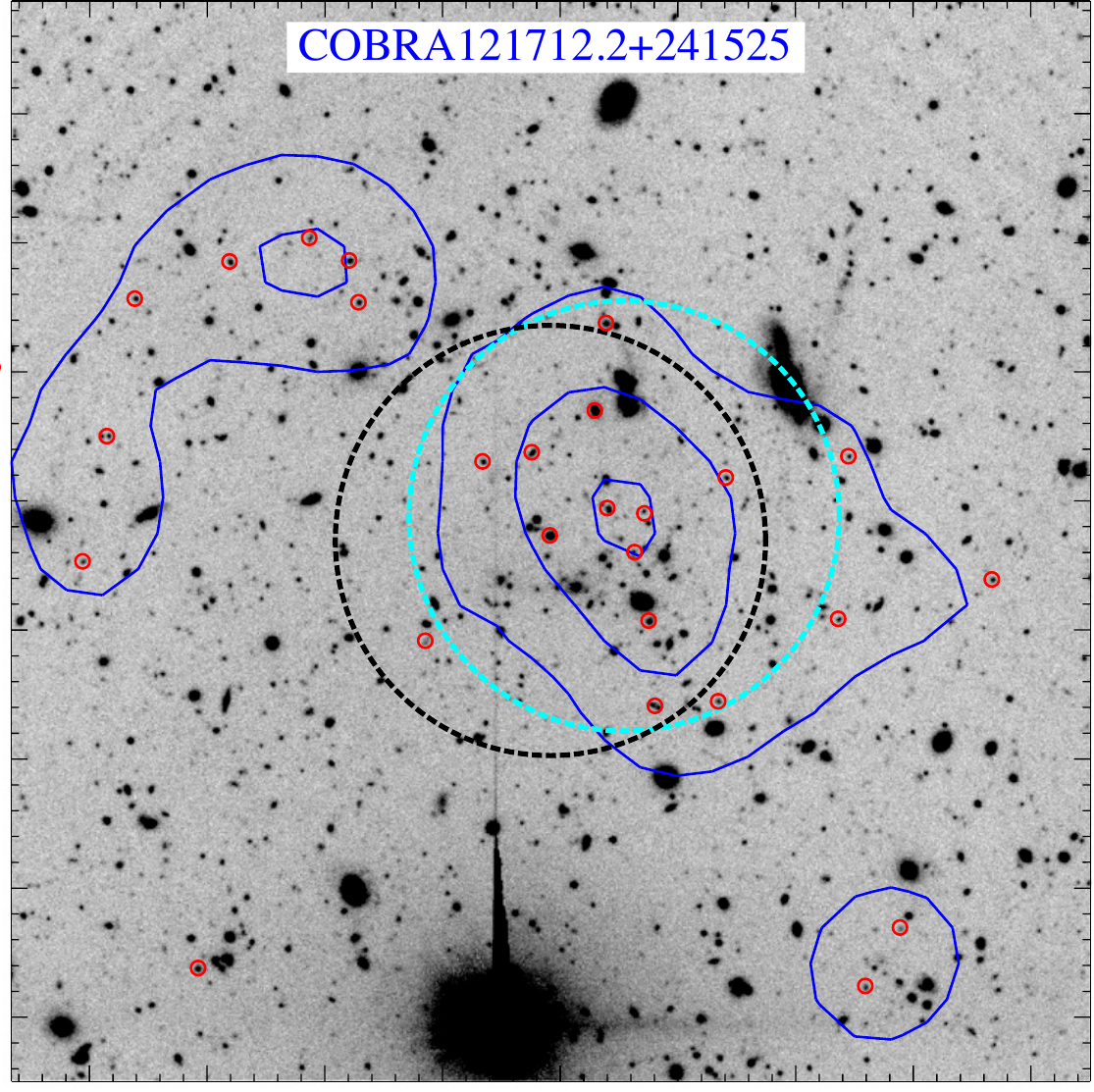}}
\subfigure{\includegraphics[scale=0.75,trim={1.87in 3.in 2.19in 3.55in},clip=true]{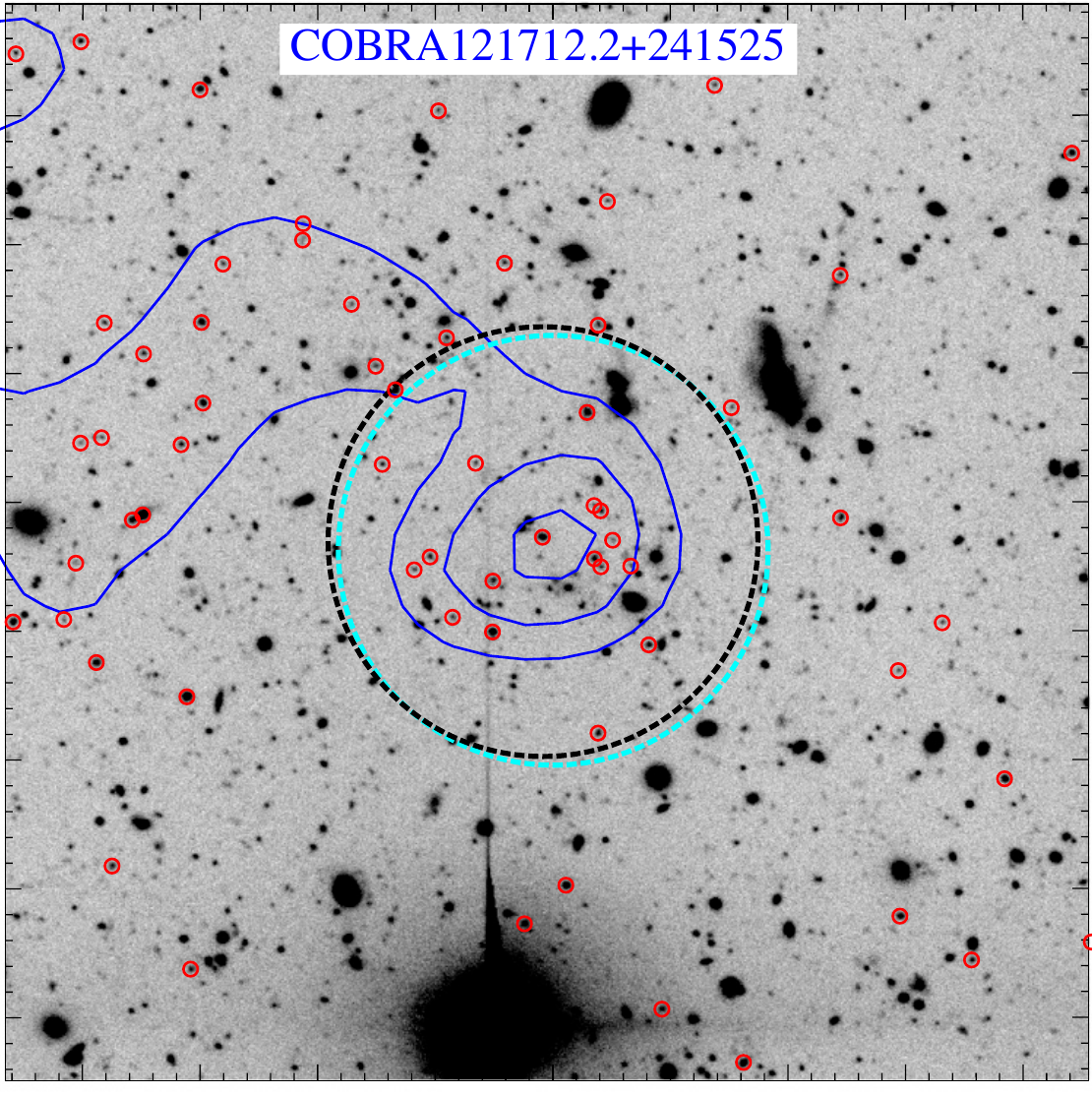}}

\caption{Examples of 5$\arcmin$ $\times$ 5$\arcmin$ cutouts of the $i - [3.6]$ and $r$ $-$ $i$ analysis for COBRA121712.2+241525 ($z$ $\approx$ 0.9).  The left image shows an $i$-band image with $i - [3.6]$ contours and the right image shows an $r$-band image with $r - i$ contours.  In each, the red circles are red sequence members (within $\pm$ 0.15\,mag of the host color for both the $i - [3.6]$ analysis the $r$ $-$ $i$ analysis).  The blue contours represent the red sequence surface density contours for each field.  The 1$\arcmin$ radius black dashed circle is centered on the radio source (with the host galaxy at the center), while the 1$\arcmin$ radius cyan dashed circle is centered on the center of the distribution of red sequence galaxies.  In $i - [3.6]$ COBRA121712.2+241525 has a 7.3$\sigma$ detection when centered on the radio source and an 8.1$\sigma$ detection when centered on the distribution of red sequence sources in $i - [3.6]$.  In the $r$ $-$ $i$ analysis, the cluster candidate shows a 5.5$\sigma$ detection when centered on the radio source and a 5.1$\sigma$ detection when centered on the distribution of red galaxies.}
\label{Fig:CONTOURS3}
\end{center}
\end{figure*}

For most of the fields, the difference in the position between the $i - [3.6]$ and $r - i$ cluster centers relative to the overall size of clusters is small, which implies both colors detect the same sources.  Figure\,\ref{Fig:CONTOURS3} shows COBRA121712.2+241525, one of the strongest red sequence candidates in both $r$ $-$ $i$ and $i - [3.6]$.  The two red sequence color ranges identify five shared galaxies in the central region based on the $i$ $-$ [3.6] analysis and identify cluster centers almost identical to one another (offset by $\approx$ 150\,kpc).  The majority of sources detected in one color but not another are either below the $r$-band magnitude limit, or are blended sources in our $Spitzer$ image.  Additionally, both pick up sources of a similar color to east of the bent AGN, which may imply that this is some kind of infalling subcluster or filament.  The similarity between the different color cuts further strengthens our resolve that these methods identify similar cluster galaxies. 

\subsection{Determining Potential Cluster Contamination and Identifying Previously Detected Clusters}
\begin{figure*}
\begin{center}
\figurenum{24}
\subfigure{\includegraphics[scale=0.75,trim={1.85in 3.in 2.19in 3.55in},clip=true]{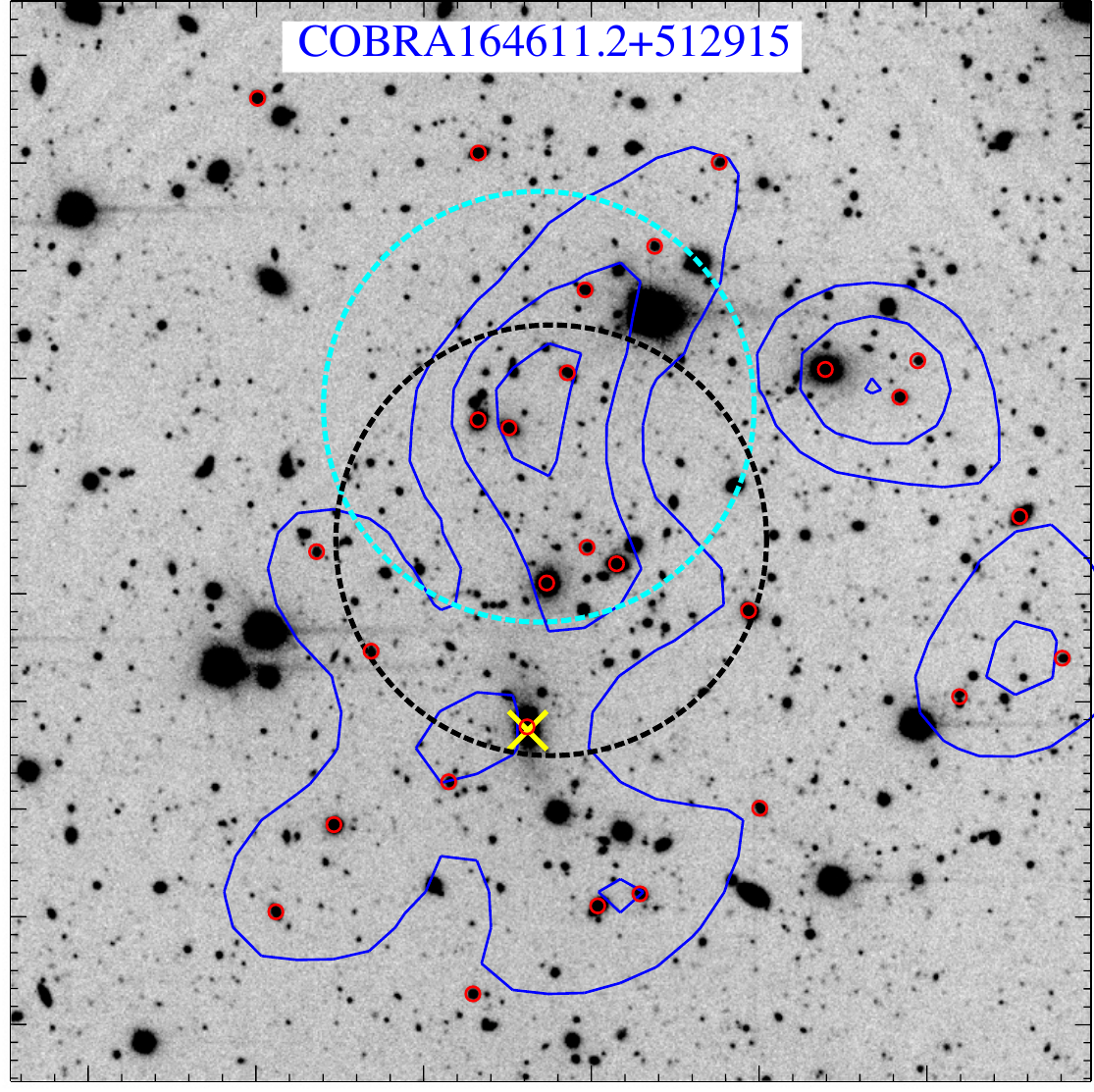}}
\subfigure{\includegraphics[scale=0.75,trim={1.85in 3.in 2.19in 3.53in},clip=true]{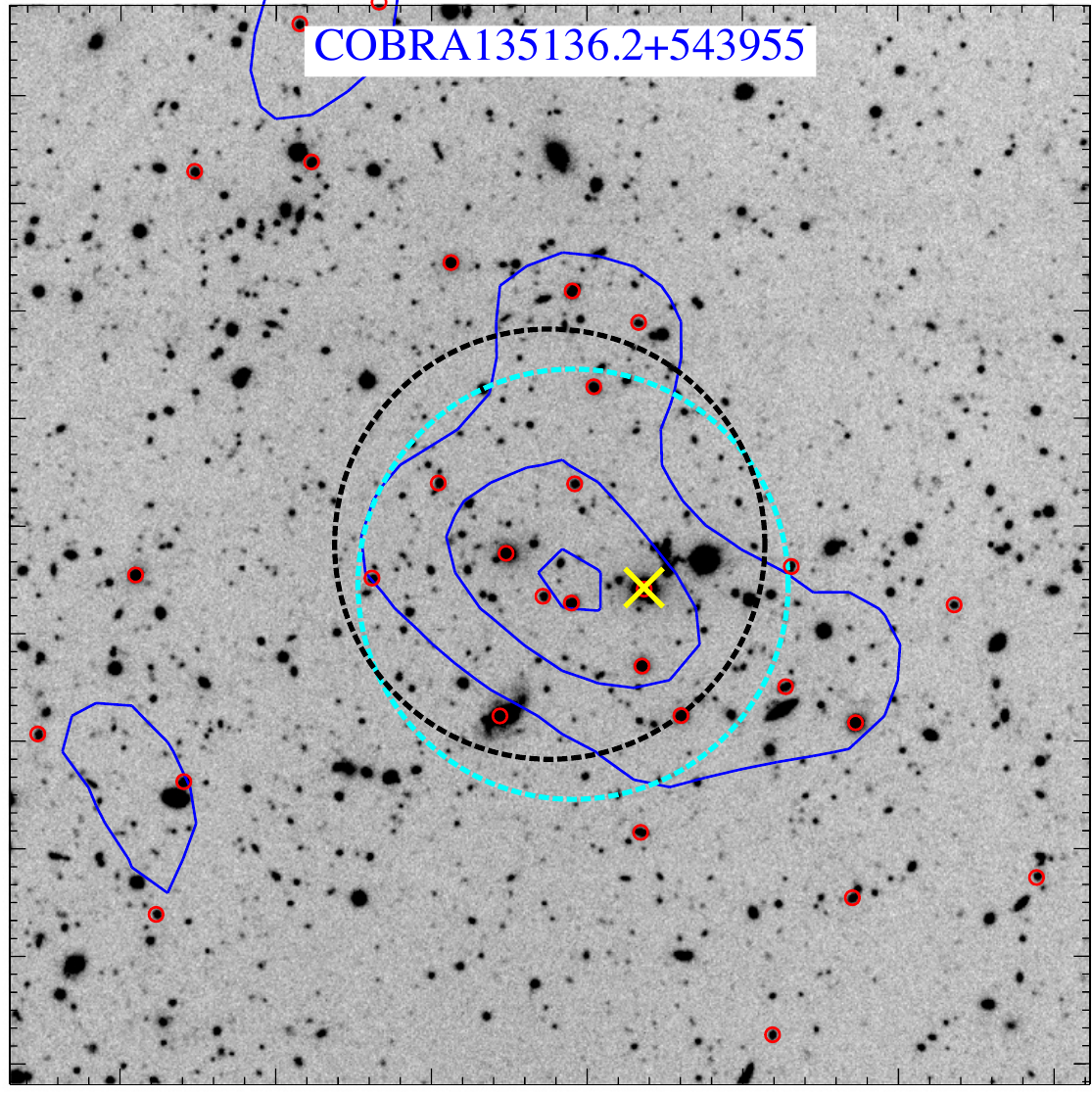}}

\caption{Examples of 5$\arcmin$ $\times$ 5$\arcmin$ cutouts of DCT $i$-band images of the two fields that are identified as known clusters in redMaPPer and NED.  The left image shows COBRA164611.2+512915 ($z$ = 0.351) and the right images shows COBRA135136.2+543955 ($z$ $\approx$ 0.55).  In each, the red circles are red sequence members (within $\pm$0.15\,mag of the host color for $i - [3.6]$ analysis).  The blue contours represent the red sequence surface density contours for each field.  The 1$\arcmin$ radius black dashed circle is centered on the radio source (with the host galaxy at the center), while the 1$\arcmin$ radius cyan dashed circle is centered on the center of the distribution of red sequence galaxies.  The yellow X identifies the location of the cluster in either redMaPPer or NED.  Both are centered on the BCG.  In $i - [3.6]$ COBRA164611.2+512915 has an 8.7$\sigma$ detection when centered on the radio source and a 7.7$\sigma$ detection when centered on the distribution of red sequence sources.  This field shows an interesting amount of structure and is distinctly non-symmetrical, when compared to those in Figure\,\ref{Fig:CONTOURS1}.  In $i - [3.6]$, COBRA135136.2+543955 has a a 6.4$\sigma$ detection when centered on the radio source and a 7.0$\sigma$ detection when centered on the distribution of red sequence sources.}
\label{Fig:CONTOURS4}
\end{center}
\end{figure*}

To strengthen our confidence that we are detecting real overdensities rather than random alignments of galaxies, we measure the fraction of $i - [3.6]$, $[3.6] - [4.5]$, and $r - i$ background regions that are above our 2$\sigma$ limit and the average separation between these fields.  The separation in 1$\arcmin$ regions between our various background regions is an approximation of the spacing between the original AGN centered search region and our new red sequence surface density centered search region.  In our $i - [3.6]$ analysis, $\approx$ 5\,$\%$ of our $i - [3.6]$ background regions are above 2$\sigma$ for the m*+1 sample, while 5 - 10\,$\%$ of $i - [3.6]$ background regions are above 2$\sigma$ in the magnitude-limited sample. The average separation between 2$\sigma$ overdensities in these $i - [3.6]$ background regions is $\approx$ 15$\arcmin$ in the m*+1 sample and 13$\arcmin$ in the magnitude-limited sample.  In our $[3.6] - [4.5]$ analysis, $\approx$ 4\,$\%$ of $[3.6] - [4.5]$ background regions are at the 2$\sigma$ level and the average distance between 2$\sigma$ regions is $\approx$ 16$\arcmin$.  Although we are not able to measure the mean spacing between 2$\sigma$ regions for our $r$ $-$ $i$ background, 6 - 12\,$\%$ of $r - i$ background regions are above the 2$\sigma$ threshold.  We identify well above 10$\%$ of our COBRA fields as cluster candidates ($\approx$ 67$\%$ for $i - [3.6]$, 35$\%$ for $[3.6] - [4.5]$, and 64$\%$ for $r$ $-$ $i$) and most fields are offset less than 1$\arcmin$ from the radio source, far less than the mean separation between 2$\sigma$ background regions in all of our samples.  These factors strengthen our confidence that we are not identifying random line of sight alignments of galaxies.

To verify that we are not identifying low-$z$ clusters, we measure the distance between each radio source and the nearest redMaPPer cluster \citep{Rykoff2014} and search the NASA/IPAC Extragalactic Database (NED) archive for all clusters within 6$\arcmin$ of the radio source (the maximum distance a cluster center could be offset from our AGN and still be in our image based on the AGN being the center of our $\approx$ 12$\arcmin$ $\times$ 12$\arcmin$ DCT LMI images).  In comparing our cluster coordinates to those of redMaPPer v6.3, we find that COBRA164611.2+51915 ($z$ $\approx$ 0.35) is the only COBRA cluster in redMaPPer (with an offset of $\approx$ 0$\farcm$89), while 4 other fields include redMaPPer clusters within 10$\arcmin$.  Of these fields, none overlap with our cluster centers.  From our NED search, we find some fields have at least one low-$z$ cluster within 6$\arcmin$ of the radio source.  To verify that we aren't detecting low-$z$ contaminants, we compare our new cluster centers to these low-$z$ clusters for all fields.   Only two fields show a slight overlap between 1$\arcmin$ radial regions centered on the new cluster center and the NED cluster, while the vast majority are significantly offset.  Additionally, only one field has a low-$z$ cluster within 2$\arcmin$ of the radio source.  As a result of this search, we find that two of our cluster candidates, COBRA135136.2+543955 ($z$ $\approx$ 0.55) and COBRA164611.2+512915 ($z$ $\approx$ 0.35; also in redMaPPer), are previously identified clusters (both redshift estimates agree with our measurements).  Interestingly, in both clusters, the bent radio source is offset by more than 0$\farcm$5 from the reported cluster center (0$\farcm$56 for COBRA135136.2+543955 and 0$\farcm$89 for COBRA164611.2+512915 - the same location as the redMapper identification).  These results are encouraging as we continue to search for overdensities offset from the radio source.

We compare the positions of our cluster centers with the locations reported in NED for these fields and find that our locations are slighty offset from the reported values ($\approx$ 0$\farcm$3 for COBRA135136.2+543955 and $\approx$ 1$\farcm$5 for COBRA164611.2+512915; See Figure\,\ref{Fig:CONTOURS4}).  The major difference between are center locations comes down to how we identify cluster centers.  We estimate the cluster center based on the overall distribution of all red sequence galaxies.  We do not bias ourselves to the BCG or proto-BCG.  However, the NED and redMaPPer center values are all the location of the BCG, creating the difference.  Although our centers are offset from the BCG, the fact that our host galaxy in these clusters is not the BCG strengthens our decision to not treat the radio source as the cluster center.  Additionally, the low-$z$ nature of both sources may allow for more non-cluster contaminants based on our color criteria.  

\subsection{Comparing COBRA to Other High-$z$ Galaxy Cluster Surveys}
Using a similar method of AGN targeting, \citet{Wylezalek2013} find that 55.3$\%$ of CARLA RLAGN are in overdense cluster environments.  When CARLA fields are analyzed with identical methodology to COBRA, \citet{Paterno-Mahler2017} find that 44$\%$ of CARLA sources and 29$\%$ of COBRA bent, double-lobed radio sources are in overdense cluster environments ($>$ 2$\sigma$).  Although we only examine a small and biased sample of COBRA with our new measurements of the cluster center and red sequence galaxies, we add four cluster candidates not previously identified in \citet{Paterno-Mahler2017} to the COBRA cluster candidate sample (the other 35 red sequence cluster candidates were all identified using the 3.6\,$\mu$m analysis in \citet{Paterno-Mahler2017} or using the $i - [3.6]$ red sequence analysis centered on the radio source), thus increasing the number of cluster candidates to 195 of 646, or 30$\%$.  Since we find these four fields in the 77 with redshift estimates that we analyze in this work, we might expect to find an additional 30 fields that are cluster candidates in the entire sample.  However, given that our sample of optical fields was selected based on stronger IRAC overdensities presented in \citet{Paterno-Mahler2017} (see Figure\,\ref{Fig:Allsources}), and in general our red sequence overdensity scales linearly with $Spitzer$ overdensity, if we assume a 30 field increase as an upper limit for the entire COBRA sample, we expect $\approx$ 210 cluster candidates.  Thus, addressing the offset in cluster position with respect to the AGN could account for some of the difference between CARLA and COBRA's ability to detect high-$z$ cluster candidates described in \textsection{1} and \citet{Paterno-Mahler2017}.  

Furthermore, \citet{Cooke2015} report the significance of their 37 high-$z$ cluster candidates with their 3.6\,$\mu$m and 4.5\,$\mu$m observations and find that most are above 3$\sigma$, with the highest being 6.3$\sigma$.  This is in contrast with our seven cluster candidates found using our $[3.6] - [4.5]$ analysis (see Table\,\ref{tb:ch1ch2AGN}), where four of the seven are between 2$\sigma$ and 3$\sigma$ and the maximum overdensity is 4.1$\sigma$.  As noted, these differences likely result from COBRA on average finding less massive cluster candidates than CARLA and CARLA observations reaching a fainter completeness limit and thus being more sensitive to high-$z$ clusters.  

Although CARLA is one of the largest AGN targeting cluster surveys, our methodology shares many similarities with other large scale cluster surveys. The Massive and Distant Clusters of $WISE$ Survey (MaDCoWS), which covers the entire sky, has found at least 11 $z$ $\ge$ 1 spectroscopically confirmed clusters by searching for rich overdensities of IR galaxies with $WISE$ and $\sl{Spitzer}$ \citep[e.g.,][]{Stanford2014,Gonzalez2015}.  With further optical observations from Pan-STARRS and the SuperCOSMOS Sky survey, \citet{Gonzalez2018} identify 2683 MADCoWS cluster candidates at $z$ $>$ 0.7.  These massive clusters have a median redshift of $z$ $\approx$ 1.0 \citep{Gonzalez2018}, thus covering a similar redshift range to our sample.  Additionally, \citet{Moravec2018} undertake a pilot study of 10 $z$ $\approx$ 1.0 MaDCoWS clusters with extended radio sources and find sources offset between 0$\arcmin$ and 1$\arcmin$ from their $Spitzer$ centers.  Although their methodology for determining cluster centers differs from our own, the similarity between this result and ours measured offset between the radio source and the cluster center highlights that radio sources can be offset from high-$z$ cluster centers.  

Beyond MaDCoWS, the IRAC Shallow Cluster Survey (ISCS) explored the 8.5 square degree Bo{\"o}tes field using a mixture of optical and $Spitzer$ observations.  Unlike this COBRA analysis, ISCS identifies clusters using photometric redshifts of all galaxies regardless of SED type to do wavelet analysis on galaxy density maps \citep{Eisenhardt2008}.  ISCS reaches similar magnitude limits in their optical and IR observations to COBRA ($m_{3.6\mu m}$=21.935 for ISCS compared to 21.4\,mag in COBRA; ISCS is complete for an L* galaxy at $z$ $\approx$ 1.3 in $I$-band, which is approximately the same depth as COBRA in $i$-band).  Of the 335 ISCS galaxy cluster candidates, 106 are at $z$ $>$ 1.0 \citep{Eisenhardt2008}, while 36 of 146 (with redshift estimates) COBRA galaxy cluster candidates are at $z$ $>$ 1.0.  We find 24.7$\%$ of cluster cadidates are at $z$ $>$ 1.0 in COBRA as compared to 26.7$\%$ in ISCS.  Thus, both surveys find higher-$z$ clusters at a similar rate.  Since there are still 49 COBRA cluster candidates without redshift estimates, this number should rise, bringing the rate of high-$z$ clusters in COBRA closer to or beyond that of ISCS.  

\begin{figure}
\begin{center}

\figurenum{25}
\includegraphics[scale=0.7,trim={1.7in 2.85in 2.0in 3.35in},clip=true]{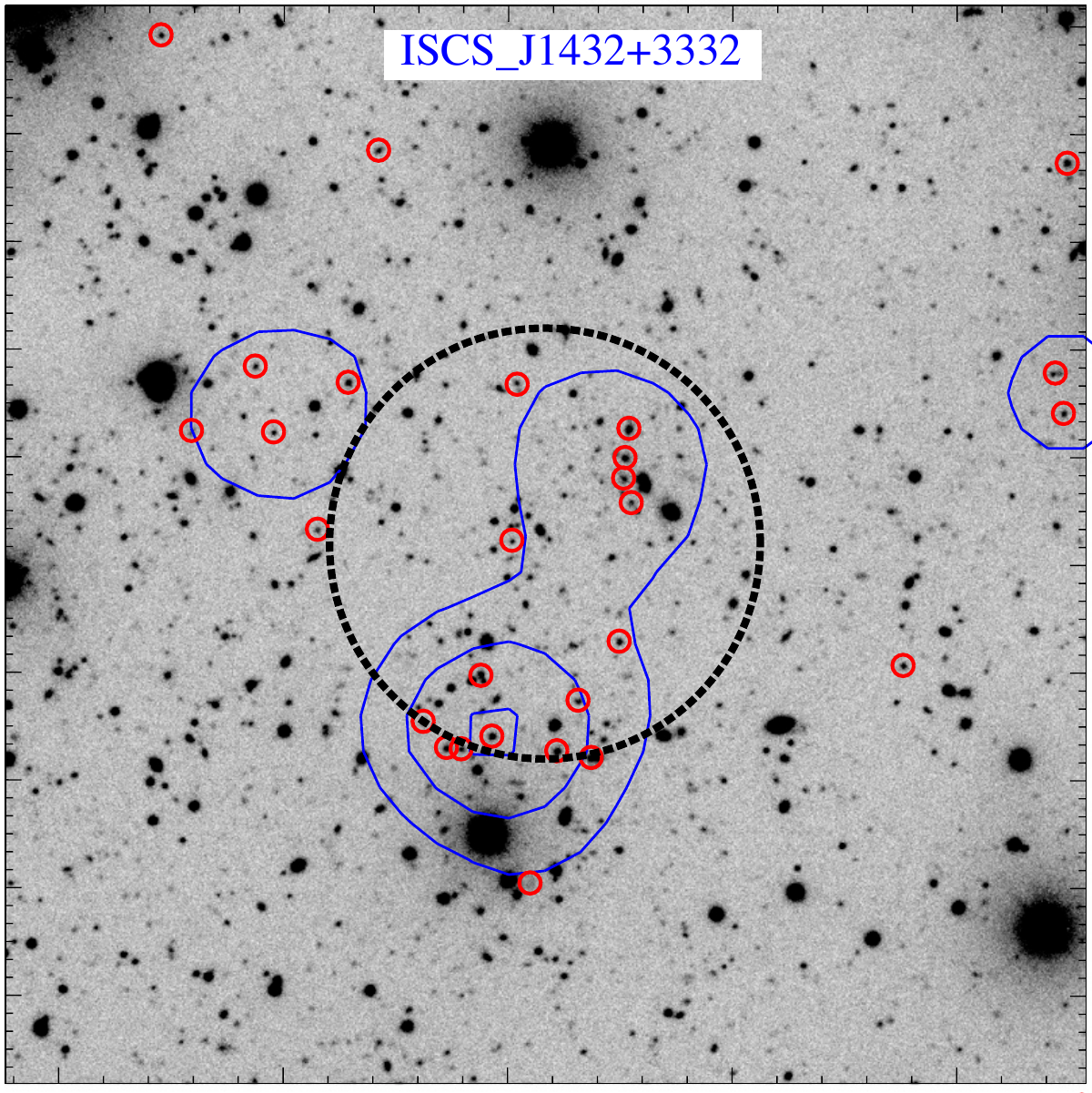}
\caption{An example of ISCS J1432+3332 ($z$=1.11).  For comparison to Fig\,\ref{Fig:CONTOURS1}, we show a 5$\arcmin$ $\times$ 5$\arcmin$ cutout of $I$-band observations \citep{Eisenhardt2008} of this fields.  The 1$\arcmin$ radius dashed black circle is centered on the IRAC centroid.  The surface density contours are shown in blue and measured in the same manner as our COBRA fields.  All red sequence galaxies detected with our red sequence color range are shown in red.  \label{Fig:ISCS}}   
\end{center}
\end{figure}

To further characterize our local red sequence surface density measurements discussed in \textsection{5} and shown in Figure\,\ref{Fig:CONTOURS1}, we measured the local surface density of ISCS J1432+3332 ($z$ = 1.11).  ISCS J1432+3332 was one of the first spectroscopically confirmed clusters at $z$ $>$ 1.0 and was observed as part of the FLAMINGOS Extragalactic Survey \citep{Elston2006}, the NAOA Deep Wide-Field Survey \citep{Brown2003}, and ISCS  \citep{Eisenhardt2008}.  It has a weak-lensing mass of M$_{200}$ = 4.9$^{+1.6}_{-1.2}$ $\times$ 10$^{14}$\,M$_\odot$ \citep{Jee2011}.  Using an identical analysis for the Spitzer observations and a magnitude limit of 23.1\,mag (Vega magnitudes) in $I$-band, we find a similar distribution of red sequence galaxies to our own cluster candidates (Figure\,\ref{Fig:ISCS}).  For comparison, we show the same 5$\arcmin$ $\times$ 5$\arcmin$ FOV.  We find similar results for the ISCS fields as for our COBRA fields, with ISCS fields showing an excess of galaxies near, but slightly offset from, their IRAC centroid (the 1$\arcmin$ radius dashed black circle is centered on the IRAC centroid).  The similarity between our cluster candidates and ISCS confirmed high-$z$ galaxy clusters solidifies our use of the surface density measurements to identify high-$z$ clusters.
    
The XMM-Large Scale Structure (XMM-LSS) survey also explores a similar range of redshifts to COBRA.  XMM-LSS detects clusters in a 9 square degree field by identifying extended X-ray sources and comparing them to optical/IR observations to look for an excess of all sources as well as red sources.  Using the XMM-LSS, \citet{Willis2013} find nine spectroscopically confirmed clusters at 0.8 $<$ $z$ $<$ 1.2 and 11 cluster candidates with photometric redshifts at 0.8 $<$ $z$ $<$ 2.2.  The XMM-LSS cluster finding method requires a high surface density of red sources within each cluster, similar to our work.  We note that \citet{Willis2013} find that although many of their clusters have developed red sequences consistent with passive evolution, a subset of clusters have bluer red sequences, implying differences in the high-$z$ stellar population.  Additionally, \citet{Willis2013} estimate a range of cluster masses of 6 $\times$ $10^{13}$ $M_\odot$ to 1 $\times$ $10^{14}$ $M_\odot$, as opposed to the more massive clusters reported for the MaDCoWS sample \citep{Brodwin2015}.  This mass range makes the XMM-LSS survey clusters similar to the lower-mass end expected in the COBRA survey because bent-sources are not a mass-biased cluster tracer.   

\citet{Obrien2018} more recently examine the environment of 46 bent-tail radio galaxies using the 86 square degree $Spitzer$-South Pole Telescope deep field.  Of their 46 sources, only 16 have known redshifts.  Of these, four are associated with clusters, all of lower mass ($<$ 3.75 $\times$ 10$^{14}$ M$\odot$).  Although \citet{Obrien2018} do not identify any high-mass clusters hosting bent sources, their sample is small and the area of the sky covered is much smaller than the area probed by the COBRA survey.  In contrast to these results, COBRA contains 646 bent, double-lobed radio sources, and based on our overdensity measurements, we find both low- and high-mass cluster candidates.

\section{Conclusion}
This is the second in a series of COBRA papers and the first introducing our optical follow-up observations.  For the 646 COBRA fields, 195 of which are cluster candidates, we use our new DCT $r$- and $i$- band imaging to measure photometric redshifts for 77 COBRA fields.  Of these 77 fields with redshift estimates, 26 are new redshift estimates for COBRA fields, allowing us to better determine the redshift distribution of COBRA sources in the universe.  By combining these redshift estimates with our determinations of which fields are cluster candidates, we increase the number of COBRA cluster candidates with redshift estimates from 125 to 146.  More importantly, we explore the idea that bent, double-lobed radio sources need not live at the cluster center by measuring the surface density of red sequence galaxies across the entire field.  This allows us to identify which bent, double-lobed radio sources live in clusters and which clusters appear to host evolved red sequence galaxies.  Additionally, we use the surface density of red sequence galaxies to estimate new cluster centers, further showing that these bent AGN need not reside at the center of clusters.  Through this red sequence analysis, we identify 39 red sequence cluster candidates, the most likely clusters within our sample.  We analyze these overdensities by accounting for the fraction of red sequence galaxies at our target redshift and find general agreement between our red sequence overdensities and our combined overdensities, which further strengthens our confidence that high-$z$ bent, double-lobed radio sources reside in overdense clusters with populations of red sequence galaxies.

 To further understand these cluster candidates, we will continue exploring the positional offset between the radio source and the cluster center with respect to the geometry of the radio source in future work.  We will further this work by analyzing the overall distribution of red sequence galaxies to analyze cluster morphology as well as estimate red sequence slopes and populations.
 
\acknowledgments
We would like to thank the referee for the very helpful comments and suggestions for improving this paper.  EGM would like to thank Jesse Golden-Marx for reading drafts of this paper and for providing useful discussion regarding our computational analysis.  EGM would also like to thank the DCT telescope operators for their help with taking observations.  Additionally, EGM would like to thank the organizers of the Early Stages of Galaxy Cluster Formation 2017 Conference and the Tracing Cosmic Evolution with Clusters of Galaxies 2019 Conference for fostering stimulating discussions that led to ideas addressed in this paper.    

This work has been supported by the National Science Foundation, grant AST-1309032.

These results made use of the Discovery Channel Telescope at Lowell Observatory. Lowell is a private, non-profit institution dedicated to astrophysical research and public appreciation of astronomy and operates the DCT in partnership with Boston University, the University of Maryland, the University of Toledo, Northern Arizona University, and Yale University. LMI construction was supported by a grant AST-1005313 from the National Science Foundation.

This work is based in part on observations made with the  $\sl{Spitzer}$ Space Telescope, which is operated by the Jet Propulsion Laboratory, California Institute of Technology under a contract with NASA. Support for this work was provided by NASA through an award issued by JPL/Caltech (NASA award RSA No. 1440385).

Funding for SDSS-III has been provided by the Alfred P. Sloan Foundation, the Participating Institutions, the National Science Foundation, and the U.S. Department of Energy Office of Science. The SDSS-III web site is http://www.sdss3.org/.

SDSS-III is managed by the Astrophysical Research Consortium for the Participating Institutions of the SDSS-III Collaboration including the University of Arizona, the Brazilian Participation Group, Brookhaven National Laboratory, Carnegie Mellon University, University of Florida, the French Participation Group, the German Participation Group, Harvard University, the Instituto de Astrofisica de Canarias, the Michigan State/Notre Dame/JINA Participation Group, Johns Hopkins University, Lawrence Berkeley National Laboratory, Max Planck Institute for Astrophysics, Max Planck Institute for Extraterrestrial Physics, New Mexico State University, New York University, Ohio State University, Pennsylvania State University, University of Portsmouth, Princeton University, the Spanish Participation Group, University of Tokyo, University of Utah, Vanderbilt University, University of Virginia, University of Washington, and Yale University.

IRAF is distributed by the National Optical Astronomy Observatory, which is operated by the Association of Universities for Research in Astronomy (AURA) under a cooperative agreement with the National Science Foundation.

\facilities{DCT, $Spitzer$, Sloan}

\clearpage
\newpage
\mbox{~}
\clearpage
\newpage
\begin{deluxetable*}{lccccccccc}
\tablenum{1}
    \tablecolumns{10}
    \tabletypesize{\footnotesize}
    \tablecaption{COBRA Follow-up Optical Observations \label{tb:1}}
    \tablewidth{0pt}
    \tabletypesize{\footnotesize}
    
    \setlength{\tabcolsep}{0.05in}
    \tablehead{
    \colhead{}&
    \multicolumn{2}{c}{Host Coordinates}& 
    \multicolumn{5}{c}{DCT}& 
    \multicolumn{2}{c}{\sl{Spitzer}} 
    \cr
    \colhead{Field}&
    \colhead{RA}&
    \colhead{DEC}&
    \colhead{Date\tablenotemark{a}}&
    \colhead{Exp $r$}&
    \colhead{Seeing}&
    \colhead{Exp $i$}&
    \colhead{Seeing}&
    \colhead{Exp 3.6\,$\mu$m}&
    \colhead{Exp 4.5\,$\mu$m}
    \cr
    \colhead{}&
    \colhead{}&
    \colhead{}&
    \colhead{}&
    \colhead{(s)}&
    \colhead{($\arcsec$)}&
    \colhead{(s)}&
    \colhead{($\arcsec$)}&
    \colhead{(s)}&
    \colhead{(s)}
    }
\startdata
COBRA003447.7$-$002137&00 34 48.07&$-$00 21 32.0&13/11/03&\nodata&\nodata&3 $\times$ 600&1.3&4 $\times$ 100&4 $\times$ 100\\
COBRA005837.2+011326&00 58 37.03&+01 13 27.8&15/10/08&3 $\times$ 600&1.4&3 $\times$ 600&1.3&4 $\times$ 100&4 $\times$ 100\\
COBRA010329.5+004055\tablenotemark{c}&01 03 29.40&+00 40 55.2&15/10/10&4 $\times$ 900&1.2&3 $\times$ 900&1.3&4 $\times$ 100&4 $\times$ 100\\
COBRA012058.9+002140&01 20 58.87&+00 21 41.7&13/11/04&4 $\times$ 600&0.8&4 $\times$ 600&0.8&7 $\times$ 30&\nodata\\
COBRA014147.2$-$092812&\nodata\tablenotemark{b}&\nodata\tablenotemark{b}&15/10/10&3 $\times$ 600&1.5&3 $\times$ 600&1.1&7 $\times$ 30&\nodata\\
COBRA014741.6$-$004706&01 47 41.73&$-$00 47 08.5&15/10/09&3 $\times$ 900&1.3&3 $\times$ 600&1.5&7 $\times$ 30&\nodata\\
COBRA015313.0$-$001018&01 53 12.96&$-$00 10 19.9&13/11/04&\nodata&\nodata&4 $\times$ 600&0.8&7 $\times$ 30&\nodata\\
COBRA020823.8$-$011536&02 08 23.30&$-$01 15 48.9&15/10/11&3 $\times$ 900&0.8&3 $\times$ 900&0.8&7$\times$ 30&\nodata\\
\enddata
\tablenotetext{a}{All dates are given as year/month/day in UT time and refer to the dates of the DCT observations.}
\tablenotetext{b}{No host coordinate is identified in \citet{Paterno-Mahler2017}.}
\tablenotetext{c}{Quasar Fields}
\tablenotetext{d}{The RA and DEC of this field were misidentified in \citet{Paterno-Mahler2017}.  Here, we report the correct RA and DEC of our identified host galaxy.}
\tablenotetext{e}{Upon further analysis, we are uncertain of the host identified in \citet{Paterno-Mahler2017} for this field and choose not to report one.  We are keeping this fields in the overall sample of 646 sources, but need to redetermine the host coordinates.}
\end{deluxetable*}

\clearpage
\newpage
\mbox{~}
\clearpage
\newpage
\begin{deluxetable*}{lcccccccccccc}
\tablenum{2}  
    \tablecolumns{13}
    \tabletypesize{\small}
    \tablecaption{COBRA Red Sequence Galaxy Excess Measurements\label{tb:2}}
    \tablewidth{0pt}
    \tabletypesize{\footnotesize}

    \setlength{\tabcolsep}{0.05in}
    \tablehead{
    \colhead{Field}&
    \multicolumn{2}{c}{Central Coordinates}&
    \multicolumn{3}{c}{$i - [3.6]$}& 
    \multicolumn{3}{c}{$[3.6] - [4.5]$}&
    \multicolumn{3}{c}{$r - i$}&
    \colhead{3.6\,$\mu$m\tablenotemark{i}}
    \cr
    \colhead{}&
    \colhead{RA}&
    \colhead{DEC}&
    \colhead{N\tablenotemark{a}}&
    \colhead{$\Delta$N\tablenotemark{b}}&
    \colhead{$\sigma$\tablenotemark{c}}&
    \colhead{N\tablenotemark{a}}&
    \colhead{$\Delta$N\tablenotemark{b}}&
    \colhead{$\sigma$\tablenotemark{c}}&
    \colhead{N\tablenotemark{a}}&
    \colhead{$\Delta$N\tablenotemark{b}}&
    \colhead{$\sigma$\tablenotemark{c}}&
    \colhead{$\sigma$\tablenotemark{c}} 
    \cr
    }
    
\startdata    
COBRA003447.7$-$002137 & 00 34 47.70 & $-$00 21 37.0 & 0\tablenotemark{f} & \nodata & \nodata & 10 & 0.2 & 0.1 & \nodata & \nodata & \nodata & $-$0.3\\
``  " & 00 34 47.70 & $-$00 21 37.0 & 0\tablenotemark{f} & \nodata & \nodata & \nodata & \nodata & \nodata & \nodata & \nodata & \nodata & \nodata \\
``  " & 00 34 46.45 & $-$00 23 33.3 & \nodata & \nodata & \nodata & 15 & 5.2 & 1.4 & \nodata & \nodata & \nodata & \nodata\\
\hline
COBRA005837.2+011326\tablenotemark{h} & 00 58 37.20 & +01 13 26.0 & 6 & 5.5 & 3.6\tablenotemark{e} & \nodata & \nodata & \nodata & 9 & 7.2 & 3.9\tablenotemark{e} & 2.2\\
``  " & 00 58 36.47 & +01 13 26.4 & 7 & 6.5 & 4.3\tablenotemark{e} & \nodata & \nodata & \nodata & \nodata & \nodata & \nodata & \nodata\\
``  " & 00 58 35.45 & +01 13 15.7 & \nodata & \nodata & \nodata & \nodata & \nodata & \nodata & 10 & 8.2 & 4.5\tablenotemark{e} & \nodata\\
\hline
COBRA010329.5+004055 & 01 03 29.50 & +00 40 55.0 & 0\tablenotemark{f} & \nodata & \nodata & 7 & $-$2.8 & $-$0.7 & \nodata & \nodata & \nodata & 0.0\\
``  " & 01 03 27.18 & +00 42 16.0 & 1\tablenotemark{f} & \nodata & \nodata & \nodata & \nodata & \nodata & \nodata & \nodata & \nodata & \nodata\\
\enddata
\end{deluxetable*}

\clearpage
\newpage
\mbox{~}
\clearpage
\newpage

\begin{deluxetable*}{lccccccc}
\tablenum{3}
    \tablecolumns{5}
    \tabletypesize{\footnotesize}
    \tablecaption{COBRA Host Apparent Magnitudes and Photometric Redshifts \label{tb:4}}

    \tablehead{
    \colhead{Field}&
    \multicolumn{4}{c}{Apparent Magnitude}&
    \multicolumn{2}{c}{Redshifts}
    \cr
    \colhead{}&
    \colhead{m$_{r}$}&
    \colhead{m$_{i}$}&
    \colhead{m$_{3.6\,\mu\,m}$}&
    \colhead{m$_{4.5\,\mu\,m}$}&
    \colhead{Photometric}&
    \colhead{Spectroscopic}
    \cr
     }
     
\startdata
COBRA003447.7$-$002137 &\nodata&\nodata&20.19&20.00& 1.75\tablenotemark{c} &\nodata \\
COBRA005837.2+011326 &21.36&19.95&18.02&18.56& 0.71\tablenotemark{a}\tablenotemark{b}\tablenotemark{c}\tablenotemark{d}&\nodata \\
COBRA010329.5+004055 &18.48&18.28&18.21&17.77& \nodata&1.433 \\
COBRA012058.9+002140 &21.37&20.14&17.96&\nodata&0.75\tablenotemark{a}\tablenotemark{b}\tablenotemark{d} &\nodata \\
COBRA014147.2$-$092812 &\nodata&\nodata&\nodata&\nodata&\nodata\tablenotemark{e} &\nodata \\
COBRA014741.6$-$004706\tablenotemark{j} &23.27&23.18&23.10&\nodata&0.60\tablenotemark{f}\tablenotemark{i} &\nodata \\
COBRA015313.0$-$001018 &\nodata&24.11& 18.13&\nodata&0.44\tablenotemark{d} &\nodata \\
COBRA020823.8$-$011536 &\nodata&25.98&19.52&\nodata&\nodata\tablenotemark{f} &\nodata \\
\enddata
\tablenotetext{}{For fields with multiple colors and redshift values, we settled on a match between two or more different estimates if the estimate from EzGal was with $\pm$ 0.05 of one another.  If two or more redshift estimates matched, but a third (or fourth) did not, we chose the redshift estimate based on the matching values.}
\tablenotetext{}{For fields with multiple colors where the redshift estimates do not match, we report both values.}
\tablenotetext{a}{Photometric redshift from $r - i$ color}
\tablenotetext{b}{Photometric redshift from $i - [3.6]$ color}
\tablenotetext{c}{Photometric redshift from $[3.6] - [4.5]$ color}
\tablenotetext{d}{Photometric redshift from SDSS}
\tablenotetext{e}{No host found}
\tablenotetext{f}{No match with the EzGal models}
\tablenotetext{g}{No host identified in $r$- or $i$-band}
\tablenotetext{h}{The redshift estimate for this field in \citet{Paterno-Mahler2017} is for a nearby foreground galaxy, but not the host galaxy.}
\tablenotetext{i}{The redshift estimate is based on the red sequence histogram, not host color.}
\tablenotetext{j}{Fields with new redshift estimates.}

\end{deluxetable*}
\clearpage
\newpage
\mbox{~}
\clearpage
\newpage

\begin{deluxetable*}{lccccccccc}
\tablenum{6}
    \tablecolumns{9}
    \tabletypesize{\footnotesize}
    \tablecaption{Red Sequence Completeness Measure \label{tb:completeness}}

    \tablehead{
    \colhead{Field}&
    \multicolumn{3}{c}{$i - [3.6]$}&
    \multicolumn{2}{c}{$[3.6] - [4.5]$}&
    \multicolumn{3}{c}{$r - i$}
    \cr
    \colhead{}&
    \colhead{RS}&
    \colhead{Bluer}&
    \colhead{Redder}&
    \colhead{RS}&
    \colhead{Bluer}&
    \colhead{RS}&
    \colhead{Bluer}&
    \colhead{Redder}
    \cr
    \colhead{}&
    \colhead{$\%$}&
    \colhead{$\%$}&
    \colhead{$\%$}&
    \colhead{$\%$}&
    \colhead{$\%$}&
    \colhead{$\%$}&
    \colhead{$\%$}&
    \colhead{$\%$}
     }
     
\startdata
COBRA003447.7$-$002137 & 7.9 & 1.2 & 0.0 & 58.1 & 6.4 & \nodata & \nodata & \nodata \\
COBRA005837.2+011326 & 48.3 &18.3 & 16.9 & \nodata & \nodata & 36.1 & 20.7 & 17.6\\
COBRA010329.5+004055 & 32.2 & 2.5 & 5.9 & 58.1 & 6.4 & \nodata & \nodata & \nodata \\
COBRA012058.9+002140 & 47.8 & 19.1 & 12.2 & \nodata & \nodata & 50.8 & 14.7 & 30.2\\
COBRA014147.2$-$092812 & \nodata & \nodata & \nodata & \nodata & \nodata & \nodata & \nodata & \nodata \\
COBRA014741.6$-$004706 & 50.5 & 19.8 & 16.8 & \nodata & \nodata & 22.7 & 21.2 & 16.4 \\
COBRA015313.0$-$001018 & 38.0 & 19.4 & 5.1 & \nodata & \nodata  & \nodata & \nodata & \nodata \\
COBRA020823.8$-$011536 & \nodata & \nodata & \nodata & \nodata & \nodata & \nodata & \nodata & \nodata \\
\enddata
\tablenotetext{}{Fields with \nodata listed for all bands have no redshift estimate.}

\end{deluxetable*}

\clearpage
\newpage
\mbox{~}
\clearpage
\newpage

\begin{deluxetable*}{lcccccccccccc}
\tablenum{7}  
    \tablecolumns{13}
    \tabletypesize{\small}
    \tablecaption{COBRA Combined Overdensity Measurements\label{tb:combo}}
    \tablewidth{0pt}
    \tabletypesize{\footnotesize}

    \setlength{\tabcolsep}{0.05in}
    \tablehead{
    \colhead{Field}&
    \multicolumn{2}{c}{Central Coordinates}&
    \multicolumn{3}{c}{$i - [3.6]$}& 
    \multicolumn{3}{c}{$[3.6] - [4.5]$}&
    \multicolumn{3}{c}{$r - i$}&
    \colhead{3.6\,$\mu$m\tablenotemark{i}}
    \cr
    \colhead{}&
    \colhead{RA}&
    \colhead{DEC}&
    \colhead{N\tablenotemark{a}}&
    \colhead{$\Delta$N\tablenotemark{b}}&
    \colhead{$\sigma$\tablenotemark{c}}&
    \colhead{N\tablenotemark{a}}&
    \colhead{$\Delta$N\tablenotemark{b}}&
    \colhead{$\sigma$\tablenotemark{c}}&
    \colhead{N\tablenotemark{a}}&
    \colhead{$\Delta$N\tablenotemark{b}}&
    \colhead{$\sigma$\tablenotemark{c}}&
    \colhead{$\sigma$\tablenotemark{c}} 
    \cr
    }
    
\startdata    
COBRA003447.7$-$002137 & 00 34 47.70 & $-$00 21 37.0 & 16\tablenotemark{f} & \nodata & \nodata & 22 & 0.1 & 0.0 & \nodata & \nodata & \nodata & $-$0.3\\
``  " & 00 34 47.70 & $-$00 21 37.0 & 16\tablenotemark{f} & \nodata & \nodata & \nodata & \nodata & \nodata & \nodata & \nodata & \nodata & \nodata \\
``  " & 00 34 46.45 & $-$00 23 33.3 & \nodata & \nodata & \nodata & 29 & 3.1 & 1.4 & \nodata & \nodata & \nodata & \nodata\\
\hline
COBRA005837.2+011326\tablenotemark{h} & 00 58 37.20 & +01 13 26.0 & 28 & 4.4 & 4.0\tablenotemark{e} & \nodata & \nodata & \nodata & 39 & 3.1 & 1.9\tablenotemark{e} & 2.2\\
``  " & 00 58 36.47 & +01 13 26.4 & 27 & 4.9 & 4.4\tablenotemark{e} & \nodata & \nodata & \nodata & \nodata & \nodata & \nodata & \nodata\\
``  " & 00 58 35.45 & +01 13 15.7 & \nodata & \nodata & \nodata & \nodata & \nodata & \nodata & 39 & 3.2 & 2.0\tablenotemark{e} & \nodata\\
\hline
COBRA010329.5+004055 & 01 03 29.50 & +00 40 55.0 & 16\tablenotemark{f} & \nodata & \nodata & 22 & $-$1.4 & $-$0.7 & \nodata & \nodata & \nodata & 0.0\\
``  " & 01 03 27.18 & +00 42 16.0 & 23\tablenotemark{f} & \nodata & \nodata & \nodata & \nodata & \nodata & \nodata & \nodata & \nodata & \nodata\\
\enddata
\end{deluxetable*}

\clearpage
\newpage
\mbox{~}

\end{document}